\documentclass[a4paper,11pt]{article}
\pdfoutput=1 

\usepackage{jcappub} 

\usepackage[T1]{fontenc}

\title{The halo model in a massive neutrino cosmology}

\author[a,b]{Elena Massara,}
\author[c,b]{Francisco Villaescusa-Navarro,}
\author[c,b]{Matteo Viel}

\affiliation[a]{SISSA- International School for Advanced Studies, Via Bonomea 265, I-34136 Trieste, Italy}
\affiliation[b]{INFN-National Institute for Nuclear Physics, via Valerio 2, I-34127 Trieste, Italy}
\affiliation[c]{INAF - Osservatorio Astronomico di Trieste, Via G. B. Tiepolo 11, I-34143 Trieste, Italy}

\emailAdd{emassara@sissa.it}
\emailAdd{villaescusa@oats.inaf.it}
\emailAdd{viel@oats.inaf.it}

\abstract{We provide a quantitative analysis of the
  halo model in the context of massive neutrino cosmologies. We
  discuss all the ingredients necessary to model the non-linear matter
  and cold dark matter power spectra and compare with the results of
  N-body simulations that incorporate massive neutrinos. Our neutrino
  halo model is able to capture the non-linear behavior of matter
  clustering with a $\sim 20\%$ accuracy up to very non-linear scales
  of $k=10~h/$Mpc (which would be affected by baryon physics). The
  largest discrepancies arise in the range $k=0.5-1~h/$Mpc where the
  1-halo and 2-halo terms are comparable and are present also in a
  massless neutrino cosmology. However, at scales $k<0.2~h/$Mpc our
  neutrino halo model agrees with the results of N-body simulations at
  the level of 8\% for total neutrino masses of $<0.3$ eV. We also model the neutrino
  non-linear density field as a sum of a linear and clustered
  component and predict the neutrino power spectrum and the cold dark
  matter-neutrino cross-power spectrum up to $k=1~h/$Mpc with $\sim$ 30\%
  accuracy. For masses below 0.15 eV the neutrino halo model captures
the neutrino induced suppression, casted in terms of matter power ratios
between massive and massless scenarios, with a 2\% agreement with the results of  N-body/neutrino simulations.
Finally, we provide a simple application of the halo model: the computation of the 
clustering of galaxies, in massless and massive neutrinos cosmologies, using a simple Halo Occupation Distribution scheme and our halo model extension.}

\begin{document}
\maketitle
\flushbottom

\section{Introduction}

Neutrino oscillation experiments have clearly demonstrated that at
least two of the three neutrino species are massive
\cite{Fogli,Tortola}. Unfortunately, those experiments can only
measure the mass square difference between the different species, and
therefore, they only inform us on the lower bound of the sum of the
neutrino masses. Upper limits on the neutrino masses from particle
physics experiments are yet not very tight, allowing the masses of the
neutrinos to vary within a wide range.

Several cosmological observables have been used as a very powerful
tool to constrain the neutrino masses and number. For instance, the
possibility of massive neutrinos being the main constituents of the
dark matter was soon ruled out from the abundance of low-mass
structures in the universe. Thus, in the current cosmology paradigm
neutrinos are believed to make up a small fraction of the total dark
matter mass in the Universe.

Massive neutrinos affect, at the linear order, the growth of matter
perturbations and the matter-radiation equality time
\cite{LesgourguesPastor,LesgourguesBook,Blas_2014}. These effects
arise mainly due to the neutrino large thermal velocities, that define a free-streaming length $\lambda_{\rm FS}$. Since $\lambda_{\rm FS}$ decreases with time during matter domination, the wavenumber of neutrinos becoming non-relativistic during this epoch presents a minimum, called $k_{\rm nr}$. Thus,
massive neutrinos leave a typical signature in the matter power spectrum: modes with $k<k_{\rm nr}$ evolve like in a pure $\Lambda$CDM cosmology, whereas the others are effected by the neutrinos free-streaming. These scales are probed by different cosmological observables
and are commonly used to put constrains on the neutrino masses
\cite{Hannestad_2003,Reid,Thomas,Swanson,Saito_2010,dePutter,Xia2012,WiggleZ,Zhao2012,Costanzi,Basse,Planck_2013,Costanzi_2014,Wyman_2013,Battye_2013,Hamann_2013,Beutler_2014,Giusarma_2014,Palanque-Delabrouille:2014jca}.

Massive neutrinos also impact on the fully non-linear regime in many
different ways and on different observables like for example:  the matter power spectrum at small scales
\cite{Agarwal2011,Bird_2011,Wagner2012,Viel_2010,Yacine-Bird,Baldi_2013},
the halo-matter bias
\cite{Paco_2013,Castorina_2013,LoVerde_2014a,LoVerde_2014b},
the clustering within dark matter halos
\cite{Ma,Wong,Brandbyge_2008,Brandbyge_haloes,Paco_2011,Paco_2012,LoVerde_2013},
 the evolution of cosmic voids
\cite{Villaescusa-Navarro_2012}, the halo mass function
\cite{Brandbyge_haloes,
  Marulli_2011,Ichiki_Takada,Paco_2013,Castorina_2013,Costanzi_2013},
 redshift-space distortions \cite{Marulli_2011}, Ly$\alpha$ forest
statistical properties of the transmitted flux \cite{Viel_2010,Villaescusa-Navarro_2012,Rossi_2014},
Sunyaev-Zel\'dovich effects in galaxy cluster surveys \cite{Roncarelli_2014}, galaxy clustering
\cite{Fontanot_2014}.

Current constraints on the total neutrino mass range from an upper
limit (2$\sigma$ C.L.) of $\sim$ 0.2-0.3 eV from galaxy clustering
data \cite{riemer}, to $\sim$ 0.2 eV by using cosmic microwave
background data in combination with baryonic acoustic oscillations
\cite{Planck_2013}. Among the different observables the tightest constraints are provided by a combination of large scale structure data that include the cosmic microwave background and the Lyman-$\alpha$ forest.
For example, \cite{2006JCAP...10..014S} obtained a 2$\sigma$ upper limit of 0.17 eV using Sloan Digital Sky Survey quasar spectra which has improved to 0.14 eV using Planck \cite{Costanzi_2014}. A more recent analysis \cite{Palanque-Delabrouille:2014jca}, based on Baryonic Oscillation Spectroscopic Survey BOSS quasar spectra and new simulations that incorporate neutrino induced non-linearities self consistently yields similar results in the range 0.14-0.15 eV (2$\sigma$ C.L. upper limits) by combining Planck and Baryonic Acoustic Oscillations with the Lyman-$\alpha$ flux power measurements of \cite{2013A&A...559A..85P}. It is also interesting to notice that non-zero
neutrino masses are invoked to reconcile
 the tension
between Planck results and other low redshift cosmological probes \cite{Battye_2013,Hamann_2013,Wyman_2013,Beutler_2014}.

The best way to study the impact of massive neutrinos on the mildly
and fully non-linear regime is via N-body simulations. However these
simulations are computationally expensive and thus, the parameter
space can not be fully sampled. The aim of this paper is to extend the
halo model \cite{Sheth-Cooray}, which is a complementary approach to some Perturbation Theories (PT), in order to be able to compute the
matter and cold dark matter power spectrum in massive neutrinos
cosmologies. The main reason to compute the cold dark matter power
spectrum is prompted by the fact it has been recently shown that this
is the fundamental quantity that sets the halo mass function and halo bias
in massive neutrino cosmologies
\cite{Ichiki_Takada,Paco_2013,Castorina_2013,Costanzi_2013}.  We
notice that previous works have tried to extend the halo model to
incorporate the effects of massive neutrinos \cite{Abazajian_2005},
although their results are in great tension with those from the
N-body. 

Having an analytic model allows to get physical insight on massive neutrino cosmologies at non-linear scales. We use the model to understand the typical spoon-shape seen in N-body simulations when computing the ratio between power spectra in massless and massive neutrinos cosmologies. Moreover, we apply it to the study of galaxy clustering.

It is important to note that on small scales baryonic processes are
very important \cite{vandaalen,fedeli_2014a,fedeli_2014b} and can (at
least partially) mimic the neutrino induced effects
\cite{Natarajan}. In this work we do not account for these important
processes since we want the isolate the effects of massive neutrinos
w.r.t. the same simulation set-up without massive neutrinos. We thus
caution the reader that on small scales the matter power spectrum has
to be modeled more carefully by incorporating baryonic physical
processes, e.g. galactic feedback, especially in view of future
missions like Euclid.

This paper is organized as follows. In section~\ref{sec:N-body} we
present the set of N-body simulations run for this work. In
section~\ref{sec:halo_model} we review the standard halo model, which
is capable of predicting, with high accuracy, the matter power spectrum
in cosmologies with massless neutrinos. Our extension of the halo
model to incorporate cosmologies with massive neutrinos is presented
in section~\ref{sec:halo_model_nu}, where we compare the results of
our extended halo model against N-body simulations. In
section~\ref{sec:ratio} we use the halo model to explain the small
scale features present in the matter power spectrum of cosmologies
with massive neutrinos. In section~\ref{sec:galaxy} we present the
galaxy clustering predicted by halo model, once a Halo Occupation
Distribution (HOD) framework has been considered, and we compare it
with measurements. Finally, the conclusions of the present
work are summarized in section~\ref{sec:conclusions}.

\section{N-body simulations}
\label{sec:N-body}

In this section we describe the suite of simulations performed. 

We have run a total of 8 large box-size N-body simulations using the
TreePM code {\sc GADGET-III} \cite{Springel_2005}. Our simulations
follow the evolution of $512^3$ CDM particles and $512^3$ neutrino
particles (only for the massive neutrinos cosmologies). The values of
the cosmological parameters are in agreement with the latest results
found by the Planck collaboration \cite{Planck_2013}: $\Omega_{\rm
  m}=\Omega_{\rm cdm}+\Omega_{\rm b}+\Omega_\nu=0.3175$, $\Omega_{\rm
  b}=0.049$, $\Omega_\Lambda=0.6825$, $h=0.6711$, $n_{\rm s}=0.9624$. We
have run simulations for four different cosmologies with different
neutrino masses (we assume three degenerate neutrino families): $\sum
m_\nu=0.00$, 0.15, 0.30 and 0.60 eV. In our simulations the value of
$\Omega_{\rm m}$ and $\Omega_{\rm b}$ is fixed, whereas the values of
$\Omega_{\rm cdm}$ and $\Omega_\nu$ depend on the neutrino masses such
as $\Omega_{\rm cdm}=\Omega_{\rm m}-\Omega_{\rm b}-\Omega_\nu$. For
each cosmological model we have run two simulations with two different
box sizes, 1000 Mpc$/h$ and 200 Mpc$/h$, in order to sample a wide
range in wave numbers. The softening length of both the CDM and
neutrino particles is set to $1/40$ of the mean linear inter-particle
distance. The initial conditions have been generated at $z=99$ by
displacing the particle positions according to the Zel'dovich
approximation. The transfer function used for the CDM field is a mass
weighted average between the transfer functions of the CDM and
baryons. The name of our simulations arises from its size (L for 1000
Mpc$/h$ and S for 200 Mpc$/h$) and from the masses of the neutrinos
(0 for 0.0 eV, 15 for 0.15 eV and so on). For instance, the simulation
L30 the is the simulation with $\sum m_\nu=0.3$ eV neutrinos and
box-size equal to 1000 Mpc$/h$. A summary of our simulation suite is
shown in table~\ref{tab:i}.

For each simulation we have computed the CDM power spectrum, the neutrino power spectrum, the CDM-neutrino cross-power spectrum and the total matter power spectrum. The amplitude of the neutrino and the total matter power spectrum has been corrected to account for the shot-noise associated to the neutrino density field.

\begin{table}
\begin{center}
\begin{tabular}{| c | c | c | c | c | c | c | c | c | c |}
\hline
Name & Box Size & $\sum m_\nu$ & $\Omega_{\rm m}$ & $\Omega_\Lambda$ & $\Omega_{\rm b}$ & $\Omega_\nu$ & $h$ & $n_s$ & $\sigma_8$\\
&  (${\rm Mpc}/h$) & (eV) & & & & & & & (z=0)\\
\hline
L0 & 1000 & 0.00 & 0.3175 & 0.6825 & 0.049 & 0.00000 & 0.6711 & 0.9624 & 0.834 \\
\hline
S0 & 200 & 0.00 & 0.3175 & 0.6825 & 0.049 & 0.00000 & 0.6711 & 0.9624 & 0.834 \\
\hline
L15 & 1000 & 0.15 & 0.3175 & 0.6825 & 0.049 & 0.00354 & 0.6711 & 0.9624 & 0.801 \\
\hline
S15 & 200 & 0.15 & 0.3175 & 0.6825 & 0.049 & 0.00354 & 0.6711 & 0.9624 & 0.801 \\
\hline
L30 & 1000 & 0.30 & 0.3175 & 0.6825 & 0.049 & 0.00708 & 0.6711 & 0.9624 & 0.764 \\
\hline
S30 & 200 & 0.30 & 0.3175 & 0.6825 & 0.049 & 0.00708 & 0.6711 & 0.9624 & 0.764 \\
\hline
L60 & 1000 & 0.60 & 0.3175 & 0.6825 & 0.049 & 0.01415 & 0.6711 & 0.9624 & 0.693 \\
\hline
S60 & 200 & 0.60 & 0.3175 & 0.6825 & 0.049 & 0.01415 & 0.6711 & 0.9624 & 0.693 \\
\hline
\end{tabular}
\caption{\label{tab:i}Names and values of the cosmological parameters of our N-body simulation set.}
\end{center}
\end{table}

\section{Halo model in pure $\Lambda$CDM cosmology}
\label{sec:halo_model}

In this section we briefly review the halo model \cite{Sheth-Cooray}, as it was built for cosmologies without massive neutrinos.

Simulations showed that the initial smooth dark matter field evolves in a network of filaments and knots, which are highly non-linear.
The halo model provides a description of the statistical properties of this evolved dark matter field, assuming that all the matter is bound up in isolated knots, called halos.
Let us call $\vec{x}_i$ the centers of these halos. Then, the matter density at position $\vec{x}$ is given by summing up the contribution from each halo
\begin{eqnarray}
\label{eq:1}
\rho(\vec{x})&=&\sum_i  \, \rho(\vec{x}-\vec{x}_i | M_i )\\
\label{eq:2}
&=& \sum_i \int dM \, \delta(M-M_i) \int d^3x' \, \delta^3(\vec{x}'-\vec{x}_i)\; M \, u(\vec{x}-\vec{x}'|M) \,,
\end{eqnarray} 
where $\rho(\vec{x}-\vec{x}_i | M_i )$ is the density around the $i-$th halo and we have assumed that it depends only on the mass $M_i $ contained in the halo, whereas $u(\vec{x}-\vec{x}_i | M_i )\equiv\rho(\vec{x}-\vec{x}_i | M_i )/M_i$ is the normalized profile. 
\newline
Let us consider the matter density contrast, which is defined as $\delta(\vec{x})=\rho(\vec{x})/ \bar{\rho}-1$, where $\bar{\rho}$ is the comoving background matter density, and the power spectrum, which is the Fourier transform of the two-point correlation function $\langle \delta (\vec{x}_1) \delta (\vec{x}_2)\rangle$, with the average taken over the ensemble. The fully non-linear matter power spectrum predicted by the halo model is given by the sum of two terms
\begin{equation}
\label{eq:3}
P(k)=P_{1h}(k)+P_{2h}(k)\,.
\end{equation}
The 1-halo term, $P_{1h}(k)$, counts for the correlations between particles that belong to the same halo and dominates on small scales, whereas the 2-halo term, $P_{2h}(k)$, describes the correlation between particles in different halos and becomes important on large scales.
Since the comoving number density of halos of mass $M$, per mass unit, at redshift $z$ is defined as
\begin{equation}
\label{eq:4}
\left\langle \sum_i  \;\delta(M-M_i)\;\delta^3(\vec{x}'-\vec{x}_i) \right\rangle \equiv n(M,z) \, ,
\end{equation}
and we assume a spherically symmetric profile $u(\vec{x}-\vec{x}_i | M_i )=u(r_i | M_i )$, the 1- and 2-halo terms are
\begin{eqnarray}
\label{eq:5}
P_{1h}(k,z)&=&\int_0^\infty dM \, n(M,z) \left(\frac{M}{\bar{\rho}}\right)^2|u(k|M)|^2\\
\label{eq:6}
P_{2h}(k,z)&=&\int_0^\infty dM'\, n(M',z) \,\frac{M' }{\bar{\rho}}\, u(k|M')\\\nonumber
&& \quad \times \int_0^\infty dM'' \, n(M'',z) \,\frac{M'' }{\bar{\rho}}\, u(k|M'') \, P_{hh}(k|M',M'',z),
\end{eqnarray}
where $P_{hh}(k|M',M'',z)$ is the power spectrum of halos of mass $M'$ and $M''$ and $u(k|M)$ is the Fourier transform of the normalized profile
\begin{equation}
\label{eq:7}
u(k|M)=\int_0^{R_v} dr\, 4\pi r^2 \,\frac{\sin(kr)}{kr}\,u(r|M)\,.
\end{equation}
The cut-off $R_v$ is called virial radius and it is the comoving radius of the spherical region containing the halo mass $M=4 \pi R_v^3\, \Delta_v  \bar{\rho}/3$ with average comoving density $\Delta_v \bar{\rho}$, where the virial overdensity $\Delta_v$ is determined by the cosmology~\cite{Bryan}
\begin{eqnarray}
\label{eq:8}
\Delta_v &=&\frac{18 \pi^2 +82 x-39 x^2}{1+x} \\[0.4 cm]
\label{eq:8b}
x &\equiv & \Omega(z) -1 \\[0.4 cm]
\label{eq:8c}
\Omega(z)&=&\frac{\Omega_{\rm c}(1+z)^3}{\Omega_{\rm c}(1+z)^3+\Omega_{\Lambda}}
\end{eqnarray}
and $\Omega_{\rm c} $ is the cold dark matter plus baryons energy density
today.  The average density profile of cold dark matter halos has been
extensively studied and it appears to be universal over a wide ranges
of masses. Up to now, the fitting formula that better reproduces the
density around halos in simulations is the Navarro-Frank-White (NFW)
profile~\cite{NFW}
 \begin{equation}
\label{eq:9}
u(r|M)=\frac{F/4\pi}{r(r+r_s)^2}.
\end{equation}
The parameter $r_s^3= 3M/(4\pi c^3 \Delta_v\bar{\rho})$ defines a characteristic radius which is a function of the halo mass $M$; $F=1/[\ln(1+c)-c/(1+c)]$, where $c=R_v/r_s$ is the concentration parameter. Simulations show that, for fixed halo mass and redshift, there is a distribution of concentrations which is well described by a log-normal distribution~\cite{Jing-c_distribution} with variance that does not depend on the halo mass and a mean value~\cite{Bullock-c_meanvalue}
\begin{equation}
\label{eq:10}
c(M,z)=9\left(\frac{M}{M_{\star}(z)}\right)^{-0.13},
\end{equation}
where $M_{\star}(z)$ is the characteristic non-linear mass scale defined such that $\nu=1$, with $\nu$ defined below. Note also that the NFW profile goes like $r^{-3}$ at large radii, therefore the mass within it diverges. In order to have a finite halo mass $M$, the profile has to be truncated at the virial radius $R_v$. 

Halos form from regions in the initial density field which were
sufficiently dense to collapse. We need first to estimate the number
density $n(M)$ of objects of mass $M$.  The peak height $\nu$ is
defined as
\begin{equation}
\label{eq:11}
\nu=\frac{\delta^2_{sc}}{\sigma^2(M,z)},
\end{equation}
where $\delta_{\rm sc}\sim1.686$ is the critical density for having the spherical collapse today and $\sigma^2(M,z)$ is the variance of the linear density field 
\begin{equation}
\label{eq:12}
\sigma^2(M,z)=\int_0^\infty \frac{dk}{2\pi^2} k^2W^2(kR)\, P^L(k,z)\, ,
\end{equation}
when smoothed with a top-hat filter at scale $R$. Here, $W(x)=(3/x^3)[\sin(x)-x \cos(x)]$ is the Fourier Transform of the filter and $P^L(k,z)$ is the linear power spectrum at redshift $z$. The relation between the smoothing scale and the mass is dictated by the choice of the filter function; for a top-hat filter it is given by
\begin{equation}
\label{eq:13}
M=\frac{4}{3}\pi\bar{\rho} R^3\, ,
\end{equation}
which shows that in this case $M$ is actually the mass of the region in the Lagrangian space with radius $R$ that collapses in a halo with same mass and radius $R_v$ in the evolved field. 
\newline
Since there is a deterministic relation between $M$, $R$, $\sigma(M,z)$ and $\nu$, the number density $n(M,z)$ can be expressed in terms of the peak height $\nu$ as
\begin{equation}
\label{eq:14}
n(M,z) \,dM =\frac{\bar{\rho}}{M}f(\nu)\,d\nu\, ,
\end{equation}
where the mass function $f(\nu)$ is a universal function of $\nu$, i.e. independent of redshift and the shape of the initial power spectrum. For what follows we will use the Sheth-Tormen (ST) mass function~\cite{ST}, which provides a good fit to the number density of halos in simulations.

Moreover, on large scales, where the 2-halo term dominates, the halo-halo power spectrum $P_{hh}(k|M',M'',z)$ in \eqref{eq:6} can be expressed in terms of the linear halo bias $b(M,z)$ with respect to the matter density field: 
\begin{equation}
\label{eq:15}
P_{hh}(k|M',M'',z) = b(M',z) b(M'',z) P^L(k,z)\, .
\end{equation}
\begin{figure}[tbp]
\centering 
\includegraphics[width=.48\textwidth,clip]{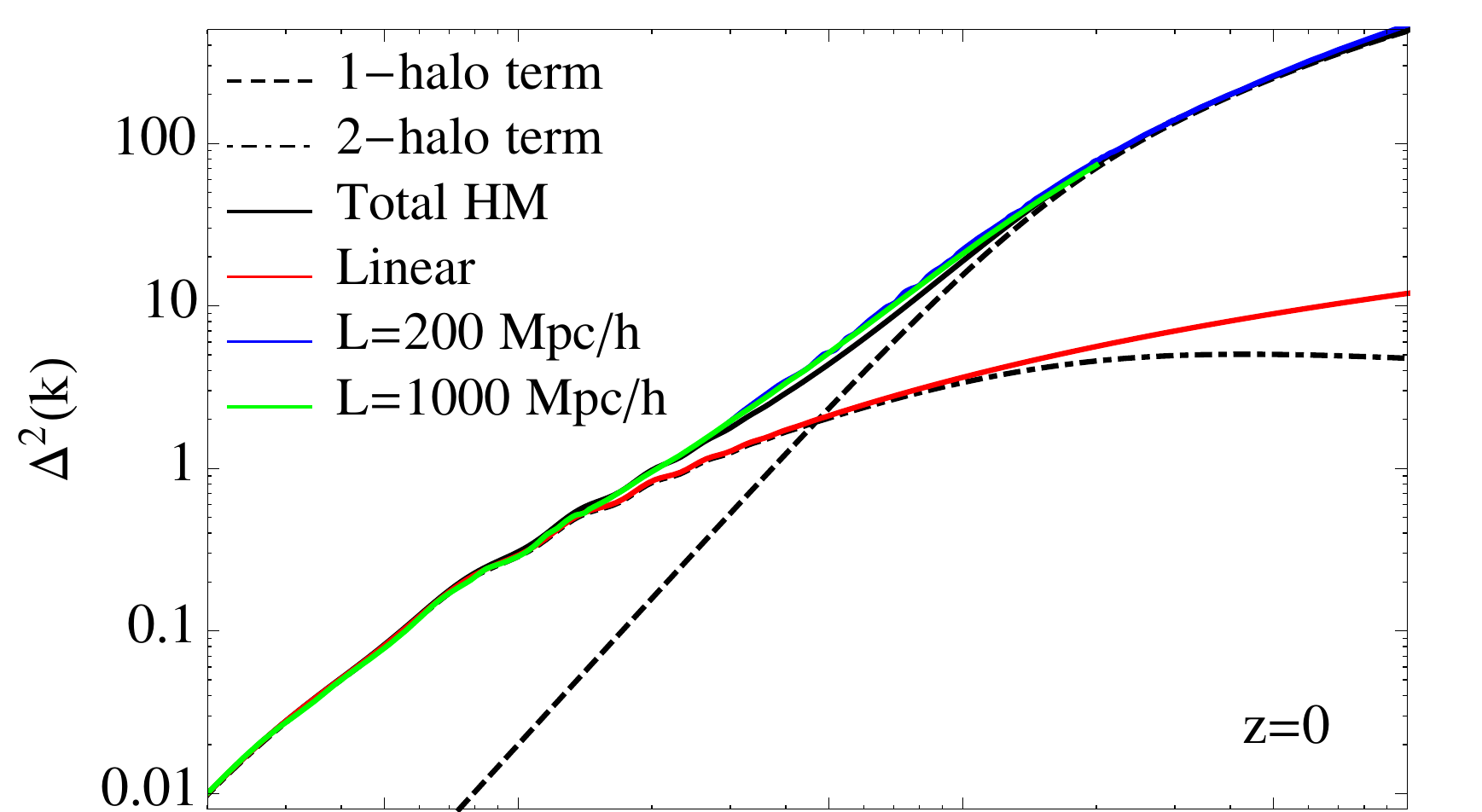}
\includegraphics[width=.48\textwidth,origin=c]{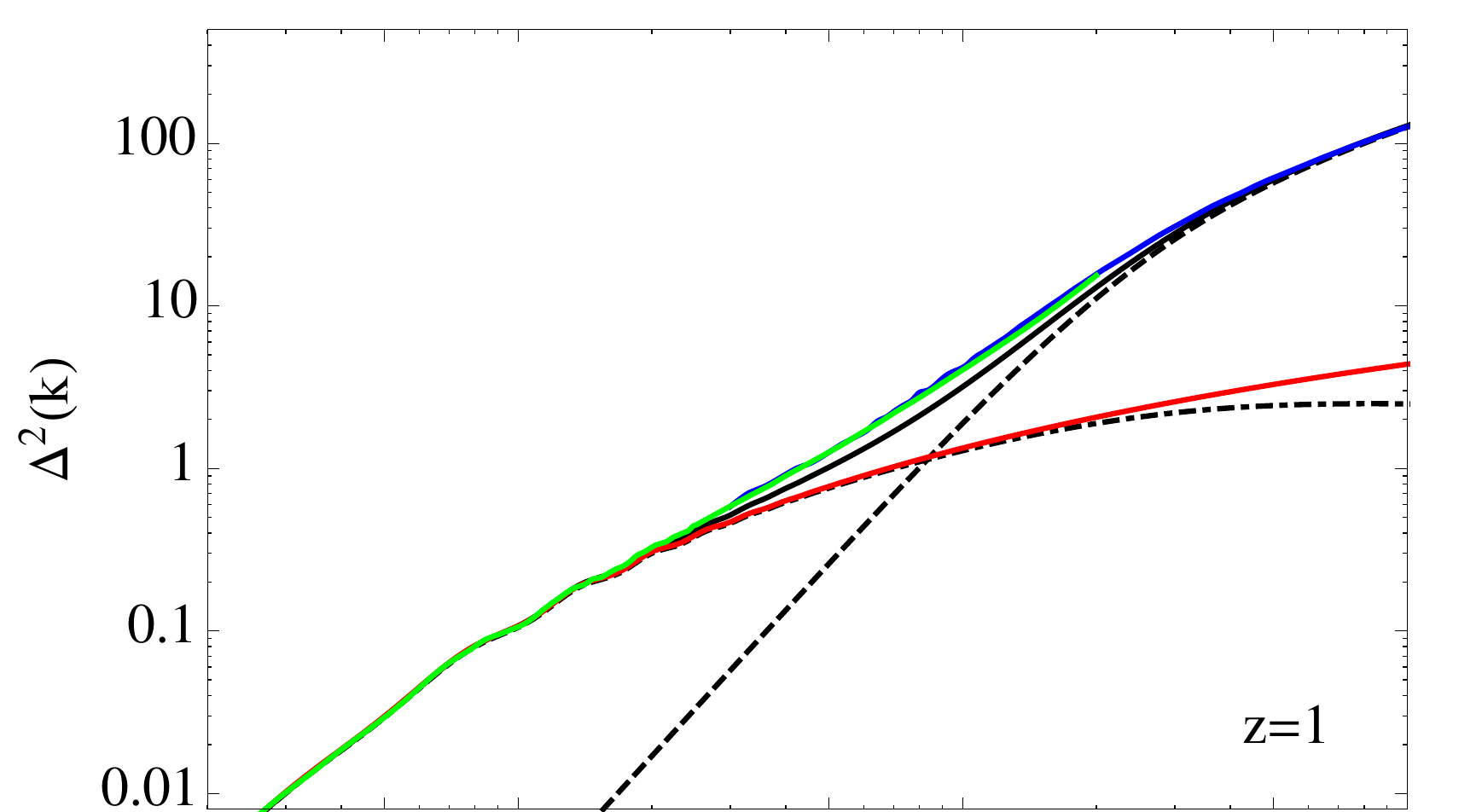}
\includegraphics[width=.48\textwidth,clip]{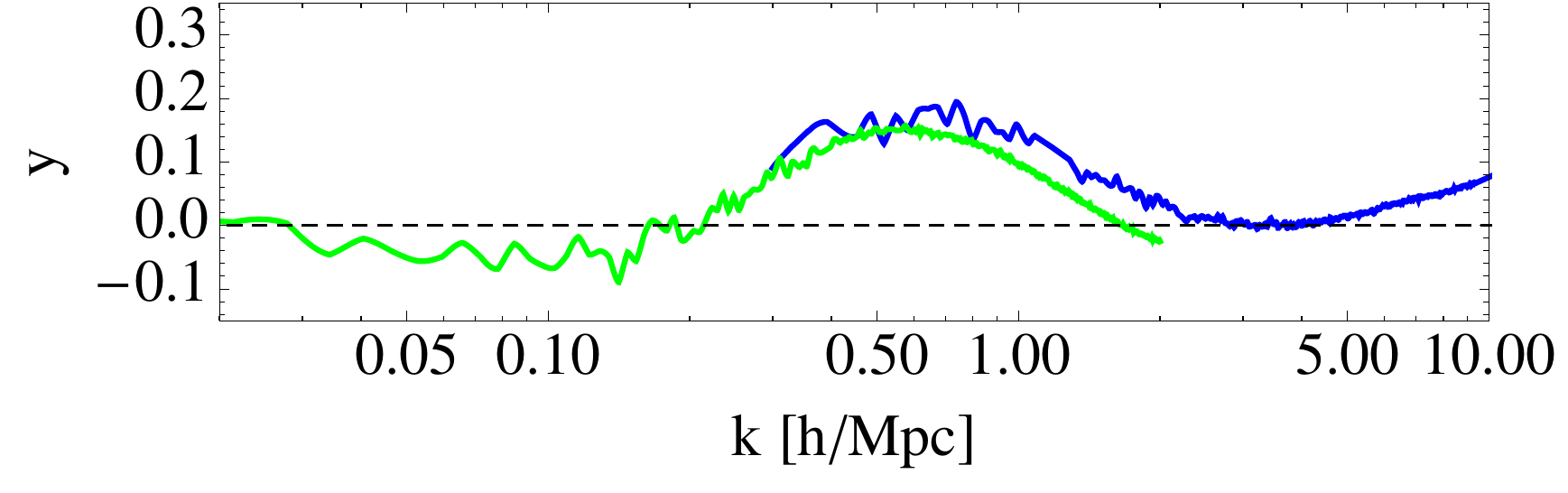}
\includegraphics[width=.48\textwidth,origin=c]{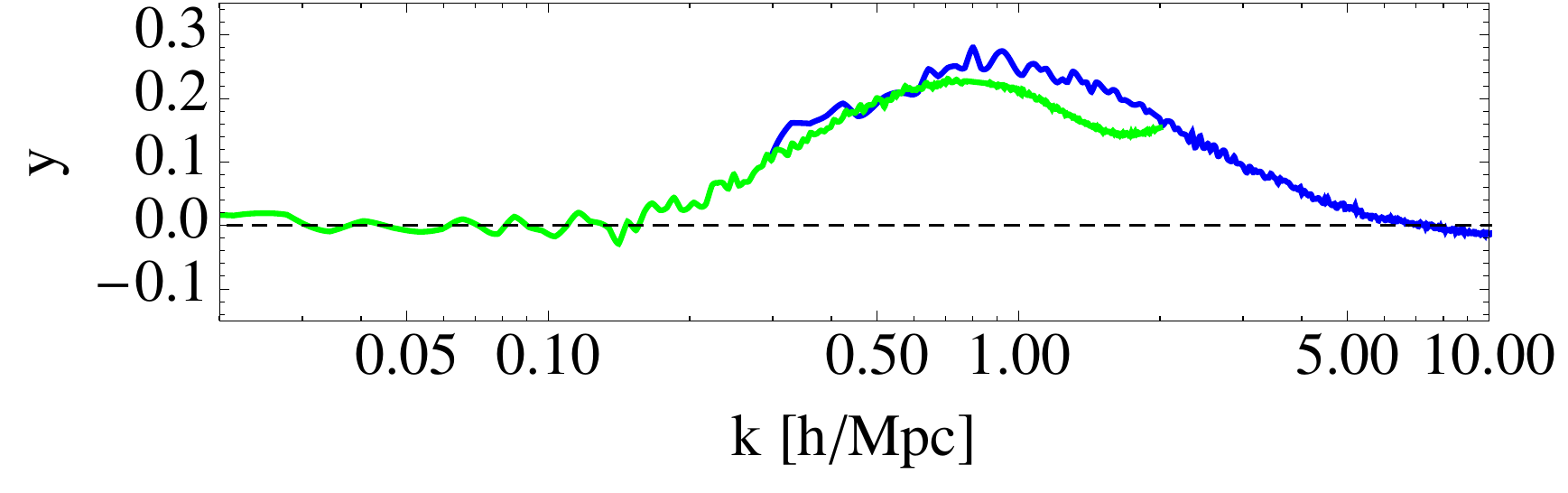}
\caption{\label{fig:1} Matter power spectrum in a $\Lambda$CDM cosmology. The left and right panels display results at redshifts $z=0$ and $z=1$, respectively. Black lines show the matter power spectrum as computed from the halo model: the dashed line is the 1-halo term, the dot-dashed one is the 2-halo term and the solid one is the sum of the two terms. Red lines show the linear predictions whereas blue and green lines are the results from N-boby simulations with box size $L=200$ Mpc/$h$ and $L=1000$ Mpc/$h$, respectively. The bottom panels show the relative difference between the power spectra from the halo model and from N-body simulations.}
\end{figure}
Therefore, using \eqref{eq:14} and \eqref{eq:15} we can rewrite \eqref{eq:5} and \eqref{eq:6} as 
\begin{eqnarray}
\label{eq:16}
P_{1h}(k,z)&=&\int_0^\infty d\nu f(\nu) \frac{M}{\bar{\rho}} \left|u(k|\nu)\right|^2\, ,\\
\label{eq:17}
P_{2h}(k,z)&=&\left[\int_0^\infty d\nu f(\nu) b(\nu)\, u(k|\nu)\right]^2 P^L(k,z)\, .
\end{eqnarray}
The ST mass function and the halo bias are normalized so that $\int_0^\infty d\nu f(\nu) b(\nu)=1$, and from \eqref{eq:9} is easy to show that $u(k\rightarrow 0,M) =\rho(k\rightarrow 0,M) /M=1$. Therefore, here the 2-halo term tends to the linear power spectrum as $k$ goes to zero, $P_{2h}(k\rightarrow 0)\rightarrow P^L(k)$, whereas they differ at high $k$ where the halo profile contributes and it is $k$-dependent. 
\newline
Now we have all the ingredients to compute the non-linear power spectrum of matter at any redshift. We consider a massless neutrinos, flat $\Lambda$CDM cosmology, with the same cosmological parameters as the N-body simulations L0 and S0 (see table~\ref{tab:i}). We use the {\sc CAMB} code~\cite{CAMB} to calculate the linear matter power spectrum $P^L(k)$. Next,  we compute the power spectrum using the halo model (HM) and we compare it with the results of the N-body simulations (S) through  
\begin{equation}
\label{eq:18}
y(k)=\frac{\Delta^2_{S}(k)-\Delta^2_{HM}(k)}{\Delta^2_{S}(k)}\, ,
\end{equation} 
where $\Delta(k)=k^3 P(k)/(2 \pi^2)$ is the dimensionless matter power spectrum. The results are presented in figure~\ref{fig:1}. The top panels show the halo model power spectrum for the pure $\Lambda$CDM cosmology, at redshift $z=0$ (left) and $z=1$ (right). The bottom panels show explicitly the comparison between the halo model and N-body simulations through the quantity $y$.
At large scales, $k<0.2\, h$/Mpc, the halo model reproduces well the prediction of simulations,  whereas at intermediate scale, $k\sim 0.2-2\,h$/Mpc, there is a disagreement below the $20\%$ level at $z=0$, and around the $20-30\%$ level at $z=1$. This region is characterized by the transition between the 1-halo and 2-halo terms' dominance and here the halo model seems not to be very accurate. On smaller scales, up to $k\sim10\, h$/Mpc, the agreement is again better than $10\%$.

\section{Halo model with massive neutrinos}
\label{sec:halo_model_nu}

In this section we discuss how to extend the standard halo model to account for the effects of massive neutrinos. From now on the quantities without subscript are related to the total matter field, whereas "${\rm c}$" denotes the cold field, which is the mass weighted average between the cold dark matter and the baryon fields and "$\nu$" indicates the neutrino component only. 

We consider three different massive neutrino cosmologies, characterized by the same amount of total matter but different neutrino masses (we have considered 3 degenerate families), which correspond to the three massive neutrino cosmologies presented in table~\ref{tab:i}.

In order to describe the neutrino density field it is important to account for the fact that massive neutrinos have large, although non-relativistic, thermal velocities at low redshift. These large thermal velocities, which set a free-streaming scale, prevent the clustering of neutrinos within dark matter halos. However, neutrinos from the low-velocity tail of the momentum distribution can cluster within the potential wells of CDM halos \cite{Ma,Wong,Paco_2011,Brandbyge_haloes,Paco_2012,LoVerde_2013}. Thus, it is useful to describe the neutrino density field as the sum of two terms
\begin{equation}
\label{eq:18a}
\delta_{\nu}=F_h\delta_{\nu}^{h}+(1-F_h)\delta_{\nu}^L,
\end{equation}
a linear one, $\delta_{\nu}^L$, and a non-linear one, $\delta_{\nu}^{h}$, which is a fraction $F_h$ of the total neutrino density field. Whereas the former will simply obey linear theory, the latter account for the fully non-linear clustering of massive neutrinos within ${\rm c}$-halos (CDM halos), forming neutrino halos ($\nu$-halos). This approach takes into account also non-linearities unlike the approach of \cite{Yacine-Bird} (see their equation 64), in which only the linear neutrino clustering is considered. The two descriptions agree up to $k \sim 0.2\, h$/Mpc and there is a 50\% extra clustering at $k\sim 0.5$ in our case due to the non-linear behavior. We also assume that the mass of the $\nu$-halos is only a function of the mass of the host ${\rm c}$-halos, $M_{\nu}=M_{\nu}(M_{c})$, and that the centers of the $\nu$- and ${\rm c}$-halos are the same. The density contrast of the total matter density field can then be written as 
\begin{equation}
\label{eq:18b}
\delta=\frac{\bar{\rho}_{\rm c}}{\bar{\rho}}\delta_{\rm c}+\frac{\bar{\rho}_{\nu}}{\bar{\rho}}\left[F_h\delta_{\nu}^{h}+(1-F_h)\delta_{\nu}^L\right]\, ,
\end{equation}
where $\rho=\bar{\rho}_{\rm c}+\bar{\rho}_{\nu}$ is the mean background matter density. The matter power spectrum is given by
\begin{equation}
\label{eq:19}
P(k)=\left(\frac{\bar{\rho}_{\rm c}}{\bar{\rho}}\right)^2P_{\rm c}(k)+2\,\frac{\bar{\rho}_{\rm c}\bar{\rho}_{\nu}}{\bar{\rho}^2}\,P_{{\rm c}\nu}(k) +\left(\frac{\bar{\rho}_{\nu}}{\bar{\rho}}\right)^2P_{\nu}(k)\, ,
\end{equation}
where $P_{\rm c}(k)$, $P_{\nu}(k)$ and $P_{{\rm c}\nu}(k)$ are the cold, neutrino and cross power spectra, respectively. 

Before presenting in detail the calculation of all these terms, we discuss the recipe to compute the mass function and the linear halo bias in a massive neutrino cosmology, which is not obvious a priori. From now on, we will not write explicitly the redshift's dependence since it can be understood from the description of the massless neutrinos $\Lambda$CDM case presented in section~\ref{sec:halo_model}.

\subsection{Matter vs. cold dark matter prescription}

Since the $\nu$-halos are located around ${\rm c}$-halos and, as we will clarify
later, their mass can be assumed to be a function of the corresponding
${\rm c}$-halos mass, there are two important consequences: their mass
function is equal to the one of the cold field,
$dM_{\nu}n(M_{\nu})=dM_{\rm c}n(M_{\rm c})$, and the linear $\nu$-halo bias is
equal to the ${\rm c}$-halo one $b(M_{\nu})=b(M_{\rm c})$. In order to make the
halo model machine working, we must express $n(M_{\rm c})$ and $b(M_{\rm c})$ in
terms of the number of regions in the Lagrangian field that are dense
enough to collapse, i.e. we have to recast it in terms of the peak
height. There are two different ways in which we can do that.

It would be natural, following the procedure adopted for the $\Lambda$CDM case, to rewrite the number density of c-halos in terms of total matter quantities,
\begin{equation}
\label{eq:23}
n(M_{\rm c})\, dM_{\rm c} =\frac{\bar{\rho}}{M}f(\nu)d\nu\, ,
\end{equation}
and the peak height as $\nu=\delta_{ sc}^2/\sigma^2(M)$, with
\begin{eqnarray}
\label{eq:24}
M&=&M_{\nu}+M_c=\frac{4}{3}\pi\bar{\rho} R^3\, ,   \\ 
\label{eq:25}
\sigma^2(M) &=& \int_0^\infty \frac{dk}{2\pi^2} k^2W^2(kR)\, P^L(k), 
\end{eqnarray}
and $P^L(k)$ being the linear total matter power spectrum. It would be natural to write the halo-halo power spectrum in terms of the halo bias $b(M_{\rm c})$ with respect to the total matter density field. This approach is the so called matter prescription \cite{Paco_2013,Castorina_2013,Costanzi_2013}. 

Even if used in the literature (e.g. in~\cite{Abazajian_2005}), this
prescription has been shown to be not fully correct by Castorina et
al.~\cite{Castorina_2013} (see also~\cite{Ichiki_Takada}), since the
resulting mass function $f(\nu)$ is not universal and the resulting
linear halo bias $b(M_{\rm c})$ is scale dependent even on large scales. The
authors argued that this is due to the wrong choice of the density
field used for computing the peak height and the halo bias, i.e. the
total matter is not the fundamental density field involved in the
clustering process. They showed that the more physical field is the
cold one and this choice goes under the name of cold dark matter
prescription. In this setup, the number density of ${\rm c}$-halos is
\begin{equation}
\label{eq:30}
n(M_{\rm c})dM_{\rm c} =\frac{\bar{\rho}_{\rm c}}{M_{\rm c}}f(\nu_{\rm c})d\nu_{\rm c},
\end{equation}
where $\nu_{\rm c}=\delta_{sc}^2/\sigma_{\rm c}^2$ with
\begin{eqnarray}
\label{eq:31}
M_{\rm c}&=&\frac{4}{3}\pi\bar{\rho}_{\rm c} R^3\, ,  \\ 
\label{eq:32}
\sigma_{\rm c}^2\equiv\sigma^2(M_{\rm c})&=& \int_0^{\infty} \frac{dk}{2\pi^2} k^2W^2(kR)\, P^L_{\rm c}(k)\, ,
\end{eqnarray}
and $P^L_{\rm c}(k)$ is the linear cold power spectrum. Moreover, this
prescription allows to express the halo-halo power spectrum in terms
of the linear cold $P^L_{\rm c}(k)$
\begin{equation}
\label{eq:33}
P_{hh}(k|M'_{\rm c},M''_{\rm c}) = b(M'_{\rm c}) b(M''_{\rm c}) P^L_{\rm c}(k)\, .
\end{equation}
Castorina et al.~\cite{Castorina_2013} showed that the resulting mass function is universal and the linear halo bias $b(M_{\rm c})$ is scale independent on large scales, as wanted. Therefore, we will use the cold dark matter prescription to build the halo model for a massive neutrino cosmology. The fact that this is the correct prescription tells us that neutrinos modify only the background in which the ${\rm c}$-field clusters, without performing any back-reaction through its density perturbations. 

Since the fraction $F_h$ of neutrinos clustered in ${\rm c}$-halos is very small, we expect that the total matter power spectrum will be well reproduced considering all the neutrinos, both linear and clustered, as driven by linear theory. Anyway, the neutrino and cross power spectra from N-body simulations are not well reproduced by the correspondent linear one, on small scales. Therefore, we will describe how to model not only the clustering of the cold field but also the clustering of neutrinos within the halo model formalism. 
\subsection{Cold dark matter Power Spectrum}
\label{sec:cold}
\begin{figure}[!tbp]
\centering 
\includegraphics[width=.45\textwidth,clip]{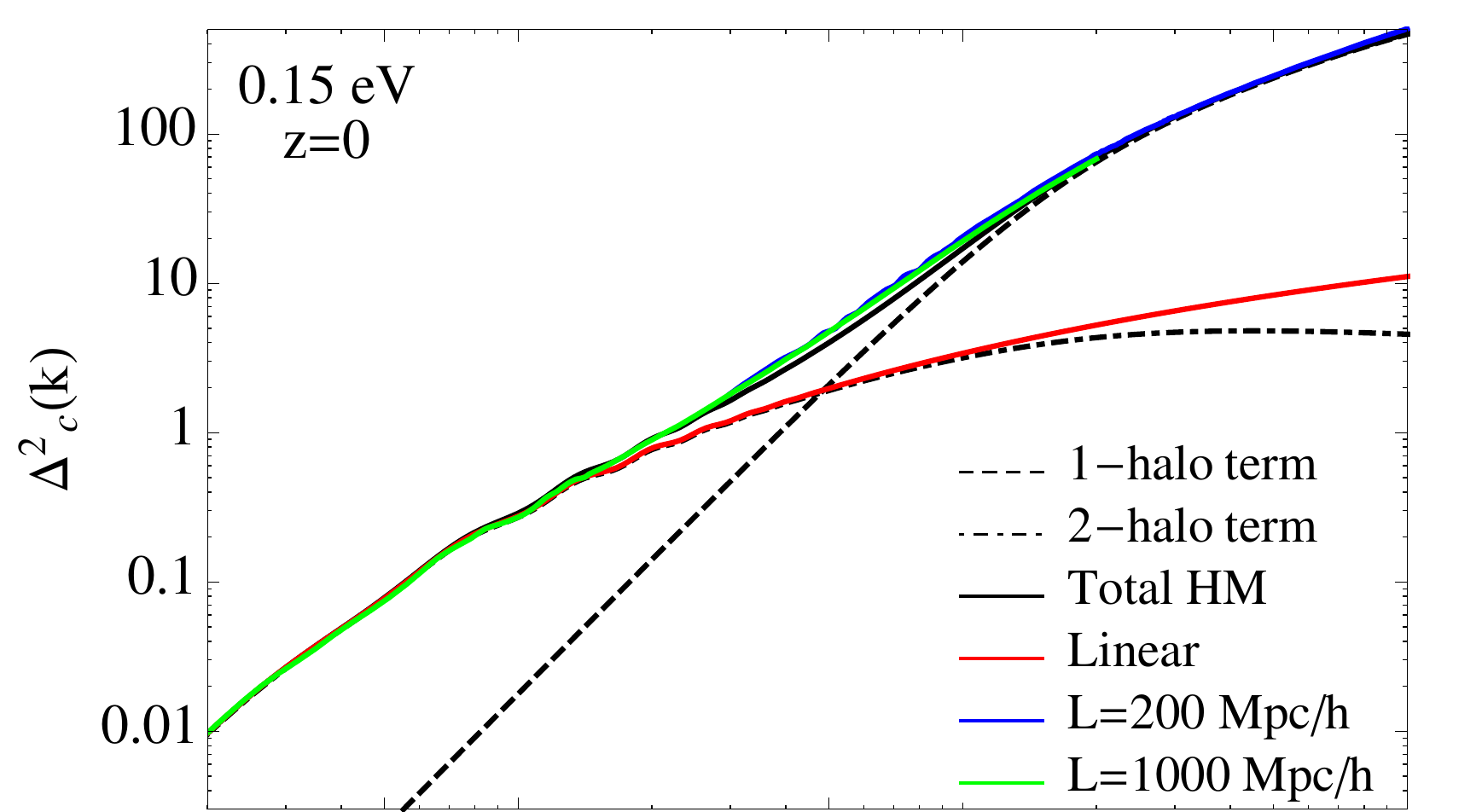}
\includegraphics[width=.45\textwidth,origin=c]{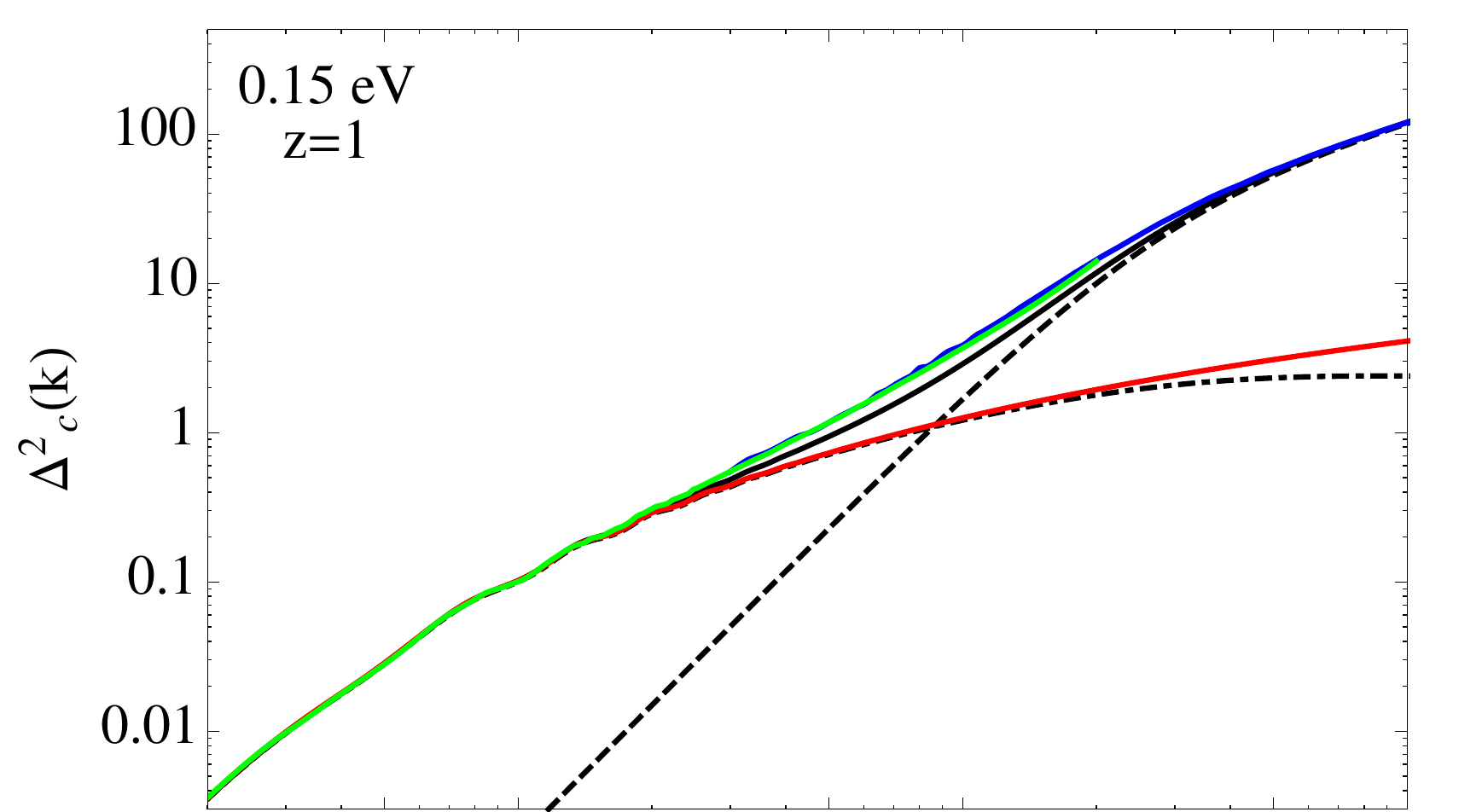}
\includegraphics[width=.45\textwidth,clip]{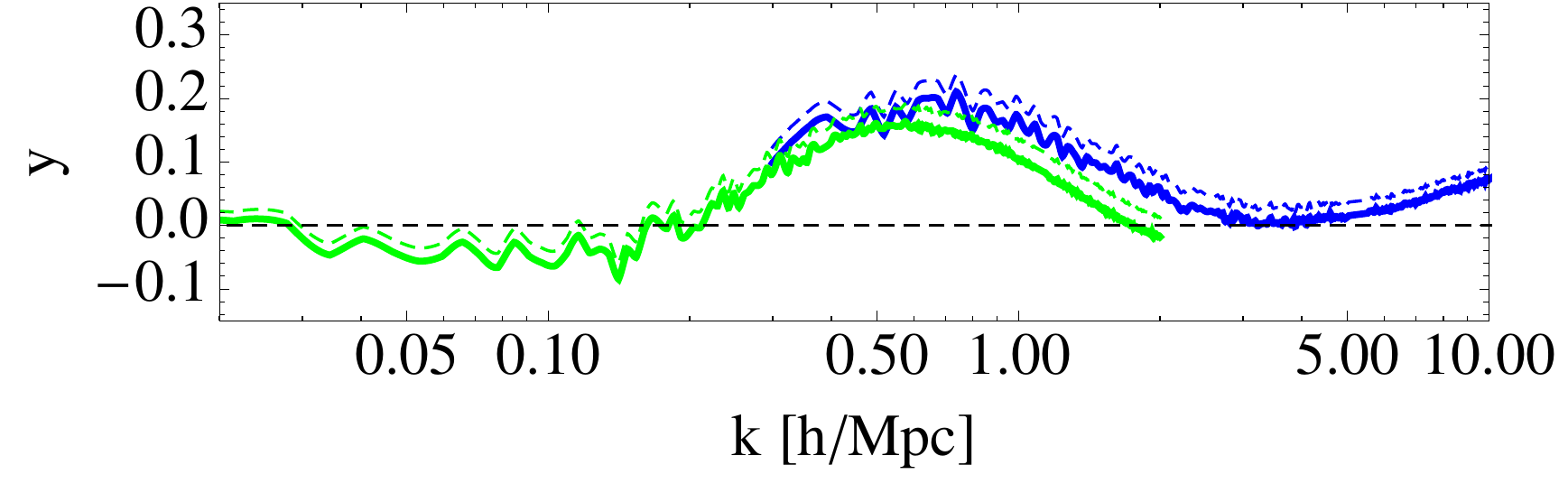}
\includegraphics[width=.45\textwidth,origin=c]{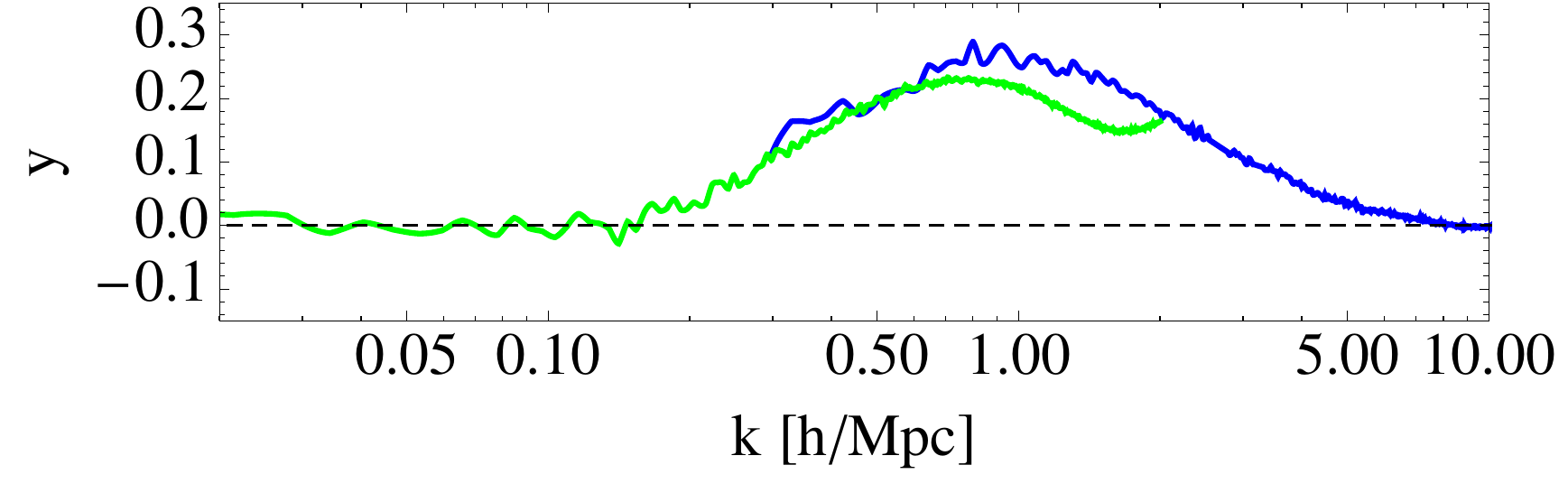}
\includegraphics[width=.45\textwidth,clip]{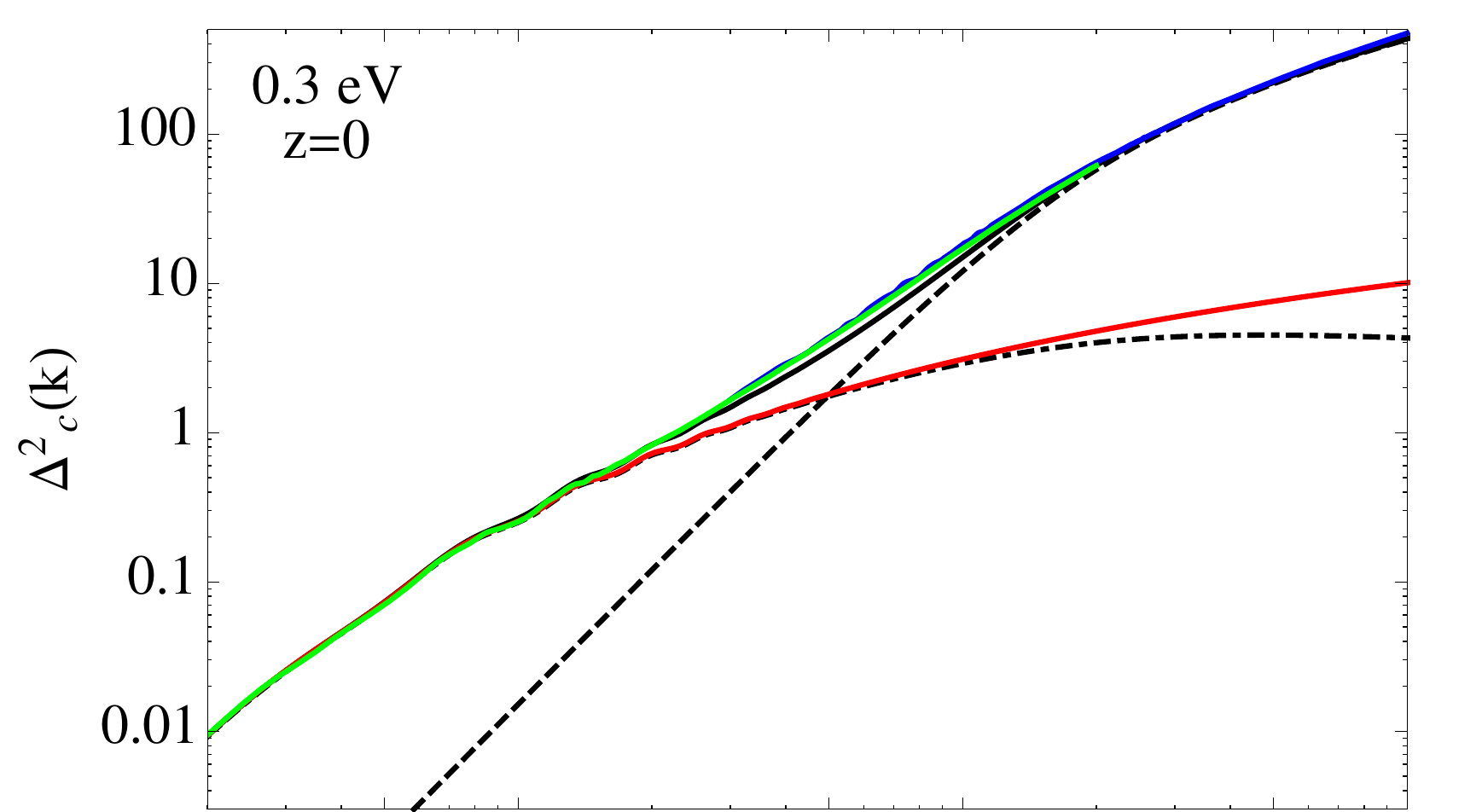}
\includegraphics[width=.45\textwidth,origin=c]{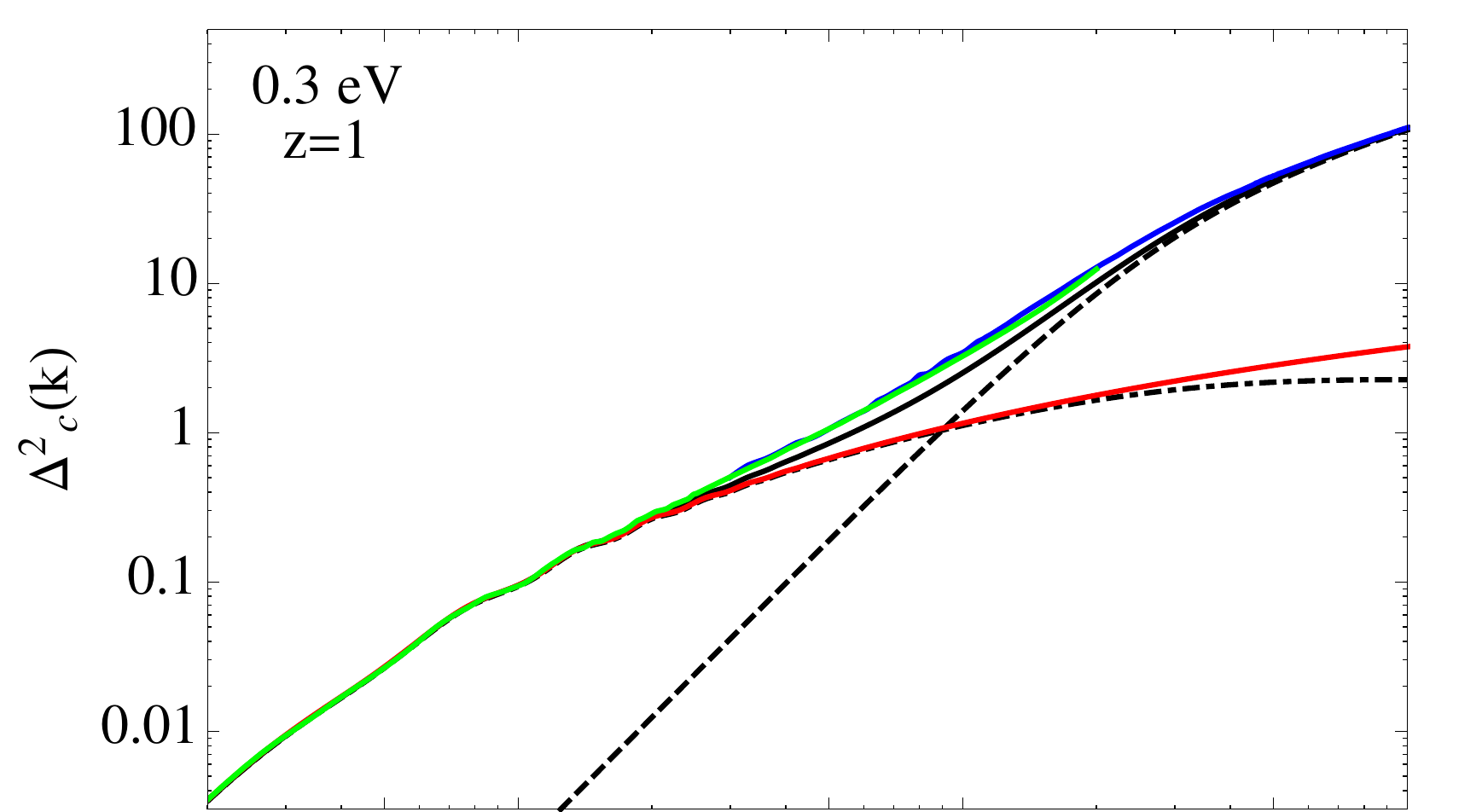}
\includegraphics[width=.45\textwidth,clip]{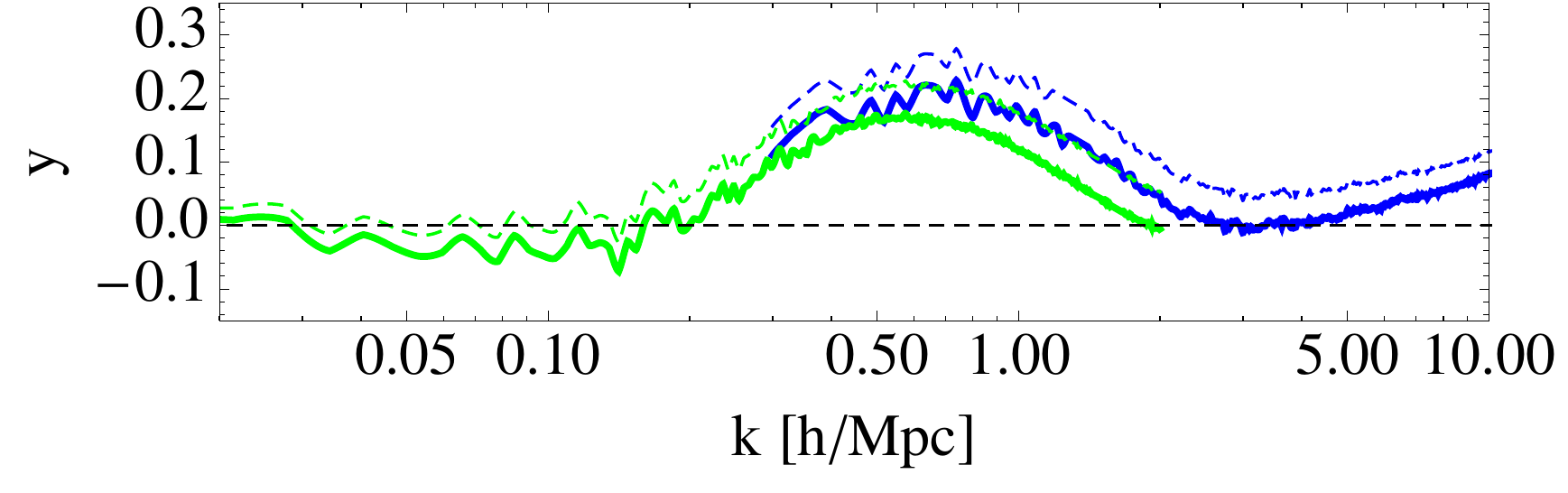}
\includegraphics[width=.45\textwidth,origin=c]{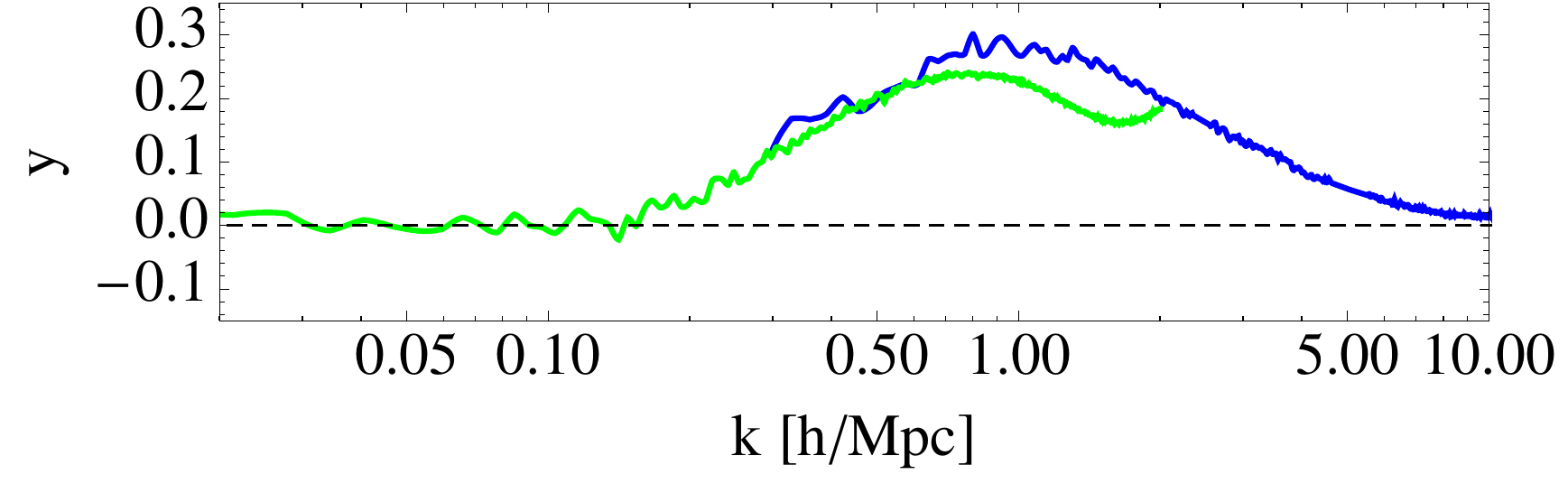}
\includegraphics[width=.45\textwidth,clip]{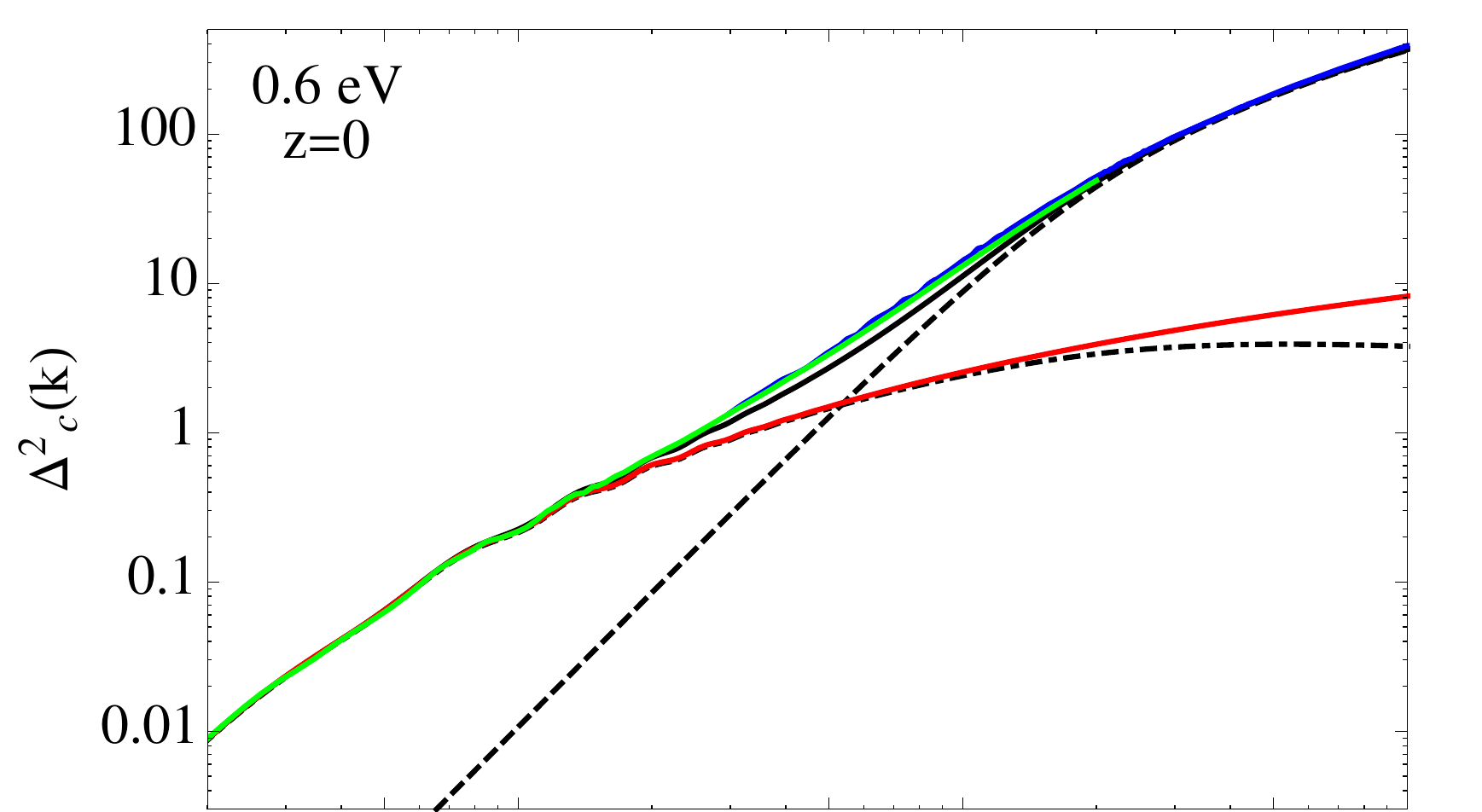}
\includegraphics[width=.45\textwidth,origin=c]{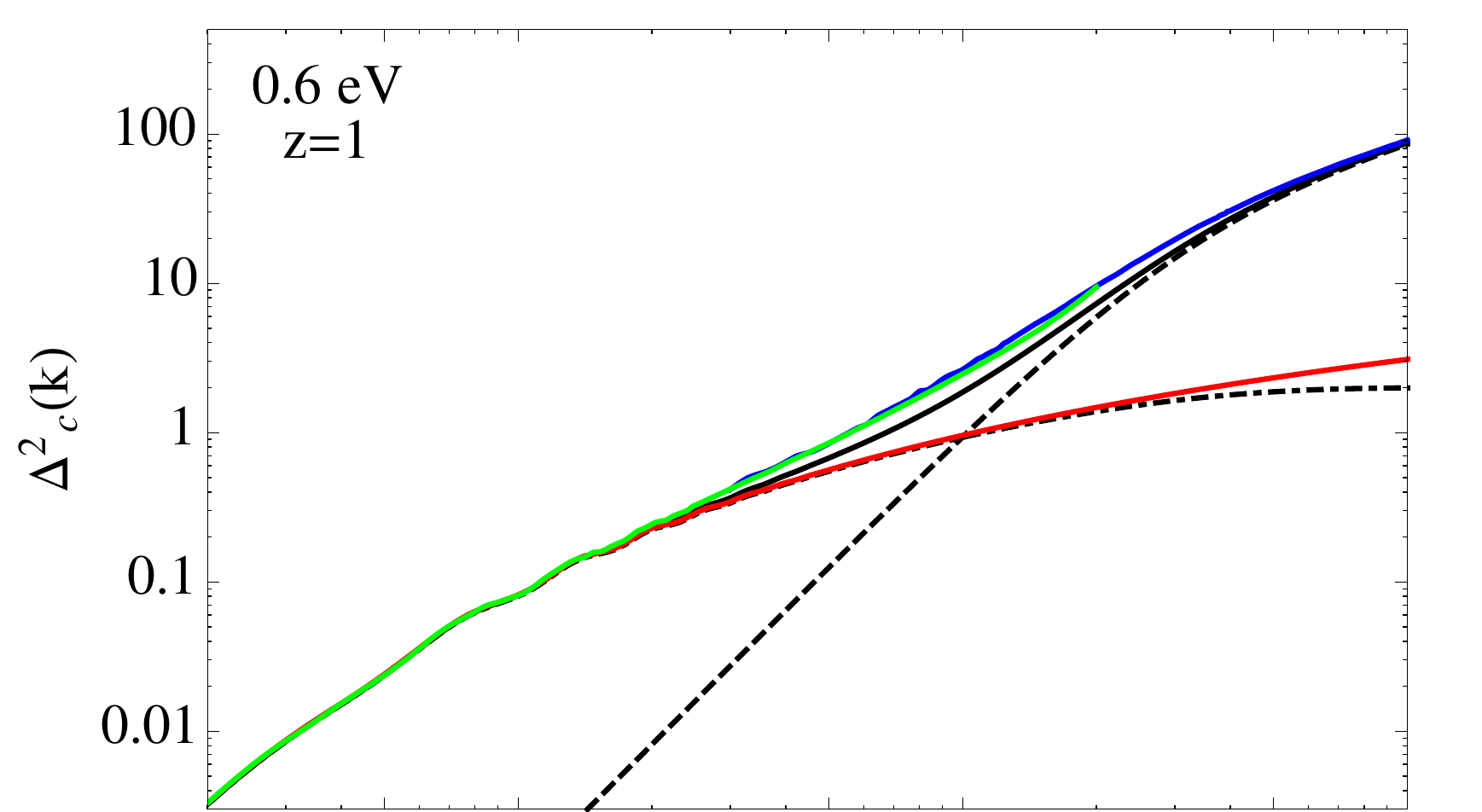}
\includegraphics[width=.45\textwidth,clip]{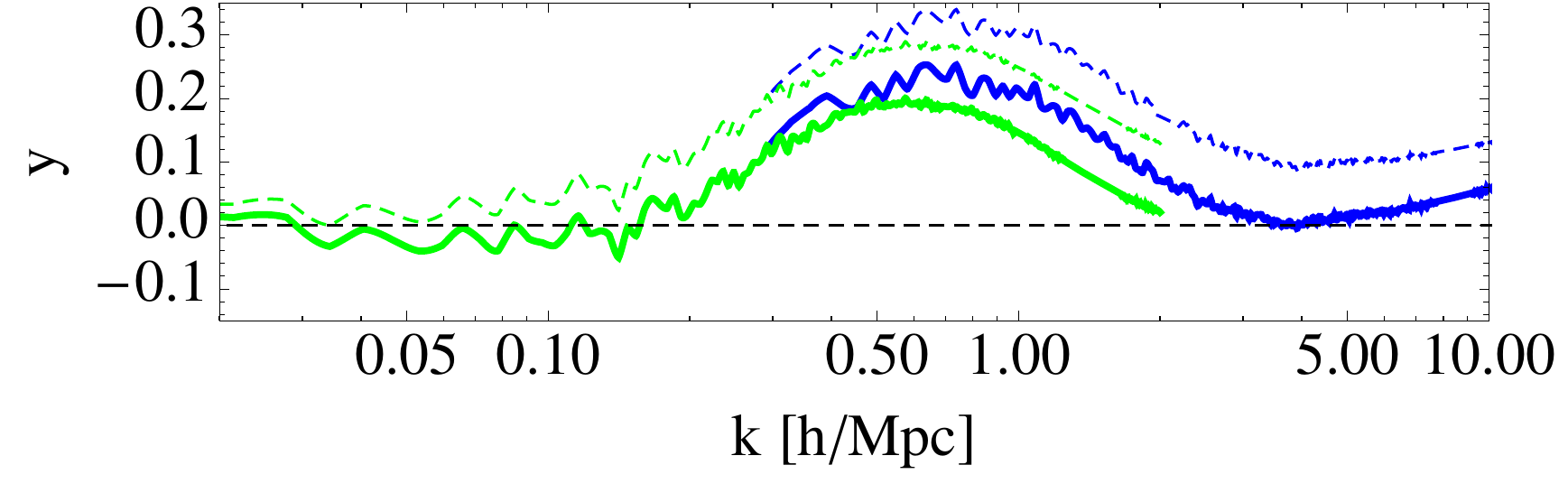}
\includegraphics[width=.45\textwidth,origin=c]{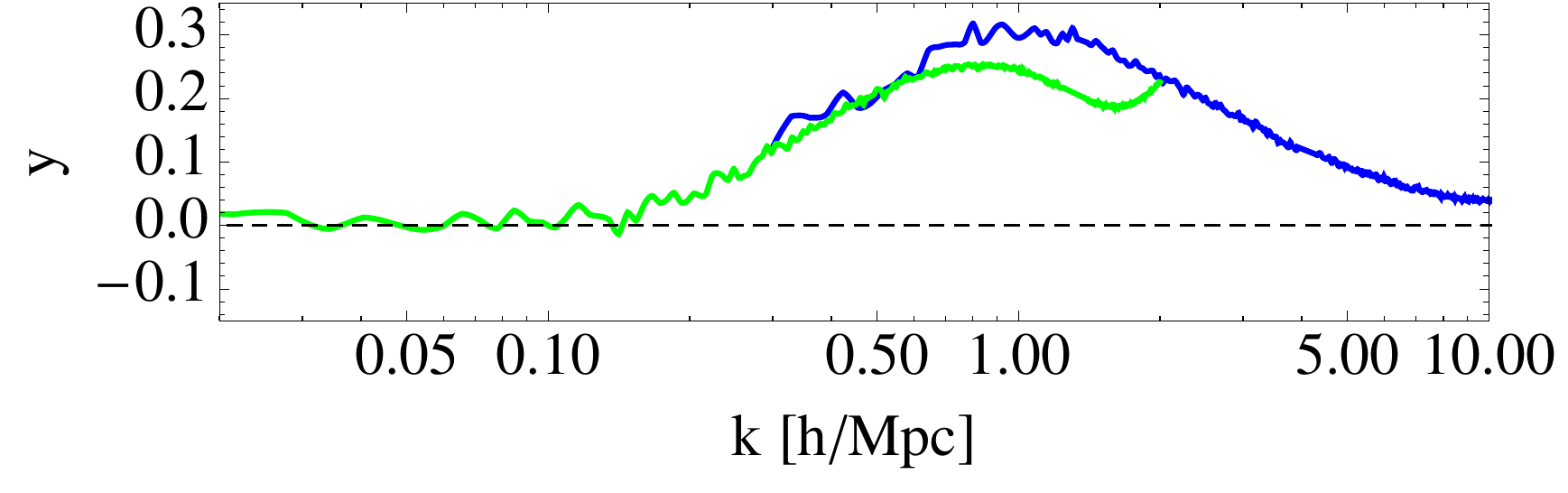}
\caption{\label{fig:2} Cold dark matter power spectrum for $\sum m_{\nu}=0.15$ eV (top), $\sum m_{\nu}=0.3$ eV (middle) and $\sum m_{\nu}=0.6$ eV (bottom) massive neutrino cosmologies. The left and right panels show the results at $z=0$ and $z=1$, respectively. Black curves display the predictions by the halo model: the dashed line is the 1-halo term, the dot-dashed one is the 2-halo term and the solid one is the sum of the two terms. Red curves show the linear predictions and blue and green curves are the results from N-boby simulations with box size $L=200$ Mpc/$h$ and $L=1000$ Mpc/$h$, respectively. The bottom part of each plot shows the residuals between results from halo model and N-body simulations.}
\end{figure}
In analogy with what we have presented in Sec.~\ref{sec:halo_model} and using the eqs. of the cold dark matter prescription \eqref{eq:30} and \eqref{eq:33}, we compute here the power spectrum of the cold field $P_{\rm c}(k)=P^{1h}_{\rm c}(k)+P^{2h}_{\rm c}(k)$, with
\begin{eqnarray}
\label{eq:37}
P^{1h}_{\rm c}(k)&=&\int_0^{\infty} d\nu_{\rm c}\,f(\nu_{\rm c}) \frac{M_{\rm c}}{\bar{\rho}_{\rm c}}|u_{\rm c}(k|M_{\rm c})|^2\, , \\
\label{eq:38}
P^{2h}_{\rm c}(k)&=&\left[\int_0^{\infty} d\nu_{\rm c}\, f(\nu_{\rm c}) b_{\rm c}(\nu_{\rm c})\, u_{\rm c}(k|M_c)\right]^2 P^L_{\rm c}(k)\,,
\end{eqnarray}
where $u_{\rm c}(k|M_{\rm c})$ is the NFW profile of a ${\rm c}$-halo of mass $M_{\rm c}$. Even if we are considering a massive neutrino cosmology, its concentration is well described by the standard formula~\eqref{eq:10} for the $\Lambda$CDM case, as our N-body simulations showed. The quantities $f(\nu_{\rm c}) $ and $b_{\rm c}(\nu_{\rm c})$ are the ST mass function and bias, which can also be used in a massive neutrino cosmology~\cite{Ichiki_Takada,Paco_2013} and guarantee that the 2-halo term is well normalized: $P^{2h}_{\rm c}(k\rightarrow 0)\rightarrow P^L_{\rm c}$. Figure~\ref{fig:2} shows the cold dark matter power spectrum, as predicted by the halo model, for three different cosmologies with massive neutrinos, $\sum m_{\nu}=0.15$ eV on top, $\sum m_{\nu}=0.3$ eV in the middle and $\sum m_{\nu}=0.6$ eV on bottom, and for two different redshifts, $z=0$ on the left and $z=1$ on the right. The relative difference between our results and N-body simulations is shown in the bottom part of each plot, where the solid curves are obtained using the cold dark matter prescription and the thin dashed ones (shown only for $z=0$) come from the matter prescription. The results are in agreement and reinforce the claim of Castorina et al., since the cold dark matter prescription agrees better with simulations. Using this right procedure we obtain a very good agreement on large scales, whereas a disagreement around $15-20\%$ level characterizes the intermediate scales $k\sim 0.2-2\,h$/Mpc at $z=0$, and it increases until $30\%$ at $z=1$. On smaller scales, up to $k\sim10\, h$/Mpc, the disagreement is below $10\%$ for all models.

\subsection{Cross Power Spectrum}
\label{sec:cross}
Here we compute the second term of~\eqref{eq:19}, i.e. the cross power spectrum $P_{{\rm c}\nu}(k)$. Following the description adopted in~\eqref{eq:18a}, the cross power is given by
\begin{equation}
\label{eq:38b}
P_{{\rm c}\nu}(k)= F_h P_{{\rm c}\nu}^h(k) + (1-F_h)P_{{\rm c}\nu}^L(k),
\end{equation}
where $P_{{\rm c}\nu}^L(k)=\sqrt{P_{\rm c}(k) P^L_{\nu}(k)}$ describes the correlation between the cold field and the linear component of the neutrino density field, once we assume that the two fields are completely correlated. This assumption is well motivated on large scales and is a good approximation at intermediate ones, where this term is supposed to be relevant \cite{Yacine-Bird,Paco_2012,Paco_2013}. The cross power spectrum between the cold and the clustered neutrino fields can be written in the language of halo model as $P_{{\rm c}\nu}^h(k) = P_{{\rm c}\nu}^{1h}(k)+P_{{\rm c}\nu}^{2h}(k)$, with
\begin{eqnarray}
\label{eq:39}
P_{{\rm c}\nu}^{1h}(k)&=&\int^{\infty}_{M_{\rm cut}} dM_{\rm c} \, n(M_{\rm c}) \frac{M_{\rm c} }{\bar{\rho}_{\rm c}}\frac{M_{\nu}}{F_h\bar{\rho}_{\nu}}\,u_{\rm c}(k|M_{\rm c})\,u_{\nu}(k|M_{\rm c})\\
\label{eq:40}
P_{{\rm c}\nu}^{2h}(k)&=&\int^{\infty}_0 dM_{\rm c}'\, n(M'_{\rm c}) \,\frac{M'_{\rm c} }{\bar{\rho}_{\rm c}}\, u_{\rm c}(k|M'_{\rm c})\\\nonumber
&& \quad \times \int^{\infty}_{M_{\rm cut}} dM''_{\rm c} \, n(M''_{\rm c})\,\frac{M_{\nu} }{F_h\bar{\rho}_{\nu}}\, u_{\nu}(k|M''_{\rm c}) \, P_{hh}(k|M'_{\rm c},M''_{\rm c}),
\end{eqnarray}
where $u_{\nu}(k|M_{\rm c})$ is the Fourier transform of the normalized density profiles, $\rho_{\nu}^h(r)/M_{\nu}$, of the $\nu$-halo with mass $M_{\nu}=M_{\nu}(M_{\rm c})$. Villaescusa-Navarro et al.~\cite{Paco_2012} measured the density contrast profile of neutrinos around ${\rm c}$-halos in simulations, for $\sum m_{\nu}=0.3,0.6$ eV cosmologies. They found that it can be well reproduced by the fitting formula
\begin{equation}
\label{eq:41}
\delta_{\nu}^{\rm sim}(r)\equiv \frac{\rho_{\nu}(r)-\bar{\rho}_{\nu}}{\bar{\rho}_{\nu}}=\frac{\rho_{\rm c}(M_{\rm c})}{1+\left[r/r_c(M_{\rm c})\right]^{\alpha(M_{\rm c})}},
\end{equation}
where $\rho_{\rm c}, \, r_{\rm c}$ and $\alpha$ are functions of the corresponding ${\rm c}$-halo mass $M_{\rm c}$ and they present different shapes depending on the chosen massive neutrino cosmology (see their figure 10). This profile was obtained considering all the neutrinos, both the linearly clustered ones (that we call linear), and the fully non-linearly clustered ones (that we call clustered). Because of our setup, we define the clustered neutrino profile as
 \begin{equation}
\label{eq:42}
\rho_{\nu}^h(r)\equiv \delta_{\nu}^{\rm sim}(r)\, \bar{\rho}_{\nu}=\rho_{\nu}(r)-\bar{\rho}_{\nu},
\end{equation}
\begin{figure}[tbp]
\centering 
\includegraphics[width=.48\textwidth,clip]{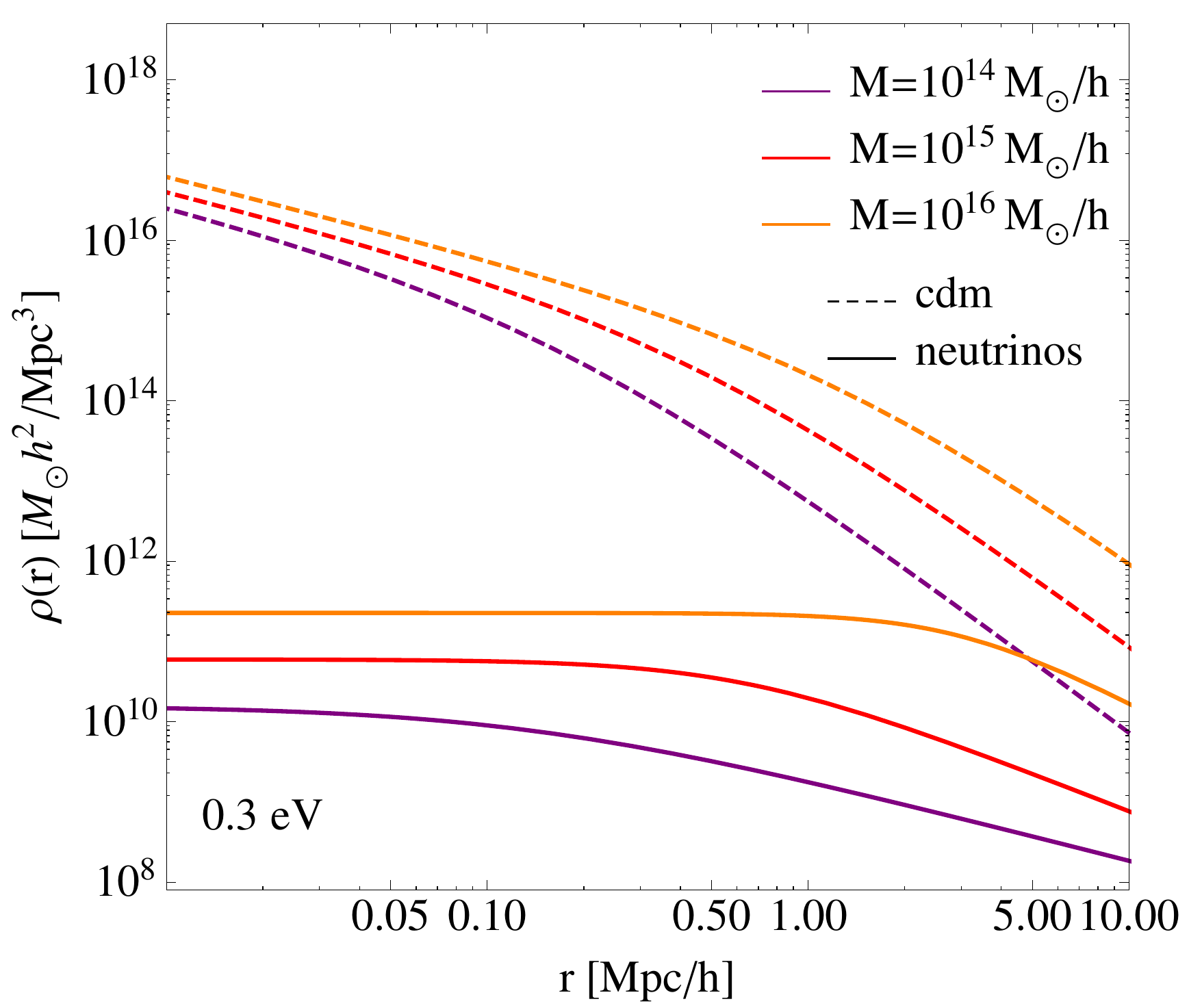}
\includegraphics[width=.48\textwidth,origin=c]{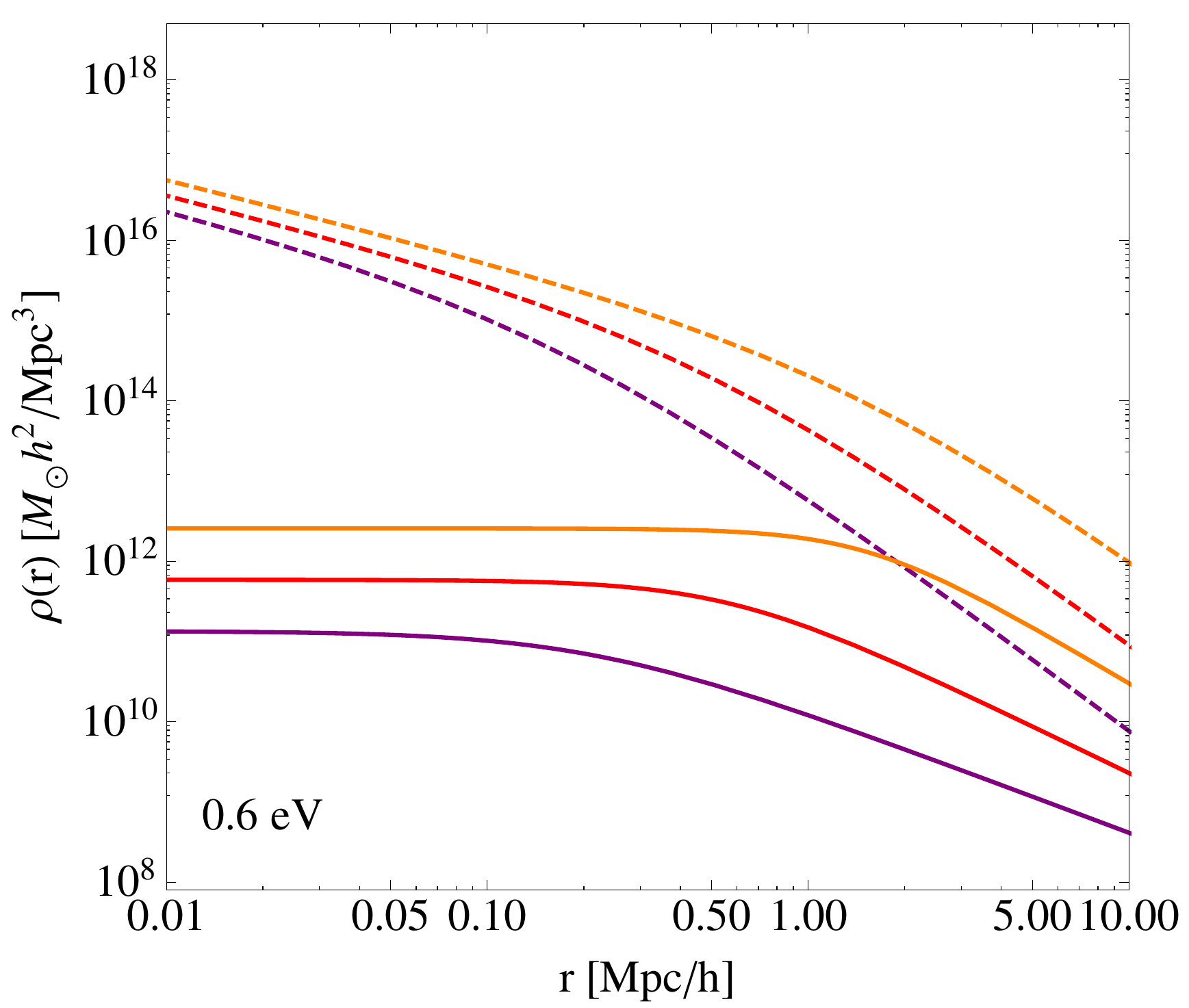}
\caption{\label{fig:3b} Density profile. The left and right panels show the $\sum m_{\nu}=0.3,\, 0.6$ eV cases, respectively. Dashed lines depict the NFW profiles of cold dark matter halos with different masses; the solid lines are the correspondent neutrino profiles.}
\end{figure}
\begin{figure}[tbp]
\centering 
\includegraphics[width=.48\textwidth,clip]{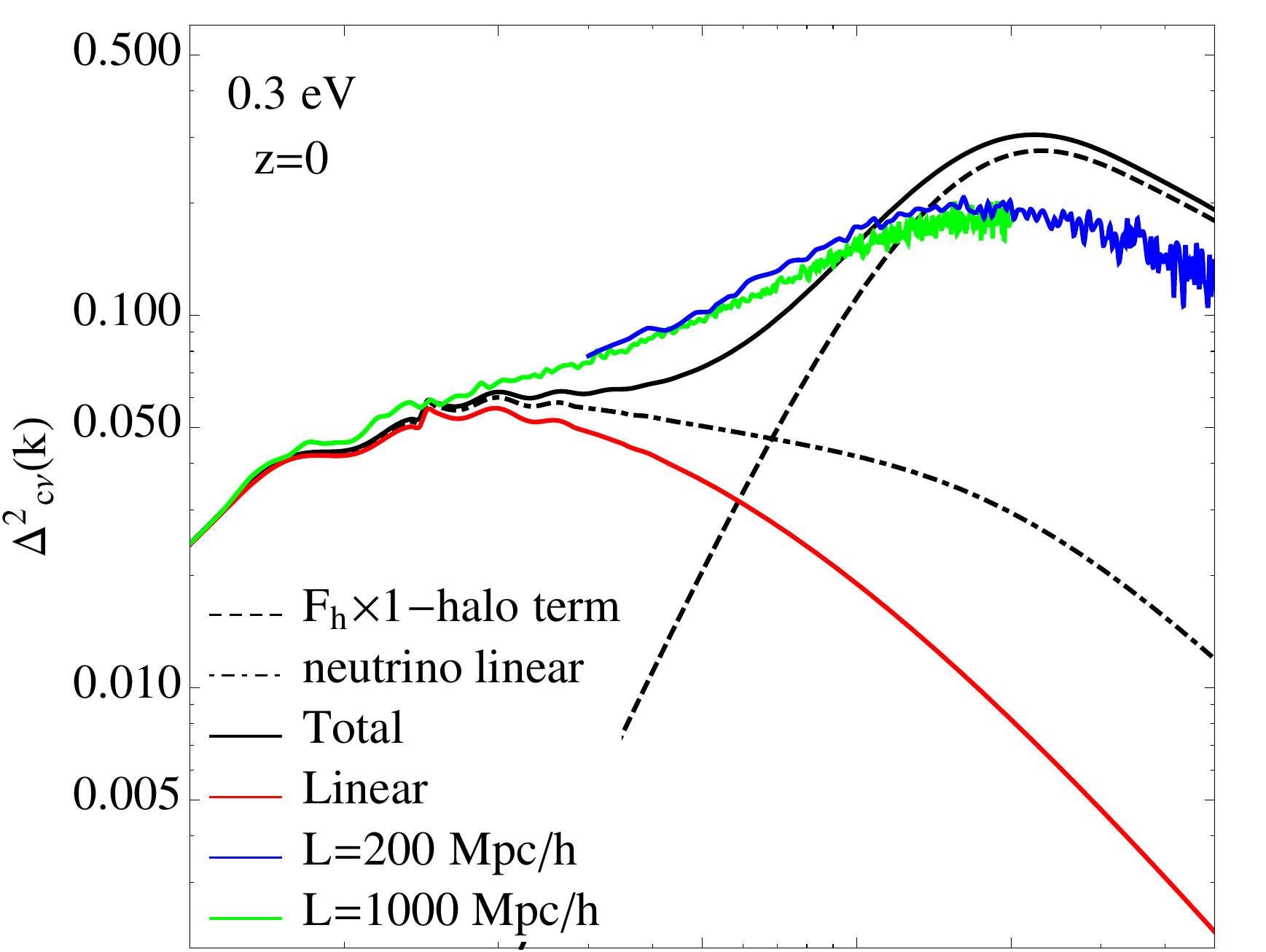}
\includegraphics[width=.48\textwidth,origin=c]{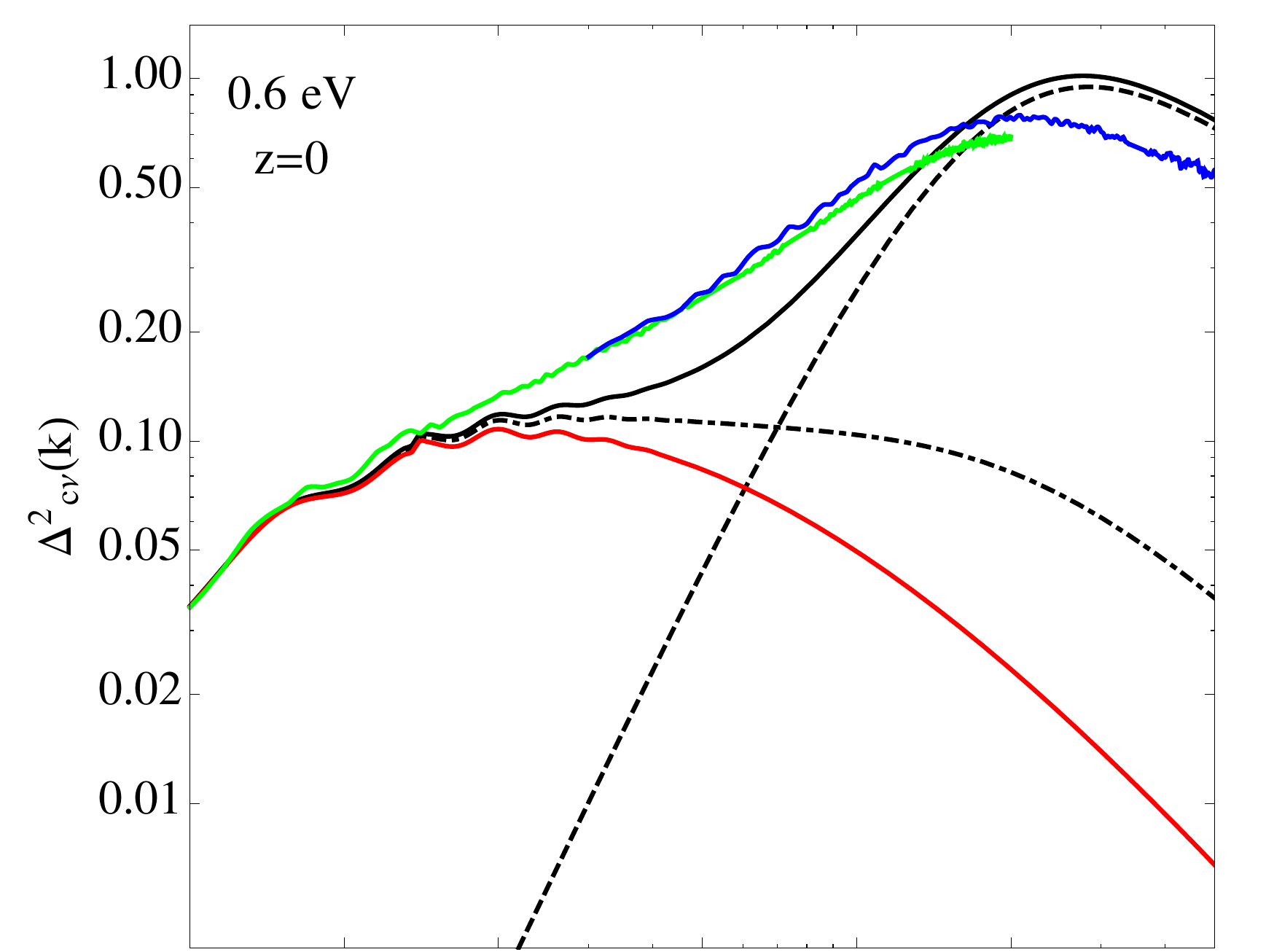}
\includegraphics[width=.48\textwidth,clip]{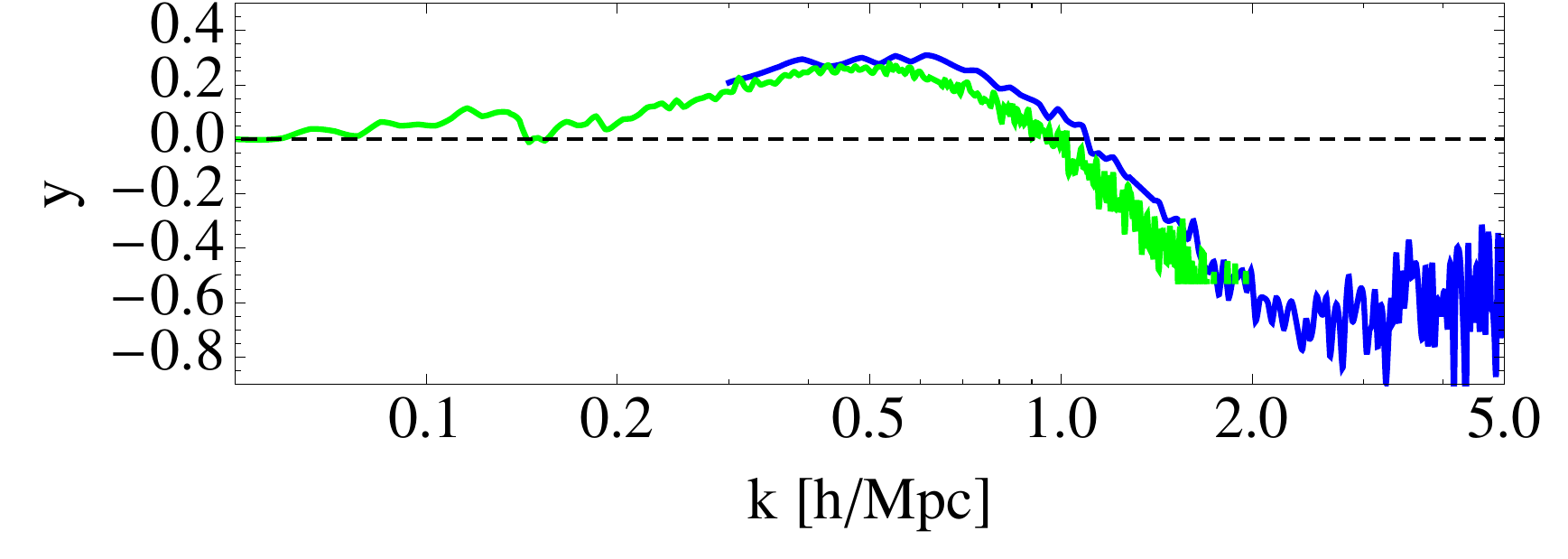}
\includegraphics[width=.48\textwidth,origin=c]{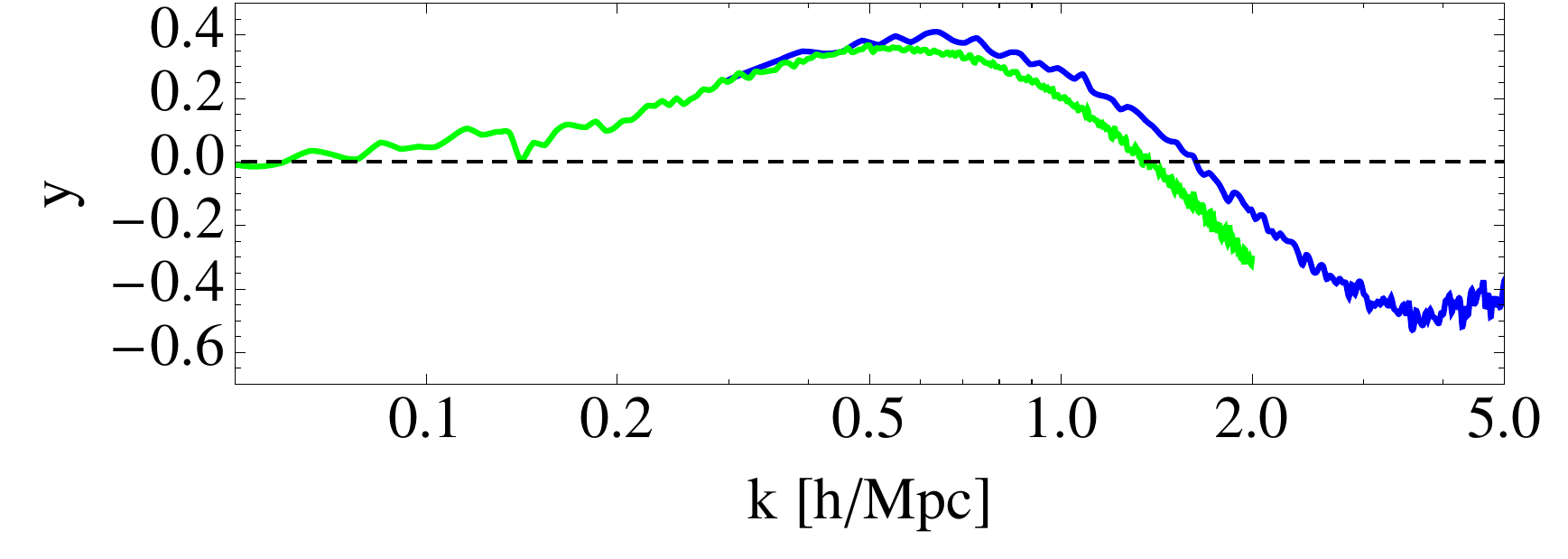}
\caption{\label{fig:3} Cold dark matter-neutrino cross power spectrum in $\sum m_{\nu}=0.3$ eV (left panel) and $\sum m_{\nu}=0.6$ eV (right panel) massive neutrino cosmologies at redshifts $z=0$. Black curves show the cross-power spectrum predicted by the halo model, red lines indicate the linear predictions and blue and green lines are the results from N-boby simulations with box size $L=200$ Mpc/$h$ and $L=1000$ Mpc/$h$, respectively. The bottom part of each plots shows the relative difference between the cross-power spectra from the halo model and from simulations.}
\end{figure}
which means that we consider as clustered the neutrinos measured around a halo, once the neutrino background, $\bar{\rho}_{\nu}$, has been subtracted. This is not the accurate procedure, but a quite good estimation and the resulting profiles are shown in figure~\ref{fig:3b} together with the NFW profile of the cold dark matter halos. From these plots we can notice that the neutrino profiles have lower amplitude than corresponding cold dark matter ones. In analogy with~\eqref{eq:7}, we define the Fourier transfer of the profile as
\begin{equation}
\label{eq:43}
\rho_{\nu}^h(k|M_{\rm c})=\int_0^{R_v} dr\, 4\pi r^2 \,\frac{\sin(kr)}{kr}\,\rho_{\nu}^h(r)\, ,
\end{equation}
where we assume that the virial radius of the ${\rm c}$- and $\nu$-halos are equal. The corresponding mass is $M_{\nu}(M_{\rm c})=\rho_{\nu}^h(k\rightarrow 0|M_{\rm c})$, which is a monotonic growing function in $M_c$. The cut-off mass $M_{\rm cut}$ in~\eqref{eq:39} and~\eqref{eq:40} is a particular ${\rm c}$-halo mass, for which the corresponding $M_{\nu}$ satisfies
\begin{equation}
\label{eq:43b}
M_{\nu}(M_{\rm cut})=0.1\times\frac{4\pi\bar{\rho}_{\nu}}{3}R_v^3(M_{\rm cut}).
\end{equation}
This means that we do not consider as clustered neutrinos the ones forming an halo with mass smaller than the 10\% of the mass of background neutrinos enclosed in the same volume. Therefore, the fraction of clustered neutrinos is given by:
\begin{equation}
\label{eq:44}
F_h=\frac{1}{\bar{\rho}_{\nu}}\int_{M_{\rm cut}}^{\infty} dM_{\rm c} \, n(M_{\rm c}) M_{\nu}(M_{\rm c})\, .
\end{equation}
It would be natural to define $M_{\rm cut}$ as the c-halo mass for which the corresponding $M_{\nu}$ is vanishing. This does not happen for the neutrinos profile defined in~\eqref{eq:42} and the definition in~\eqref{eq:43b} gives a convergent value for $F_h$, i.e. the mass in neutrinos contained in smaller halos is negligible. This fraction turns out to be very small: $F_h=9.5\times10^{-4},\,2.6\times10^{-3} $ for $\sum m_{\nu}=0.3,0.6$ eV, respectively. However, even if small, this neutrino component is very important for having a good prediction for the cross and neutrinos power spectra at small scales, as we shall see below.

We use the eqs.~\eqref{eq:30} and~\eqref{eq:33} of the cold dark matter prescription to rewrite $P_{{\rm c}\nu}^{1h}(k)$ and $P_{{\rm c}\nu}^{2h}(k)$ in terms of the peak height
\begin{eqnarray}
\label{eq:45}
P_{{\rm c}\nu}^{1h}(k)&=&\int_{M_{\rm cut}}^{\infty} d\nu_{\rm c} \, f(\nu_{\rm c}) \frac{M_{\nu}}{F_h \bar{\rho}_{\nu}}\,u_{\rm c}(k|M_{\rm c})\,u_{\nu}(k|M_{\rm c})\\
\label{eq:46}
P_{{\rm c}\nu}^{2h}(k)&=&\int_0^{\infty} d\nu'_{\rm c}\, f(\nu'_{\rm c}) \, b(\nu'_{\rm c})\, u_c(k|M'_{\rm c})\\\nonumber
&& \quad \times \int_{M_{\rm cut}} ^{\infty}d\nu''_{\rm c} \, f(\nu''_{\rm c}c) \, b(\nu''_{\rm c})\,\frac{M_{\nu} }{M''_{\rm c}}\frac{\bar{\rho}_{\rm c}}{F_h\bar{\rho}_{\nu}}\, u_{\nu}(k|M''_{\rm c}) \, P^L_{\rm c}(k)\, ,
\end{eqnarray}
where the mass function and bias are the usual Sheth-Tormen (ST) ones. Substituting the last expressions in~\eqref{eq:38b} we compute the cross power spectrum for the two different massive neutrino cosmologies. The results at redshift $z=0$ are shown in figure~\ref{fig:3}. Neither the linear cross power spectrum (red lines) nor the cross power spectrum between the clustered cold field and the unclustered component of neutrinos (dot-dashed black lines) can reproduce simulations at intermediate ($k\sim 0.2\,h$/Mpc) and up to small scales, for the two neutrino masses. Instead, our extension of the halo model (solid black line), which accounts for the clustered component of neutrinos, can describe the main behavior of N-body simulations at scales smaller than $k\sim 5\,h$/Mpc. We can notice that the main contribution to the power spectrum comes from the unclustered component of the neutrino field via $P_{{\rm c}\nu}^L(k)$ (dot-dashed line) at large scales and from the 1-halo term $P_{{\rm c}\nu}^{1h}(k)$ of the clustered neutrino component at small scales. The 2-halo term $P_{{\rm c}\nu}^{2h}(k)$ is not shown because it is small and not relevant at any scales. To conclude, our model predicts the cross power spectrum from simulation with $30\%$ accuracy until $k\sim 1\,h$/Mpc in the $\sum m_{\nu}=0.3$ eV case (left panel). In the  $\sum m_{\nu}=0.6$ eV case (right panel), the accuracy is at the $40\%$ level on scales $k< 5\,h$/Mpc. 

\subsection{Neutrino Power Spectrum}
\label{sec:neutrino}
\begin{figure}[tbp]
\centering 
\includegraphics[width=.48\textwidth,clip]{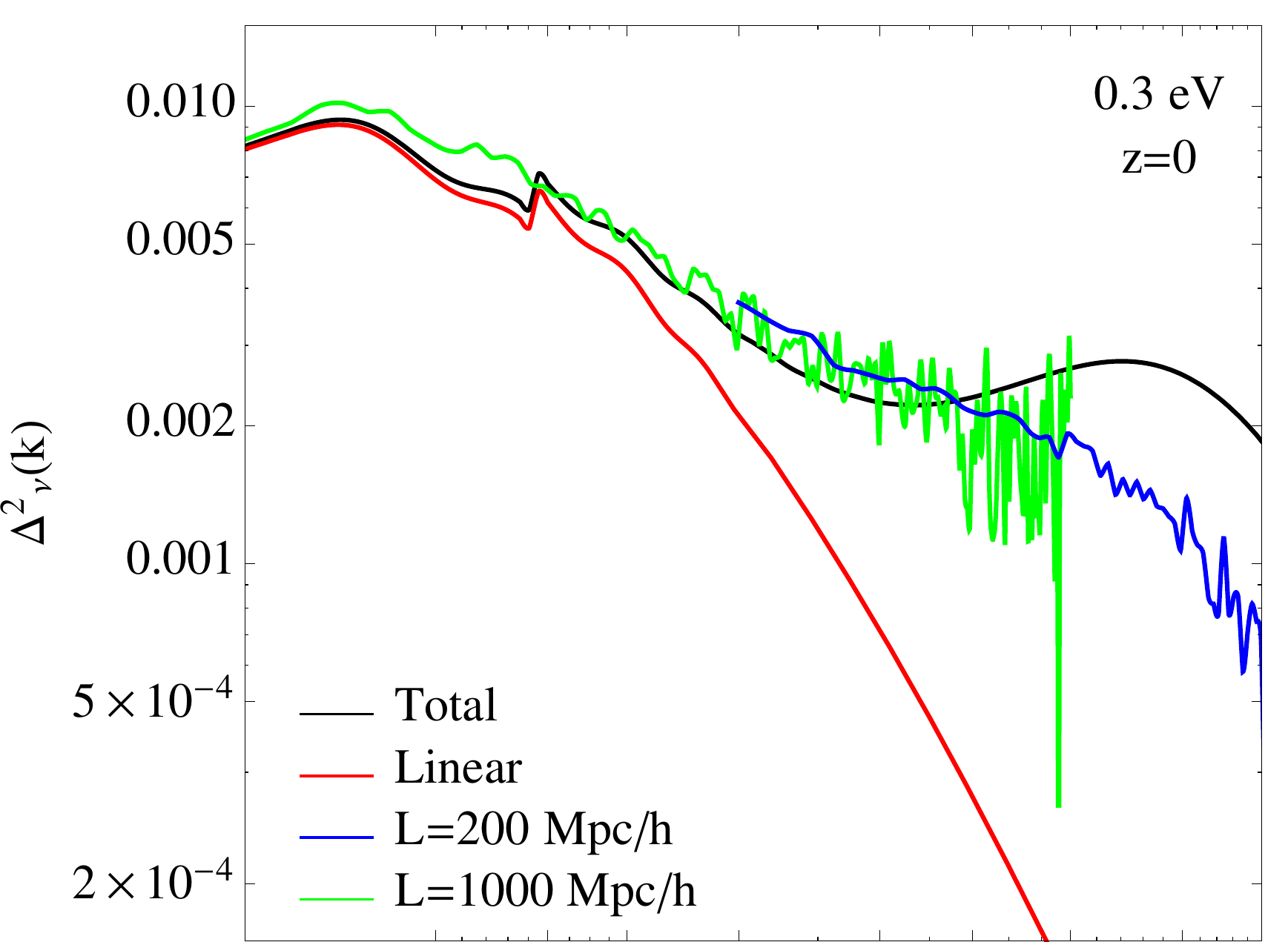}
\includegraphics[width=.48\textwidth,origin=c]{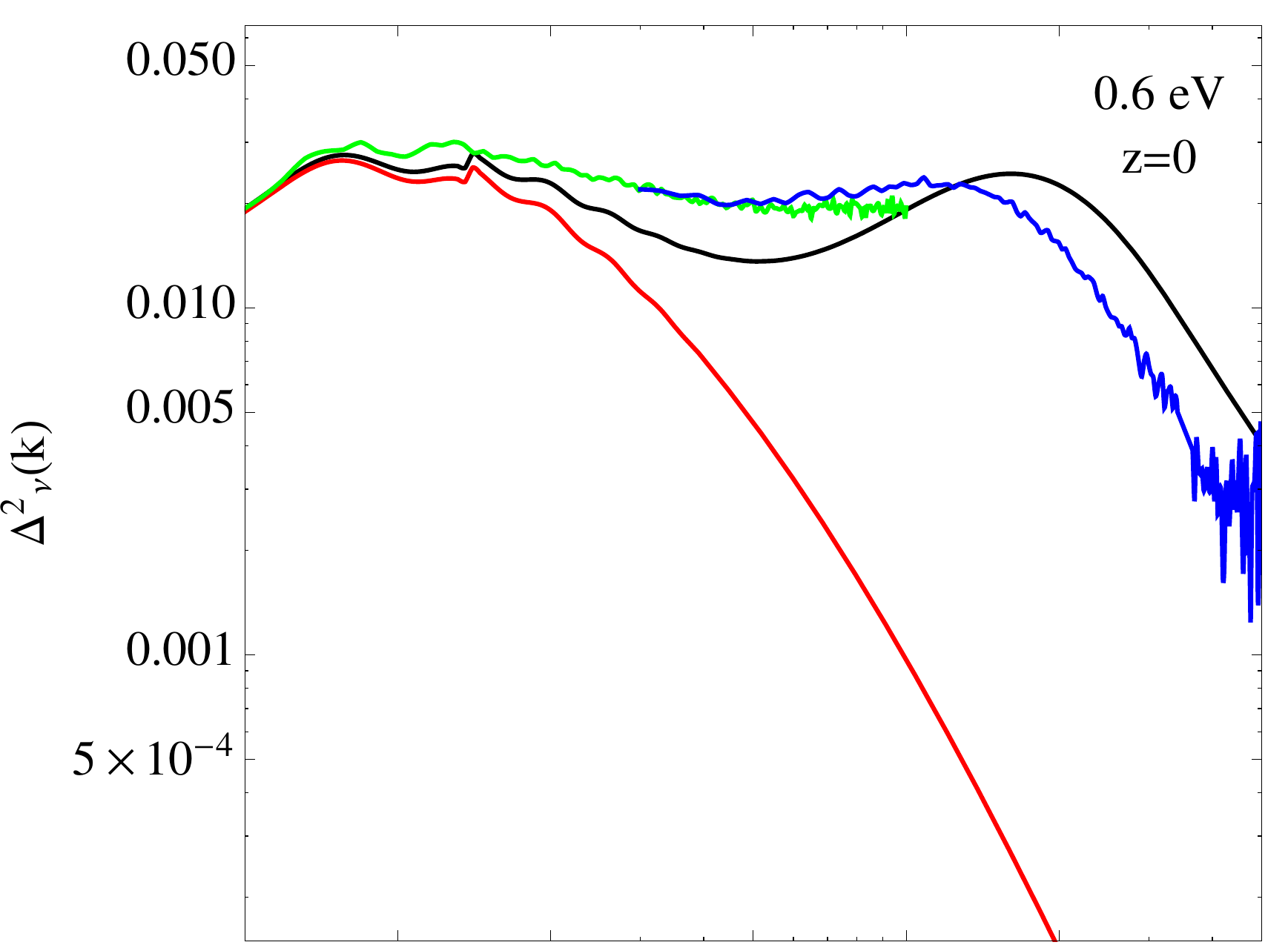}
\includegraphics[width=.48\textwidth,clip]{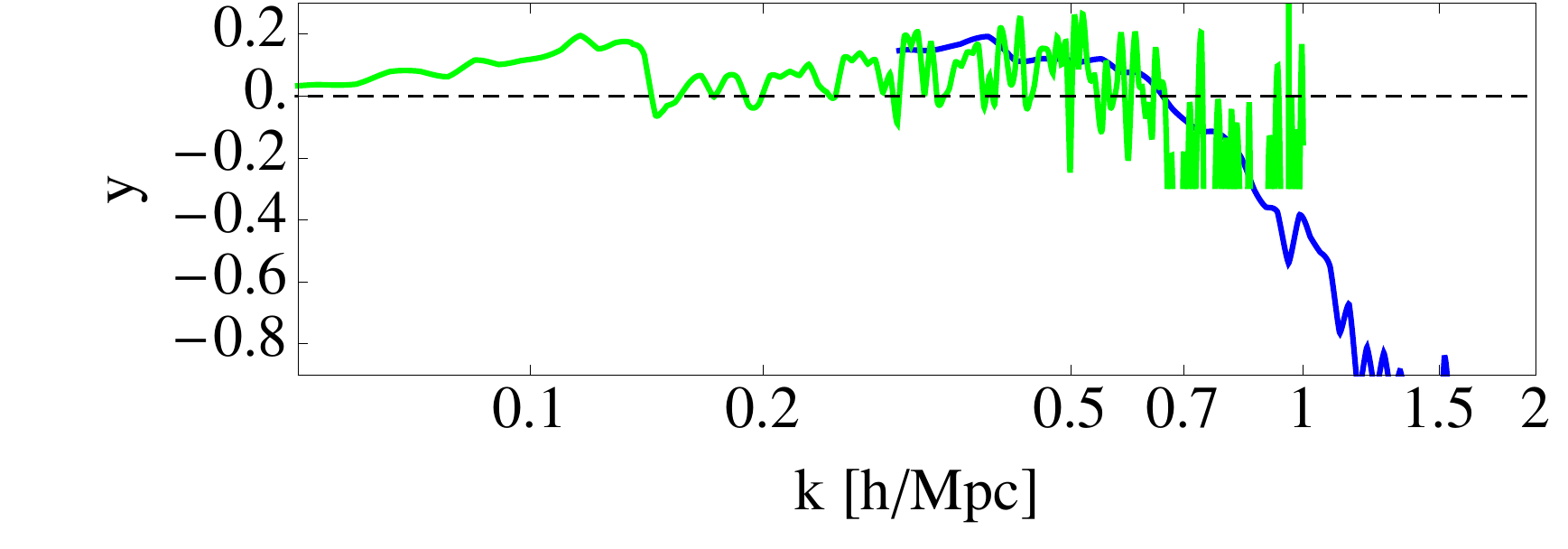}
\includegraphics[width=.48\textwidth,origin=c]{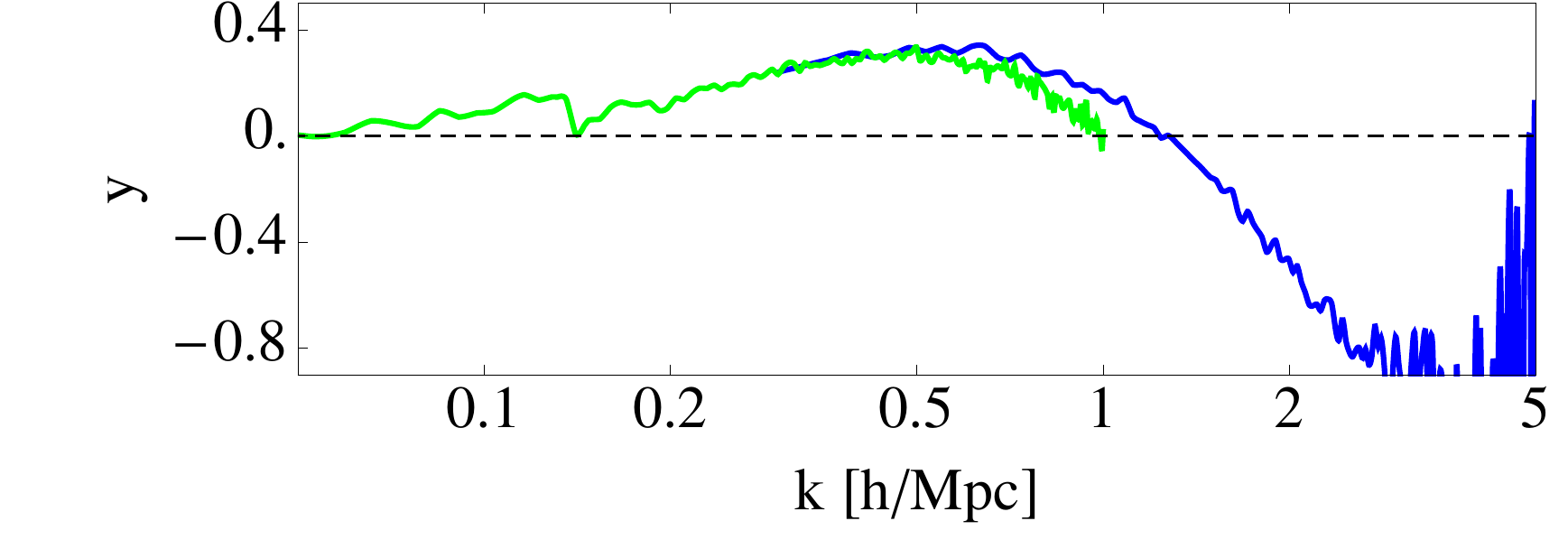}
\caption{\label{fig:4} Neutrino power spectrum in $\sum m_{\nu}=0.3$ eV (left panel) and $\sum m_{\nu}=0.6$ eV (right panel) massive neutrino cosmologies at redshift $z=0$. Black curves show the neutrino power spectrum predicted by the halo model, red curves indicate the linear predictions and blue and green curves are the results from N-boby simulations with box size $L=200$ Mpc/$h$ and $L=1000$ Mpc/$h$, respectively. The bottom part of each plots shows the relative difference between the power spectra from the halo model and from simulations.}
\end{figure}
Using the definition of the neutrino density field in equation~\eqref{eq:18a}, we write the neutrino power spectrum as
\begin{equation}
\label{eq:47}
P_{\nu}(k)=F_h^2 P_{\nu}^h(k) +2 F_h (1-F_h)P_{\nu}^{hL}(k)+(1-F_h)^2P_{\nu}^{L}(k)\, ,
\end{equation}
where the auto-power spectrum of the linear component is just the linear power $P_{\nu}^{L}(k)$ and the cross term can be expressed as $P_{\nu}^{hL}(k)=\sqrt{P_{\nu}^h(k)P_{\nu}^L(k)}$, once we assumed that the clustered and smoothed fields are completely correlated. As for the other fields, the power spectrum of the non-linearly clustered component can be split in two terms, $P_{\nu}^h(k)=P_{\nu}^{1h}(k)+P_{\nu}^{2h}(k)$, with
\begin{eqnarray}
\label{eq:48}
P_{\nu}^{1h}(k)&=&\int^{\infty}_{M_{\rm cut}} dM_{\rm c} \, n(M_{\rm c}) \left(\frac{M_{\nu}}{F_h\bar{\rho}_{\nu}}\right)^2 |u_{\nu}(k|M_{\rm c})|^2\\
\label{eq:49}
P_{\nu}^{2h}(k)&=&\int^{\infty}_{M_{\rm cut}} dM_{\rm c}'\, n(M'_{\rm c}) \,\frac{M_{\nu}}{F_h\bar{\rho}_{\nu}}\, u_{\nu}(k|M'_{\rm c})\\\nonumber
&& \quad \times \int^{\infty}_{M_{\rm cut}} dM''_{\rm c} \, n(M''_{\rm c})\,\frac{M_{\nu} }{F_h\bar{\rho}_{\nu}}\, u_{\nu}(k|M''_{\rm c}) \, P_{hh}(k|M'_{\rm c},M''_{\rm c})\,,
\end{eqnarray}
where all the quantities have already been defined in Sec.~\ref{sec:cross}. Once again we apply the cold dark matter prescription yielding
\begin{eqnarray}
\label{eq:50}
P_{\nu}^{1h}(k)&=&\int_{M_{\rm cut}}^{\infty} d\nu_{\rm c} \, f(\nu_{\rm c})\left(\frac{M_{\nu}}{F_h\bar{\rho}_{\nu}}\right)^2 \frac{\bar{\rho}_{\rm c}}{M_{\rm c}}\,|u_{\nu}(k|M_{\rm c})|^2\\
\label{eq:51}
P_{\nu}^{2h}(k)&=&\left[\int_{M_{\rm cut}}^{\infty} d\nu_{\rm c} \, f(\nu_{\rm c}) \, b(\nu_{\rm c})\,\frac{M_{\nu} }{M_{\rm c}}\frac{\bar{\rho}_{\rm c}}{F_h\bar{\rho}_{\nu}}\, u_{\nu}(k|M_{\rm c}) \right]^2 P^L_{\rm c}(k)\, .
\end{eqnarray}
Next, we compute the neutrino power spectrum $P_{\nu}(k)$ at redshift $z=0$ for two massive neutrino cosmologies with $\sum m_{\nu}=0.3$ and 0.6 eV. The encouraging results are shown in figure~\ref{fig:4}: the disagreement with simulations is below $20\%$ until $k\sim0.7~h$/Mpc for the $\sum m_{\nu}=0.3$ eV case (left panel), whereas it is under $30\%$ until $k\sim1.5~h$/Mpc for the $\sum m_{\nu}=0.6$ eV case (right panel).

\subsection{Matter Power Spectrum}
\label{sec:matter}
In the previous subsections~\ref{sec:cold},~\ref{sec:cross},~\ref{sec:neutrino} we have presented all the terms needed to compute the total matter power spectrum in a massive neutrino cosmology. However, looking at~\eqref{eq:19} one can notice that the cross and the neutrino power spectra are multiplied by $\bar{\rho}_{\rm c}\bar{\rho}_{\nu}/\bar{\rho}^2$ and $(\bar{\rho}_{\nu}/\bar{\rho})^2$, respectively. These two terms are much smaller than $1$ for light neutrinos, as the ones considered here.
\begin{figure}[tbp]
\centering 
\includegraphics[width=.45\textwidth,clip]{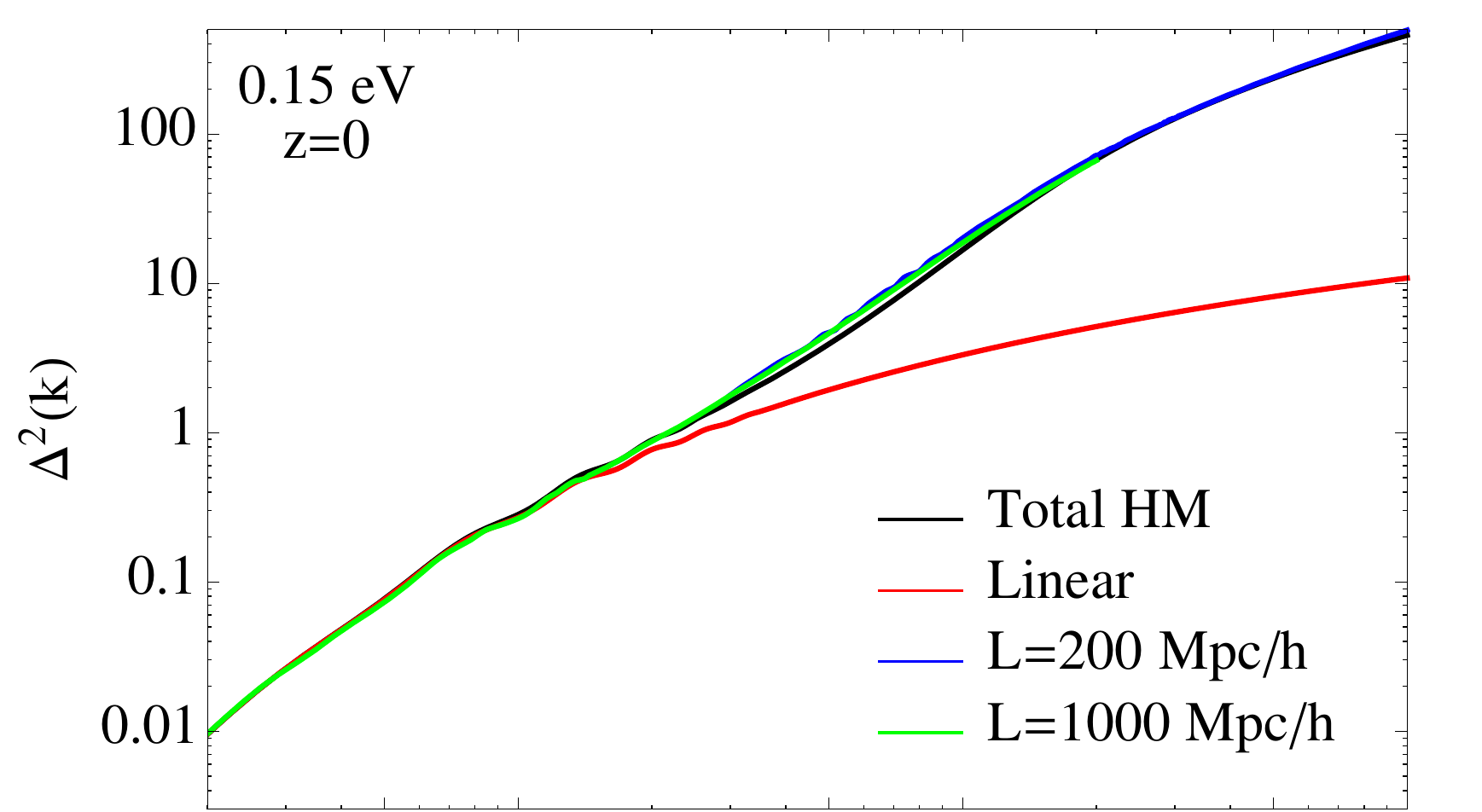}
\includegraphics[width=.45\textwidth,origin=c]{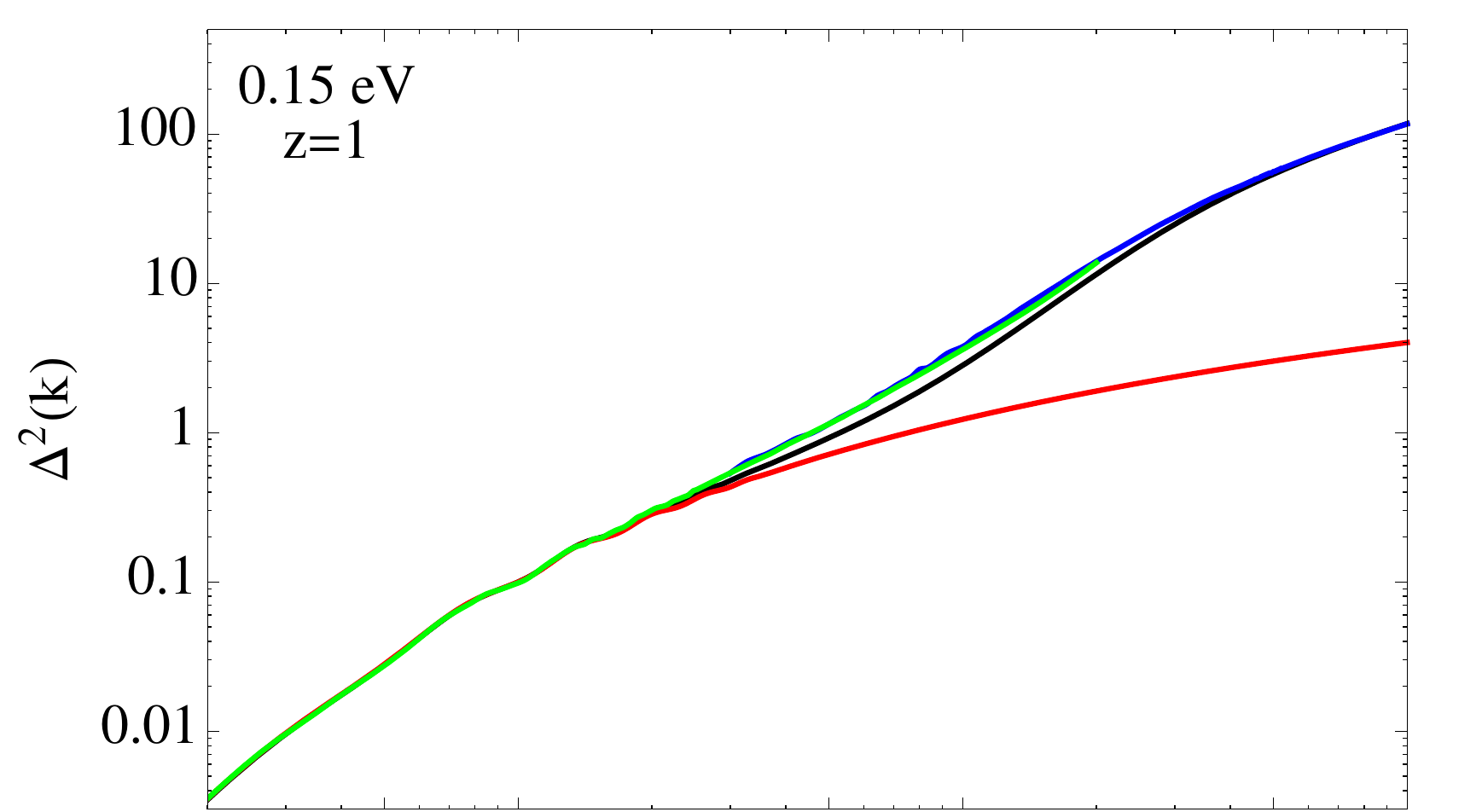}
\includegraphics[width=.45\textwidth,clip]{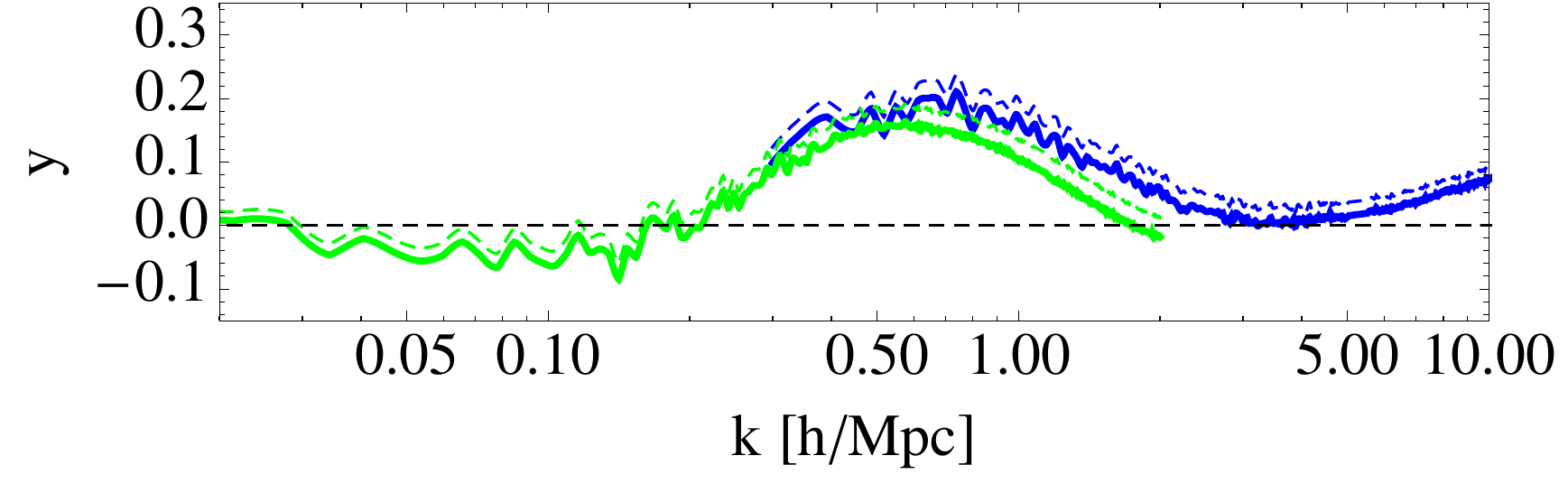}
\includegraphics[width=.45\textwidth,origin=c]{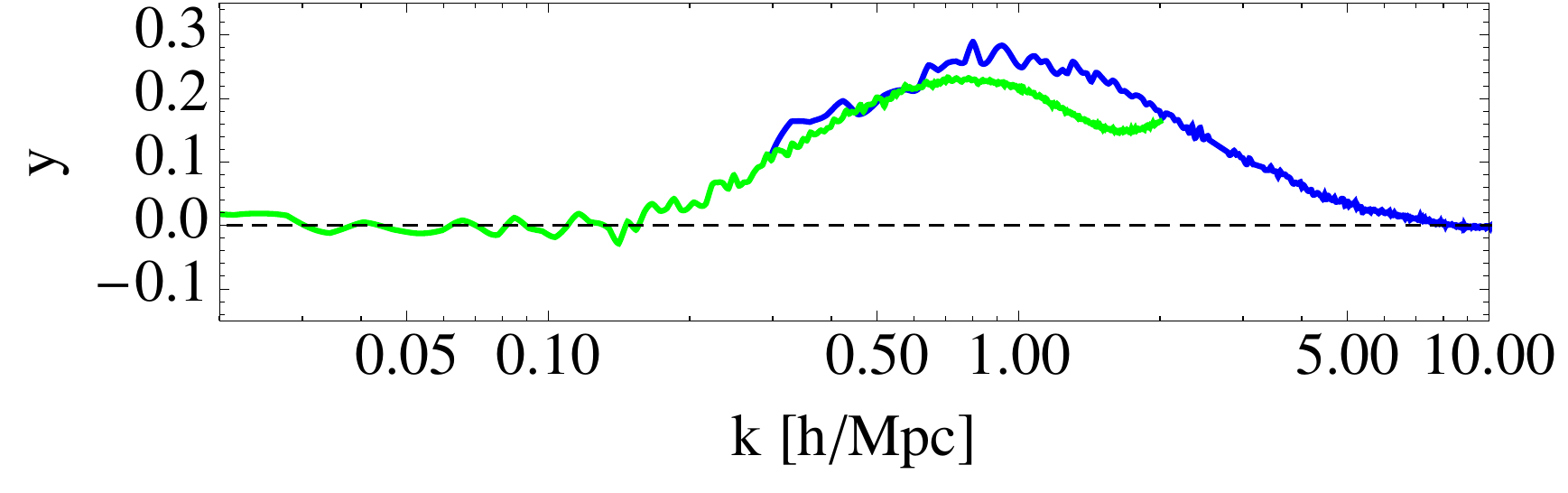}
\includegraphics[width=.45\textwidth,clip]{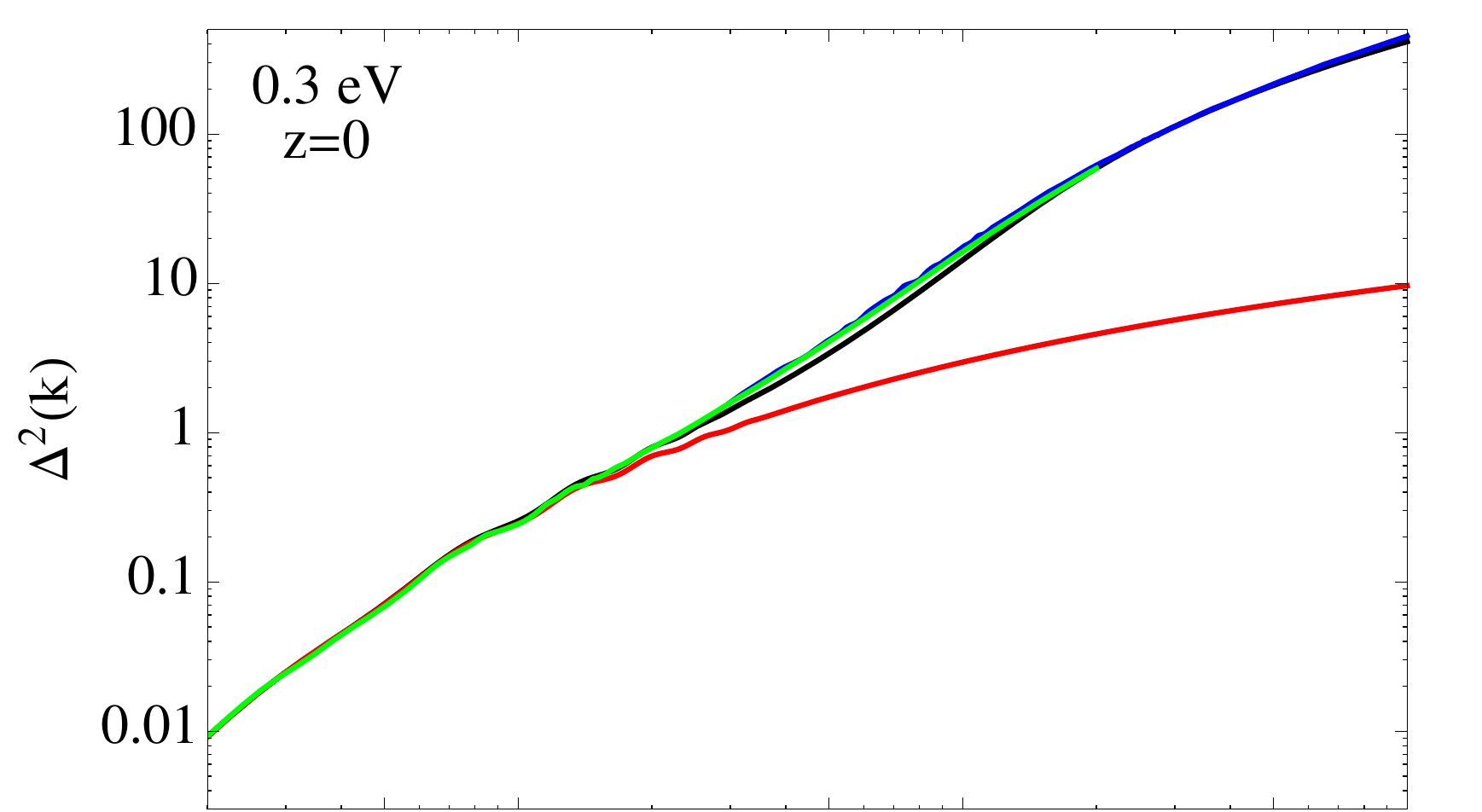}
\includegraphics[width=.45\textwidth,origin=c]{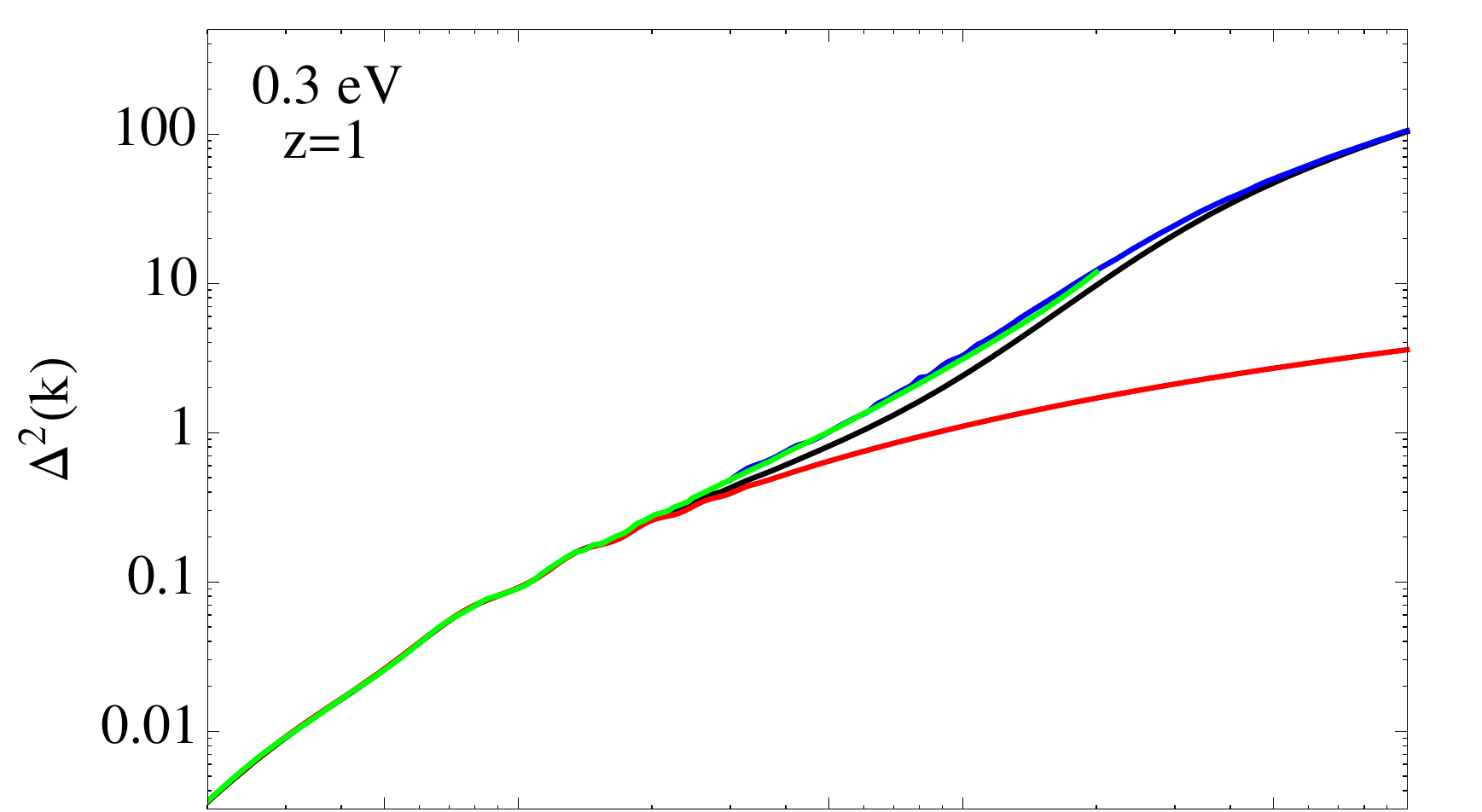}
\includegraphics[width=.45\textwidth,clip]{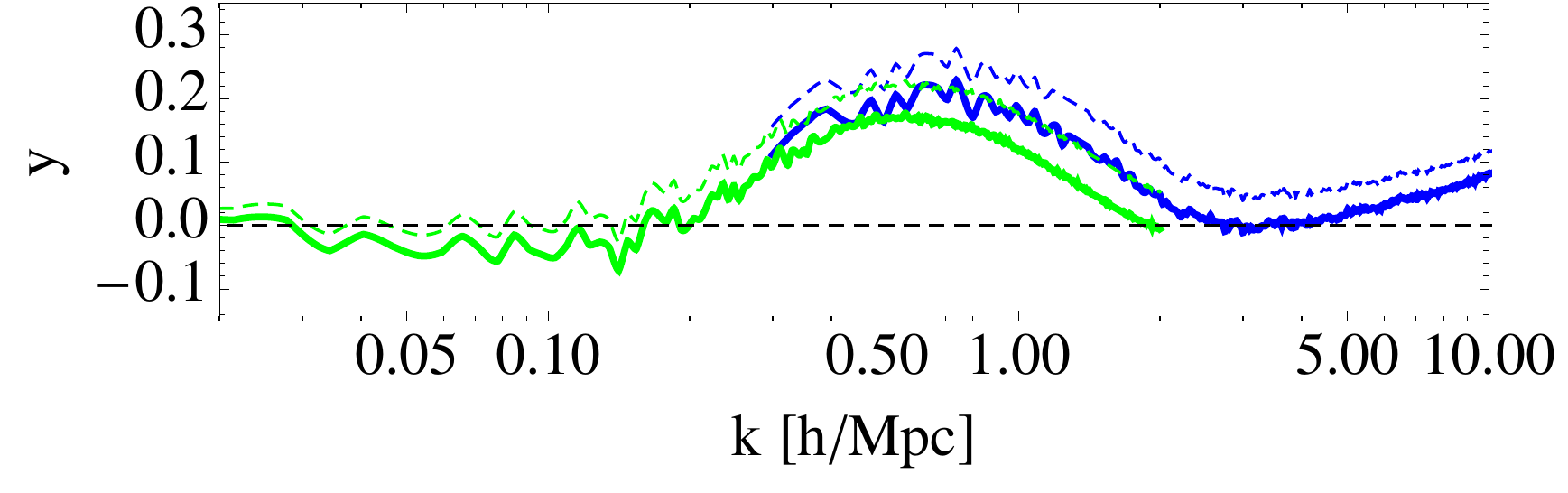}
\includegraphics[width=.45\textwidth,origin=c]{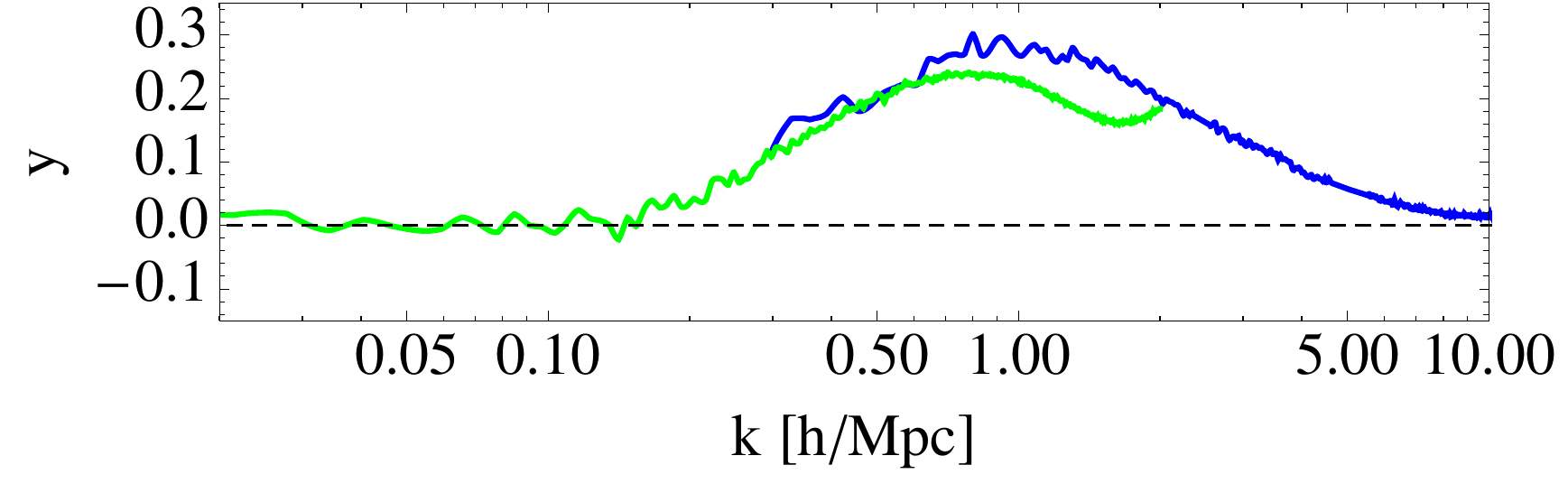}
\includegraphics[width=.45\textwidth,clip]{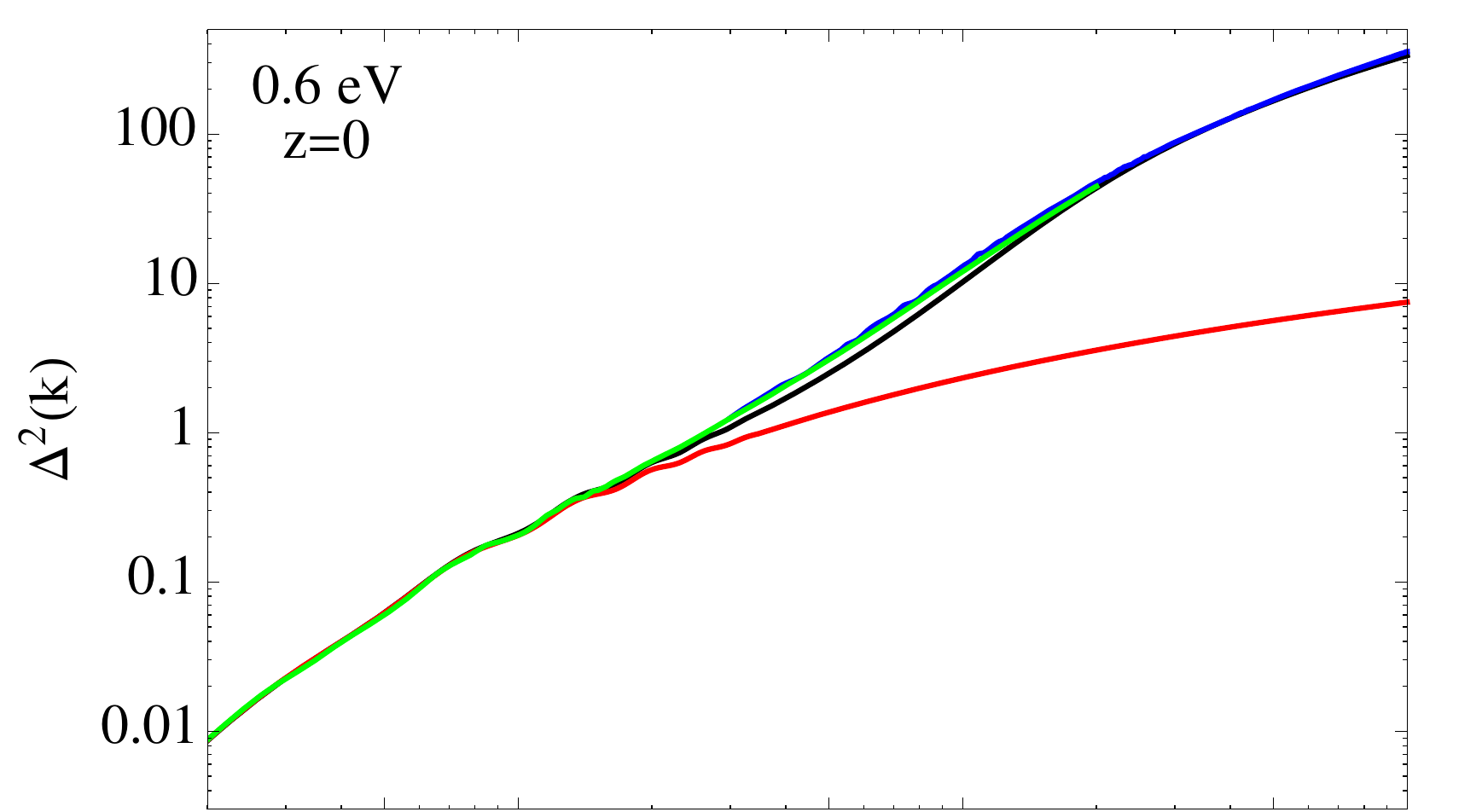}
\includegraphics[width=.45\textwidth,origin=c]{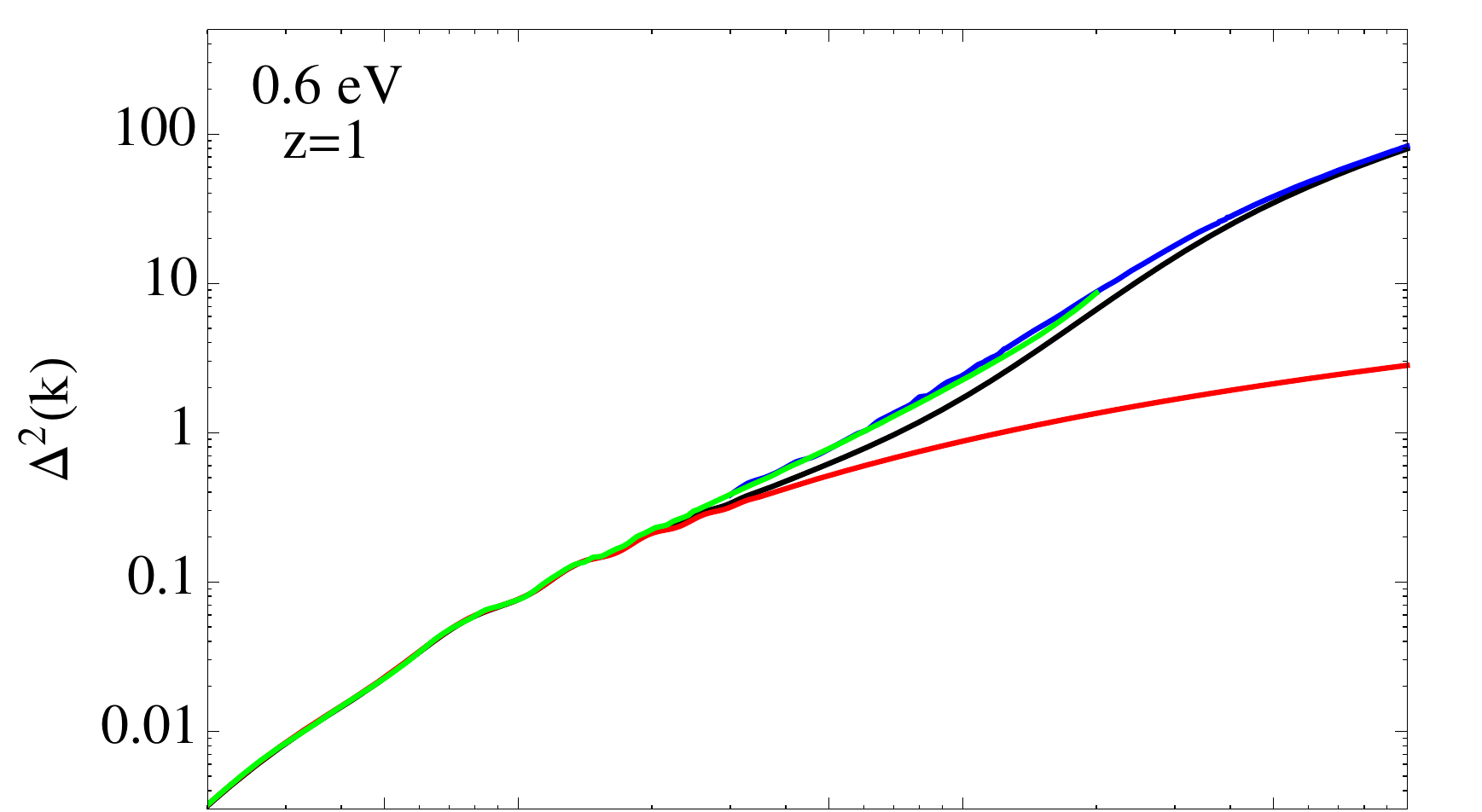}
\includegraphics[width=.45\textwidth,clip]{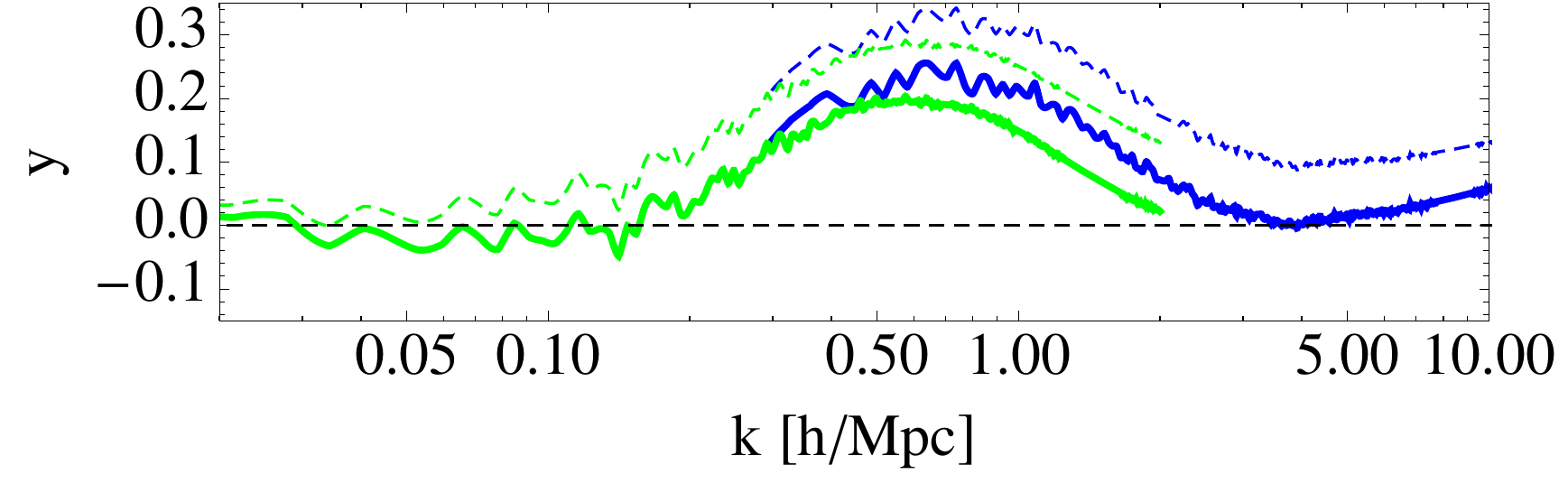}
\includegraphics[width=.45\textwidth,origin=c]{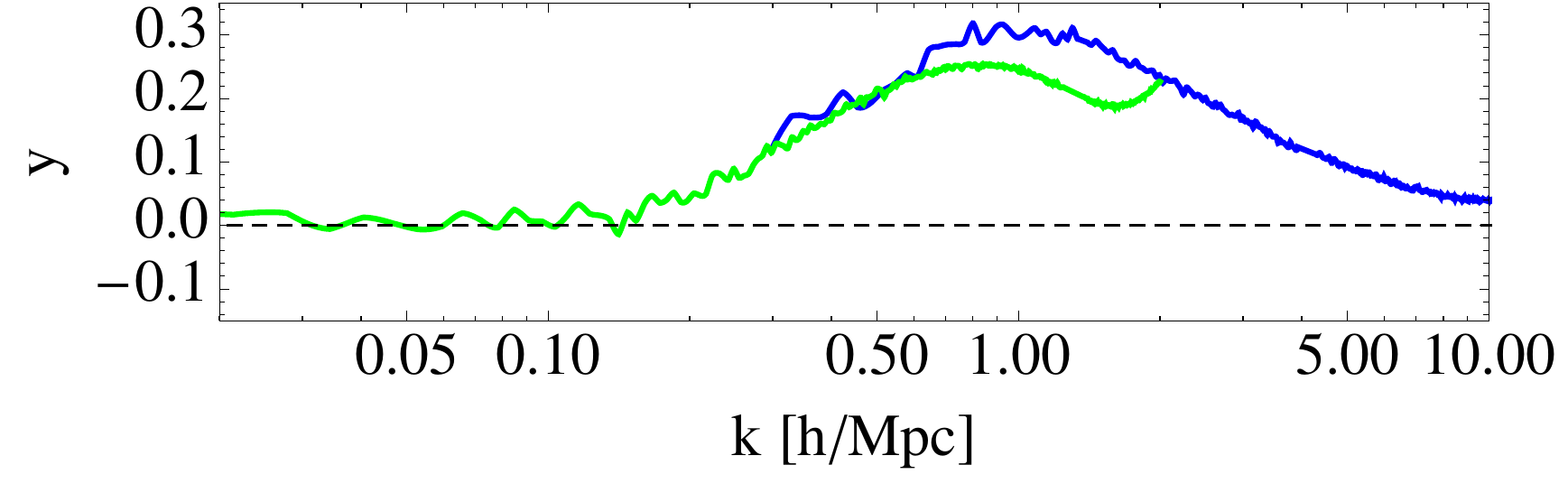}
\caption{\label{fig:5} Matter power spectrum in $\sum m_{\nu}=0.15$ eV (top), $\sum m_{\nu}=0.3$ eV (middle) and $\sum m_{\nu}=0.6$ eV (bottom) massive neutrino cosmologies. The left and right panels show the results at redshifts $z=0$ and $z=1$, respectively. Black curves display the matter power spectrum as predicted by the halo model, red curves show the linear predictions and blue and green curves are the results from N-boby simulations with box size $L=200$ Mpc/$h$ and $L=1000$ Mpc/$h$, respectively. The bottom part of each plots shows the relative difference between the power spectrum from the halo model and from simulations.}
\end{figure}
Therefore, we expect that the improvements given by computing these terms with halo model will be highly suppressed once we compute the total matter power spectrum, which should be well reproduced using just the linear cross and neutrino power spectra. And this is the case: we computed the total matter power spectrum using both the fully non-linear and the linear cross and neutrino power spectra, finding that their difference is well below the $1\%$ level for all the cosmologies studied in this paper.

Then, we present here the resulting total matter power spectra in massive neutrinos cosmologies, computed using the linear neutrino and cross power spectra and the fully non-linear cold dark matter one, at redshifts $z=0$
and $z=1$. Figure~\ref{fig:5} shows the $\sum m_{\nu}=0.15,0.3,0.6$ eV
cosmologies in the top, middle and bottom panels, respectively. Once
again, the halo model (solid black curves) reproduces well the
simulations on small and on large scales; on intermediate scales a
disagreement $<20\%$ is present at $z=0$ and $<30\%$ at $z=1$. We also
show the comparison between simulations and the halo model computed
with the matter prescription, which is represented by the thin dashed
curves. We can again confirm that this is not the ideal prescription
since it reproduces worse the results from the N-body/neutrino simulations.

\section{The ratio $\Delta^2_{\nu}(k)/\Delta^2_{\Lambda \rm CDM}(k)$}
\label{sec:ratio}
\subsection{Halo model and N-body simulations}
\label{sec:ratiosim}

\begin{figure}[tbp]
\centering 
\includegraphics[width=.48\textwidth,clip]{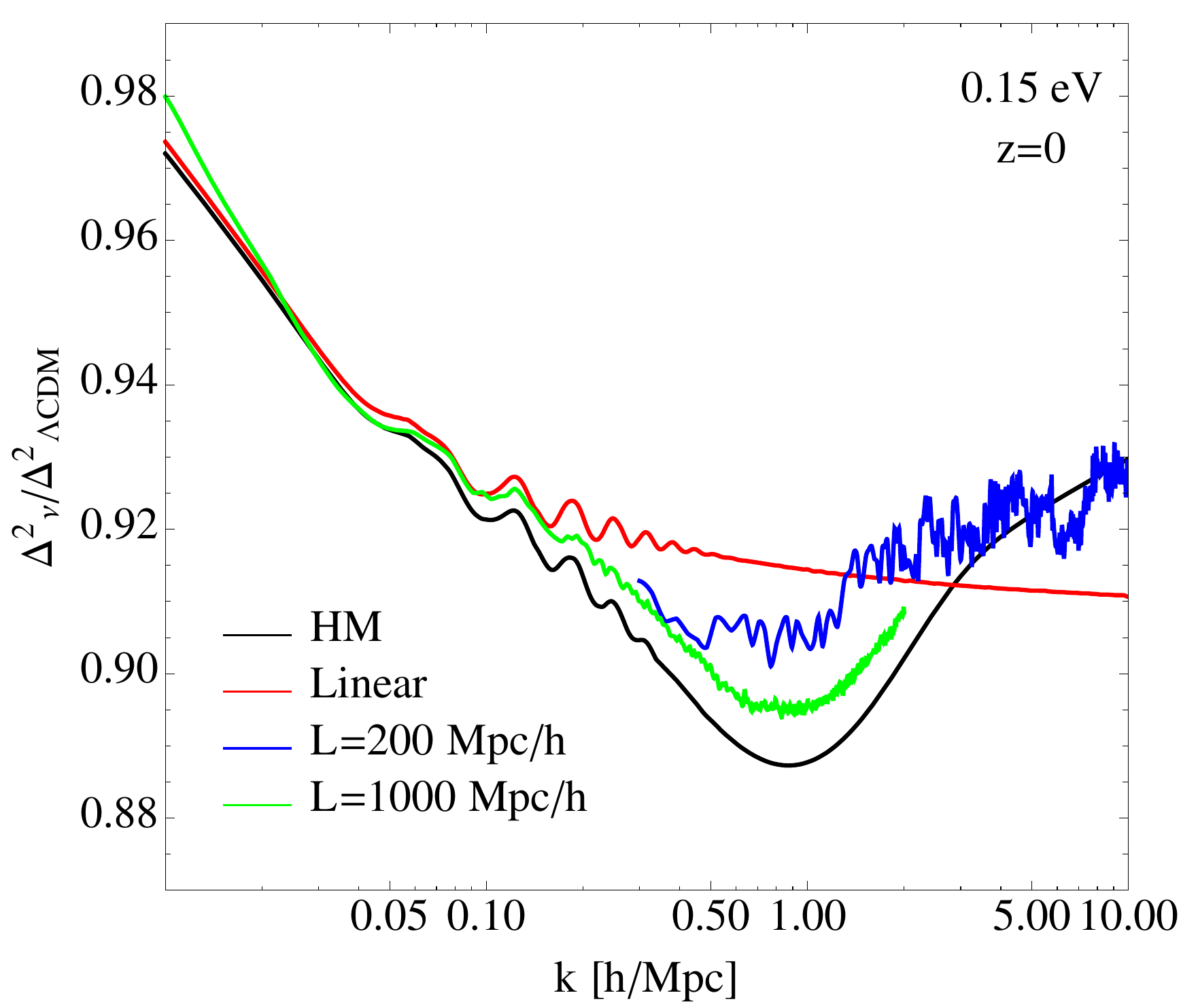}
\includegraphics[width=.48\textwidth,origin=c]{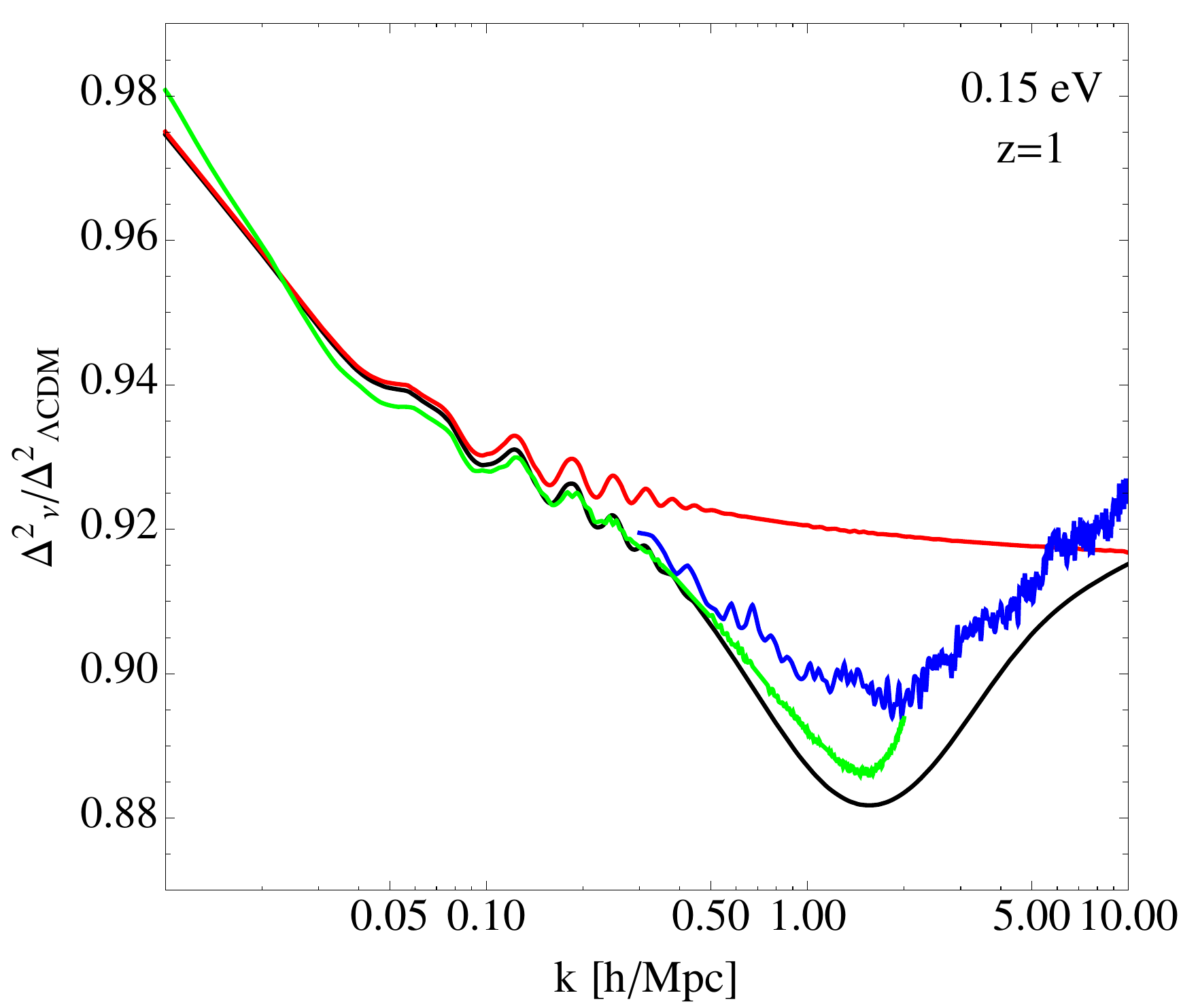}
\includegraphics[width=.48\textwidth,clip]{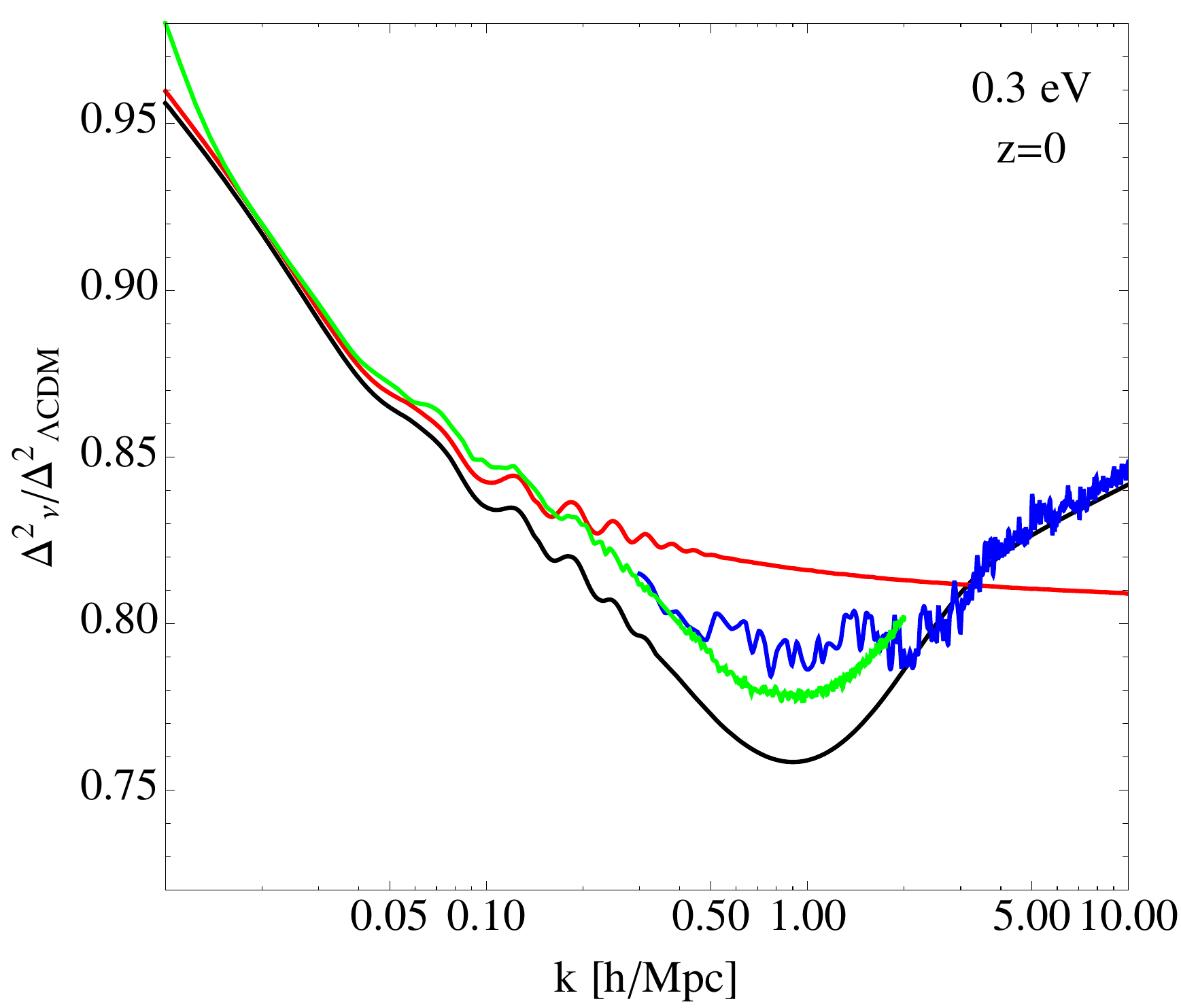}
\includegraphics[width=.48\textwidth,origin=c]{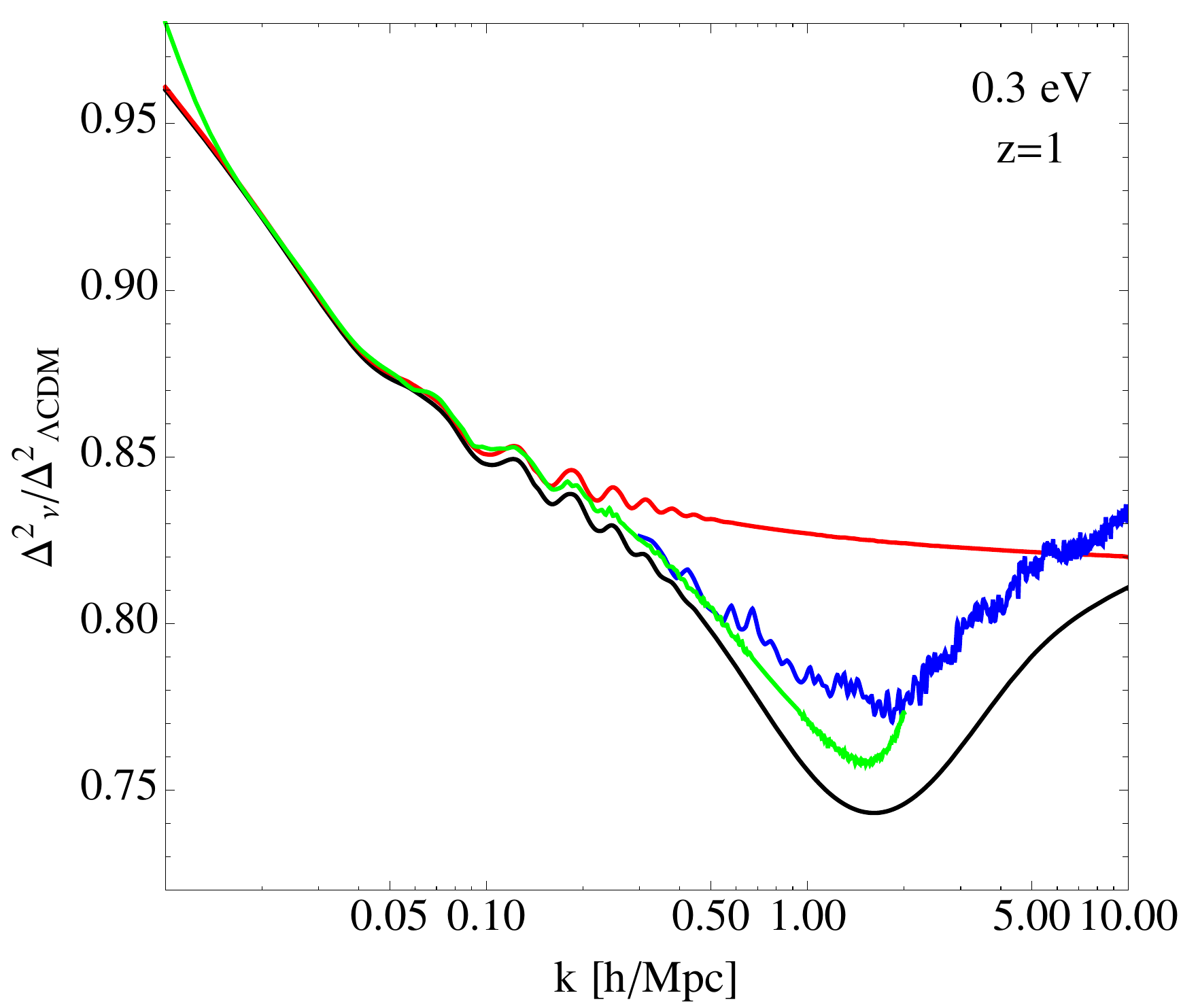}
\includegraphics[width=.48\textwidth,clip]{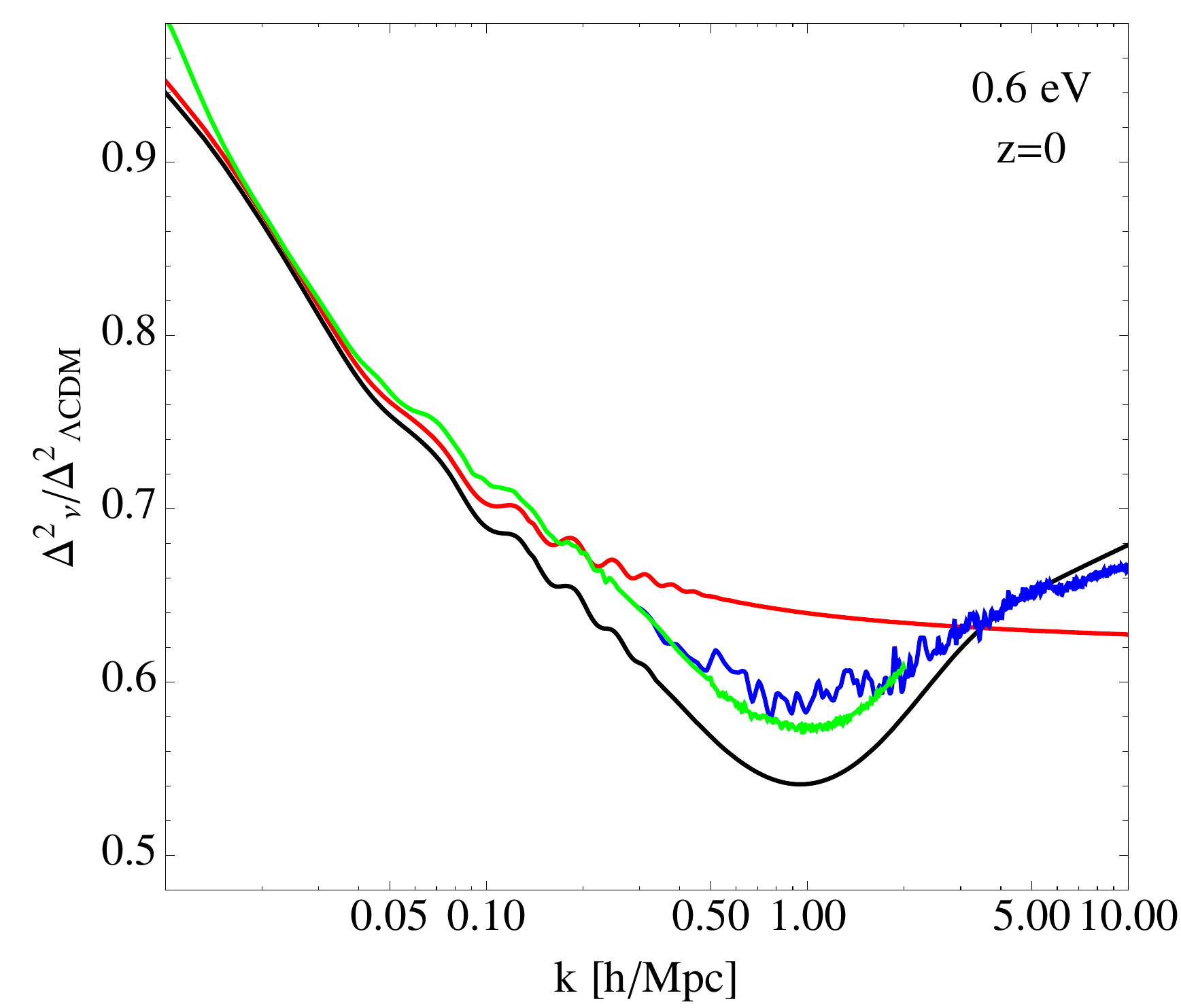}
\includegraphics[width=.48\textwidth,origin=c]{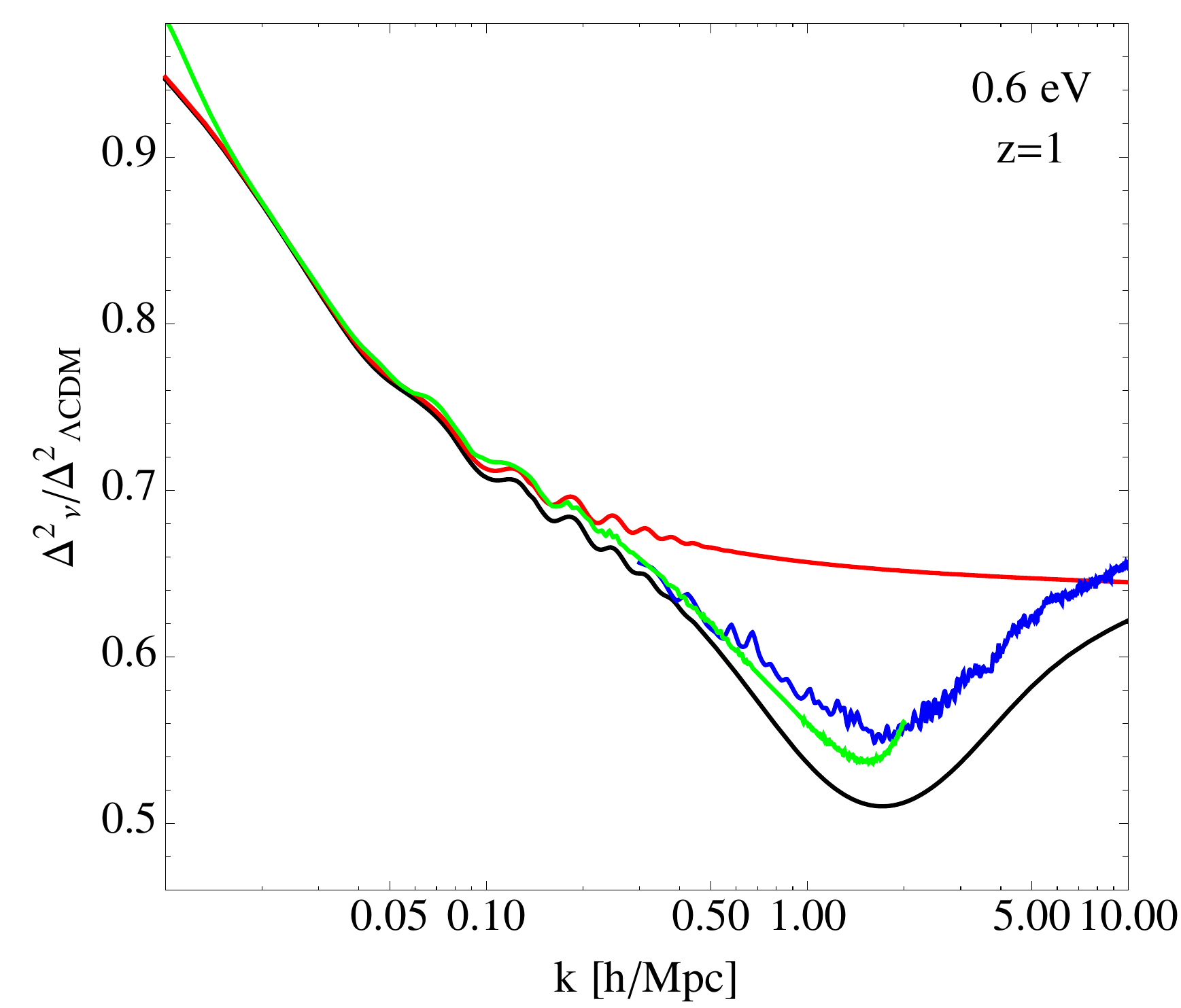}
\caption{\label{fig:6} Ratio $\Delta^2_{\nu}(k)/\Delta^2_{\Lambda \rm CDM}(k)$ for $\sum m_{\nu}=0.15$ eV (top), $\sum m_{\nu}=0.3$ eV (middle) and $\sum m_{\nu}=0.6$ eV (bottom). The left and right panels show results at redshifts $z=0$ and $z=1$, respectively. Black lines show the ratio computed from halo model, red lines show the linear predictions, blue and green lines are the results from N-boby simulations with box size $L=200$ Mpc/$h$ and $L=1000$ Mpc/$h$, respectively.}
\end{figure}

It is interesting to plot the ratio $\Delta^2_{\nu}(k)/\Delta^2_{\Lambda \rm CDM}(k)$, where the subscripts $\nu$ and $\Lambda \rm CDM$ indicate a massive and massless neutrinos cosmology, respectively. Figure~\ref{fig:6} shows this quantity for the $\sum m_{\nu}=0.15$ eV (top), $\sum m_{\nu}=0.3$ eV (middle)  and $\sum m_{\nu}=0.6$ eV (bottom) cosmological models at redshifts $z=0$ (left panels) and $z=1$ (right panels); we emphasize that here the unit scale varies for different cosmologies. The ratio obtained from N-body simulations presents a well known spoon-shape around $k\sim 1\,h$/Mpc \cite{Brandbyge_2008,Viel_2010, Agarwal2011, Bird_2011, Wagner2012}, which is not captured by linear theory. Interestingly, halo model reproduces this feature and can help us to understand its physical meaning.

First of all, we must remember that in the range $0.1<k\, (h/$ Mpc)$ <1$
there is the transition between the 1- and the 2-halo terms where they
are comparable, whereas on smaller scales, $k>1\,h$/Mpc, the 1-halo
term dominates. Then, in order to study in more detail the spoon-shape
trend, we can focus on the 1-halo term only. Moreover, on these scales the contribution of the neutrino and the cross power spectra to the total matter one is negligible. Therefore, in this analysis we consider just the cold dark matter power contributing to the 1-halo term of the total matter power spectrum in equation~\eqref{eq:19}. The 1-halo term accounts
for the correlations between particles that belong to the same halo,
therefore, only halos with size larger than the scale associated with the
given $k$ can contribute. This means that on intermediate scales only
relatively large halos give power to the 1-halo term, whereas for
$k>1\,h$/Mpc both small and large halos can in principle
contribute. However, the number of small halos is much larger than the
number of big ones; thus, on small scales the power comes primarily from
small halos.

\begin{figure}[tbp]
\centering 
\includegraphics[width=.48\textwidth,clip]{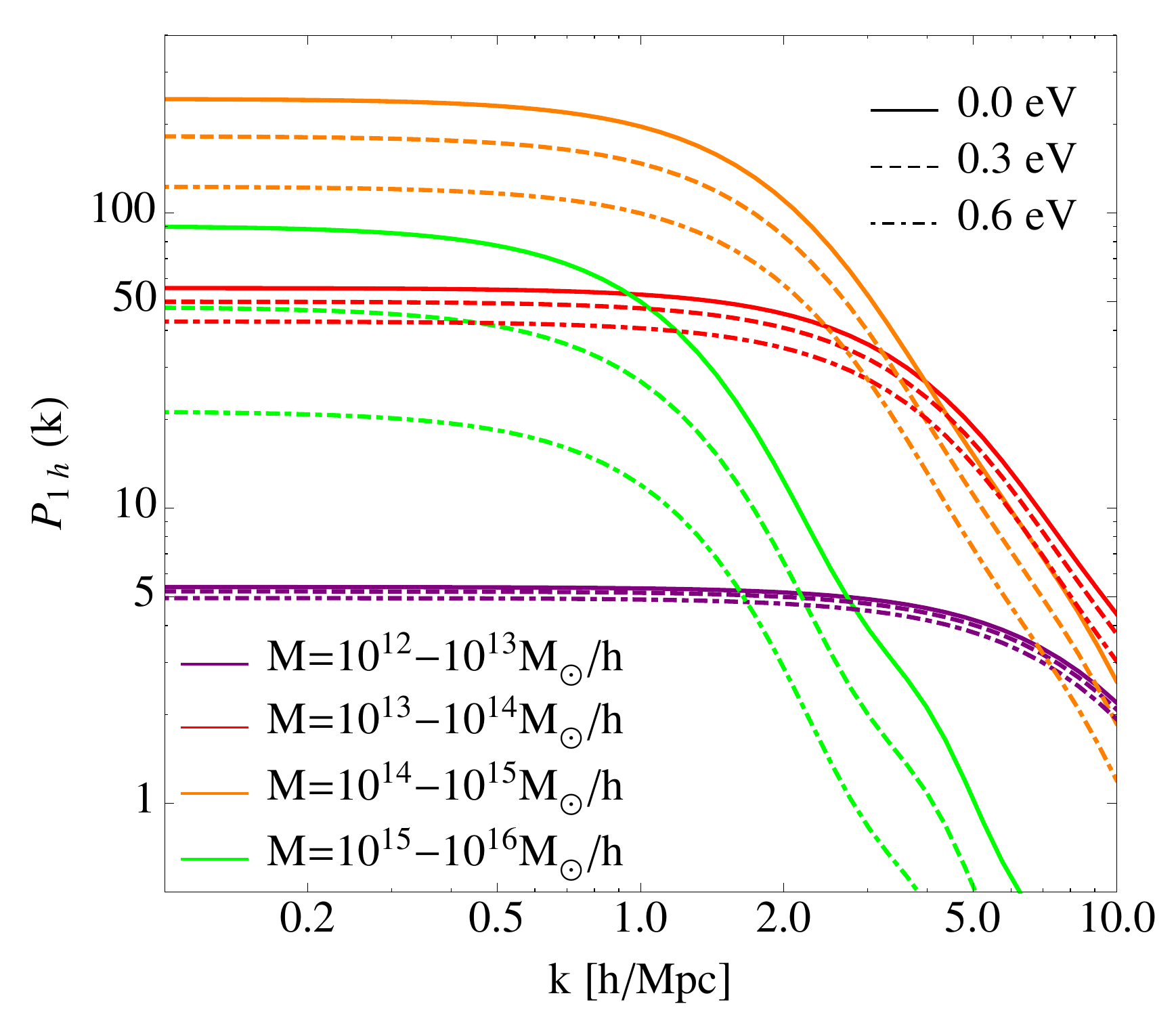}
\includegraphics[width=.48\textwidth,origin=c]{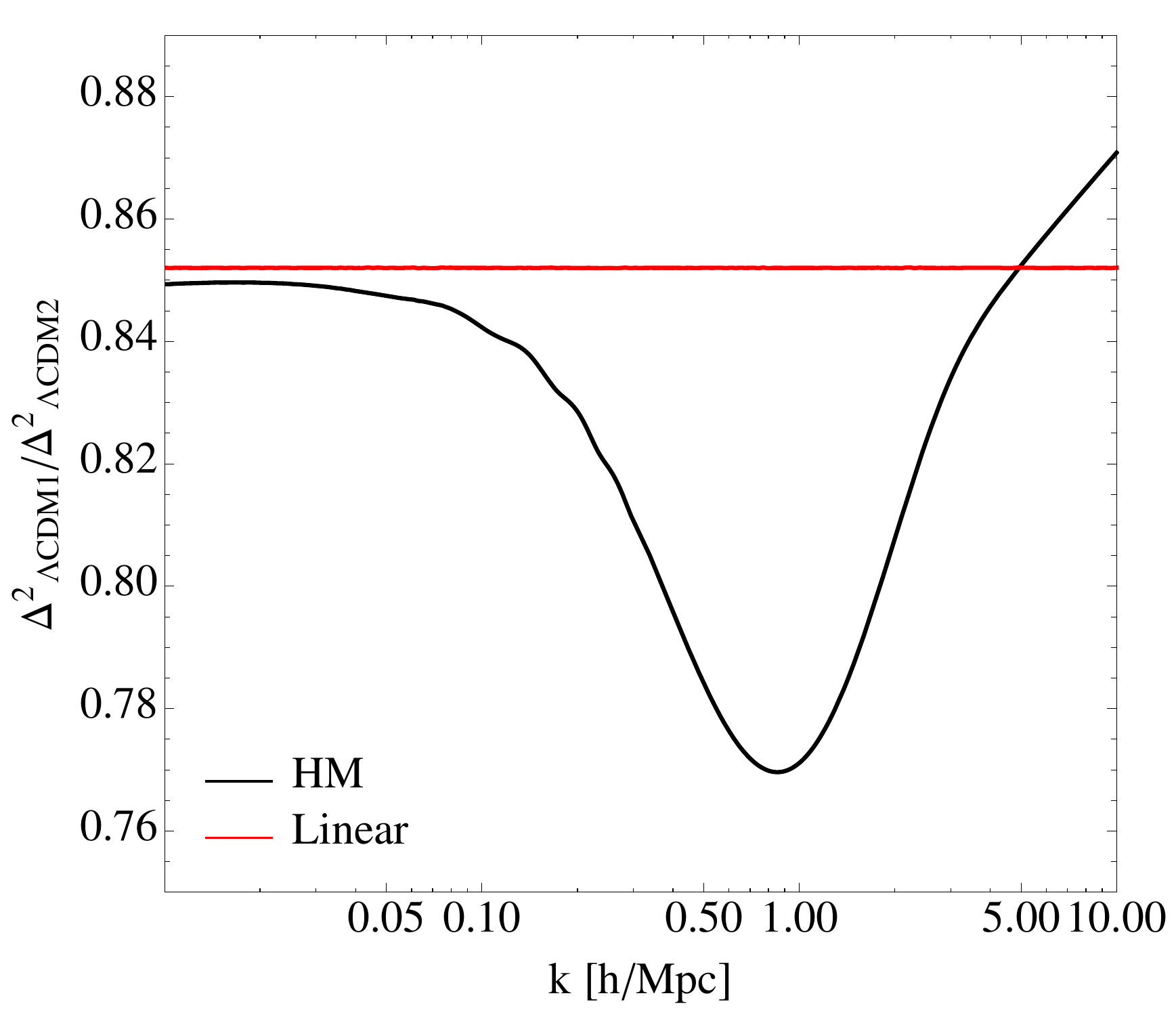}
\caption{\label{fig:7} Left panel: 1-halo term computed by integrating in different bins of mass. The different colors indicate different mass-bin, the solid, dashed and dot-dashed lines show the results for the $\sum m_{\nu}=0.0,\, 0.3,\,0.6$ eV cosmologies, respectively. Right panel: Ratio $\Delta^2_{\Lambda \rm CDM1}(k)/\Delta^2_{\Lambda \rm CDM2}(k)$ for two $\Lambda$CDM massless neutrinos cosmologies that differ only for the value of $\sigma_8$. The red line shows the linear prediction, the black one shows the results predicted by halo model, as described in section~\ref{sec:matter}.}
\end{figure}

The left panel of figure~\ref{fig:7} shows the 1-halo term (see equation~\ref{eq:37}) once the integral is computed for different mass-intervals, 
\begin{equation}
\label{eq:52}
P^{1h}_i(k)=\int_{\nu_{\rm c}(M_{\rm c}^{i})}^{\nu_{\rm c}(M_{\rm c}^{i}+\Delta M_{\rm c})} d\nu_{\rm c} \,f(\nu_{\rm c}) \frac{M_{\rm c}}{\bar{\rho}_{\rm c}}|u_{\rm c}(k|M_{\rm c})|^2\, ,
\end{equation}
for the $\sum m_{\nu}=0.0, 0.3,0.6$ eV cosmologies that we are considering in this paper, at redshift $z=0$. As we expect, small halo-masses give power at large $k$. What is more interesting is that the ratios $[P^{1h}_i(k)]_{\nu}/ [P^{1h}_i(k)]_{\Lambda\rm CDM}$ between 1-halo terms of massive and massless neutrino cosmologies computed in the same mass-bin $i$ are almost independent of $k$. Then, they can be well approximated by the ratios between the limits of $P^{1h}_i(k)$ on large scales,
\begin{equation}
\label{eq:53}
P^{1h}_i(k\rightarrow 0)=\int_{\nu_{\rm c}(M_{\rm c}^{i})}^{\nu_{\rm c}(M_{\rm c}^{i}+\Delta M_{\rm c})} d\nu_{\rm c} \,f(\nu_{\rm c}) \frac{M_{\rm c}}{\bar{\rho}_{\rm c}}\, ,
\end{equation}
which are independent of the halo profile.
This tells us that the main features of the spoon-shape are given by the mass function through the quantity $n(M_{\rm c}) M_{\rm c}^2$. In the four cosmologies considered in this work, this quantity is known to be quite similar for small halo-masses (around $10^{12}$M$_{\odot}$) and very different for big ones ($\,>10^{14}$M$_{\odot}$). 

We can now understand what creates the spoon-shape in $\Delta^2_{\nu}(k)/\Delta^2_{\Lambda \rm CDM}(k)$: the drop at intermediate scales, $0.1<k<1\,h$/Mpc, is due to the fact that the fraction of big halos is very different in the two cosmologies, whereas the rising comes from the fact that the fraction of small halos is very similar in the two cosmologies. In support of this, the right panel of figure~\ref{fig:7} shows that the spoon-shape is present also when the ratio is taken between two identical $\Lambda$CDM cosmologies, but with different $\sigma_8\equiv\sigma(R=8~\mbox{Mpc}/h)$. Indeed, this suggests that the spoon-shape is due to different relations between the peak height and the halo mass, which is what is needed to build the mass function.

Now that we understand the reason for this particular shape in the ratio of the matter power spectra, we want to stress the following point. Figure~\ref{fig:5} points out that halo model can reproduce the non-linear power spectrum from N-body simulations with $20\%$ accuracy at $z=0$ and $30\%$ accuracy at $z=1$. However, it works much better in predicting the ratio $\Delta^2_{\nu}(k)/\Delta^2_{\Lambda \rm CDM}(k)$, as figure~\ref{fig:6} demonstrates. In this case the disagreement between halo model and simulations is below $2\%,\,5\%,\,10\%$ for $\sum m_{\nu}=0.15,\, 0.3,\,0.6$ eV massive neutrinos cosmologies, respectively, at both redshifts and for the whole set of scales considered here ($k<10\,h/$Mpc).

\subsection{Halo model and {\sc HALOFIT}}
\label{sec:ratioHF}
Here we compare the predictions from our extension of the halo model against {\sc HALOFIT}~\cite{HALOFIT1}, which is a fitting formula that provides the non-linear power spectrum given a linear one and it is partially based on the halo model. The discrepancy between the new version of {\sc HALOFIT}~\cite{HALOFIT2} and N-body simulations is claimed to be below $10\%$ for $k<10\,h$/Mpc. This made {\sc HALOFIT} a useful and popular tool to compute the non-linear power spectrum, without running any simulation. Therefore we think that it is important to show a comparison also between our model for massive neutrino cosmologies and the extension of {\sc HALOFIT} presented by Bird et al. \cite{Bird_2011}.

\begin{figure}[tbp]
\centering 
\includegraphics[width=.48\textwidth,clip]{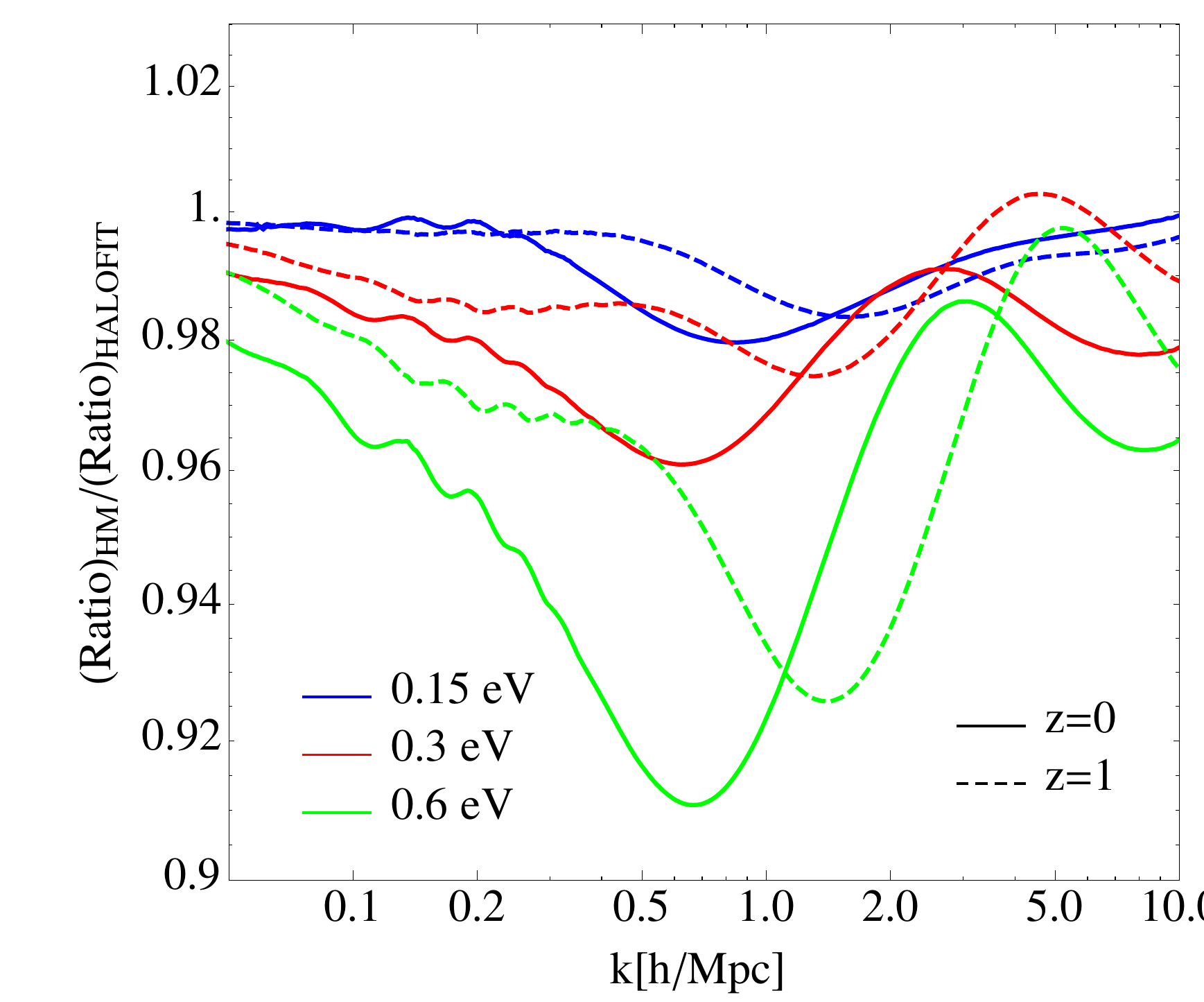}
\caption{\label{fig:8} Ratio between $\Delta^2_{\nu}(k)/\Delta^2_{\Lambda \rm CDM}(k)$ computed with halo model and HALOFIT. Different colors indicate different massive neutrino cosmologies: $\sum m_{\nu}=0.15$ eV in blue, $\sum m_{\nu}=0.3$ eV in red, $\sum m_{\nu}=0.6$ eV in green. Solid and dashed lines show results at $z=0$ and at $z=1$, respectively.}
\end{figure}

We compute the quantity $\Delta^2_{\nu}(k)/\Delta^2_{\Lambda \rm CDM}(k)$ with {\sc HALOFIT}, for all the cosmologies considered in this paper. The comparison with halo model is shown in Figure~\ref{fig:8}, where we plot the ratio between $\Delta^2_{\nu}(k)/\Delta^2_{\Lambda \rm CDM}(k)$ computed with halo model (see section~\ref{sec:ratiosim}) and {\sc HALOFIT}.
The disagreement is below $2\%$ for $\sum m_{\nu}=0.15$ eV, $4\%$ for $\sum m_{\nu}=0.3$ eV and $10\%$ for $\sum m_{\nu}=0.6$ eV.

\section{Galaxy clustering}
\label{sec:galaxy}

As an application of our halo model extension we study the clustering of galaxies in massless and massive neutrinos cosmologies. In Villaescusa et al. \cite{Paco_2013} authors populated with galaxies the dark matter halos of N-body simulations using a simple Halo Occupation Distribution (HOD) model. For a given cosmological model, the authors calibrated the values of the HOD parameters to reproduce the clustering properties of the galaxies in the main sample of the Sloan Digital Sky Survey (SDSS) II Data Release 7 \cite{Zehavi_2011}. Our purpose here is to use those HOD parameters and see whether our extension of the halo model is able to reproduce the clustering properties of the SDSS galaxies.

In this section we present the HOD model used in \cite{Paco_2013}, we describe the formalism needed to compute the galaxy clustering using the halo model  ingredients and we show the results for models with massless and massive neutrinos.

An HOD model requires two ingredients: 1) the probability distribution $p(N|M)$ of having $N$ galaxies inside a ${\rm c}$-halo of mass $M$ (in this section we drop the subscript "${\rm c}$" for indicating the cold dark matter field, since all the quantities related to halos corresponds to cdm ones) and 2) the way galaxy positions and velocities are related with those of the underlying matter within halos. The HOD model adopted in \cite{Paco_2013}, which has three free parameters, $M_{\rm min}$, $\alpha$ and $M_1$, works as follows. The first HOD ingredient is modeled assuming that halos with masses below $M_{\rm min}$ do not host any galaxy, whereas halos with masses above $M_{\rm min}$ host one central galaxy (c) and a number of satellites (s) following a Poissonian distribution with a mean equal to $(M/M_1)^\alpha$. Mathematically this can be written as
\begin{equation}
\label{eq:54}
\langle N_c|M \rangle =
\begin{cases} 1 & \mbox{if } M\ge M_{\rm min} \\ 
0 & \mbox{if }M<M_{\rm min}\, 
\end{cases}
\; \; \; \; \; \; \; \; \; \; \; \; \;
\langle N_s|M \rangle =
\begin{cases} (M/M_1)^{\alpha} & \mbox{if } M\ge M_{\rm min} \\ 
\qquad 0 & \mbox{if }M<M_{\rm min}\, .
\end{cases}
\end{equation}
The second ingredient of the HOD states that the central galaxy resides in the center of the halo whereas satellites follow the distribution of the underlying cold dark matter within the halo. The value of the HOD parameters, for the cosmological models with $\sum m_\nu=0.0,\, 0.3$ and 0.6 eV, obtained by \cite{Paco_2013} for galaxies with magnitudes $M_r - 5 \log_{10} h = -21.0$, are shown in table~\ref{tab:ii}.

\begin{table}
\begin{center}
\begin{tabular}{| c | c | c | c |}
\hline
$\sum m_{\nu}$ & $M_1$ & $\alpha$ & $M_{\rm min}$\\
(eV) &($M_{\odot}/h$) & &($M_{\odot}/h$)\\
\hline
$0.0$ & $1.15\times 10^{14}$ & $1.27$ & $5.33 \times 10^{12}$ \\
$0.3$ & $1.02 \times 10^{14}$ & $1.32$ & $4.91\times 10^{12}$\\
$0.6$ & $8.90 \times 10^{13}$ & $1.36$ & $4.47\times 10^{12}$\\
\hline
\end{tabular}
\caption{\label{tab:ii}Values of the HOD parameters, for two different cosmologies and for galaxies with magnitudes $M_r - 5 \log_{10} h = -21.0$ (from \cite{Paco_2013}).}
\end{center}
\end{table}

Given the above HOD model, and the values of the HOD parameters in table \ref{tab:ii}, we can 
compute the clustering of galaxies with magnitudes $M_r - 5 \log_{10} h = -21.0$ using the halo model. We begin depicting the required formalism. The 1-halo term of the halo model describes the correlation between particles belonging to the same halo. Therefore, it must be proportional to the average number of galaxy pairs $\langle N(N-1)|M \rangle$ in a halo of mass $M$, where $N=N_c+N_s$ indicates the total number of galaxies. This quantity can be written in terms of central and satellite galaxies as
\begin{equation}
\label{eq:56}
\langle N(N-1)|M \rangle F(r) =2\langle N_cN_s|M \rangle F_{cs}(r)+\langle N_s(N_s-1)|M \rangle F_{ss}(r)\, ,
\end{equation}
where $F(r)$ is the cumulative radial distribution of galaxy pairs and $F_{cs}(r)$ and $F_{ss}(r)$ are restricted to central-satellite and satellite-satellite pairs, respectively. Since the above HOD model assumes that the central galaxy is located in the halo center and that the distribution of satellites follow the underlying CDM, $F_{cs}(r)$ is given by the normalized NFW profile and $F_{ss}(r)$ is a convolution of two normalized NFW profiles. The term $\langle N_s(N_s-1)|M \rangle$ can be simplified taking into account that satellites follow a Poisson distribution, i.e. $\langle N_s(N_s-1)|M \rangle=\langle N_s|M \rangle^2$, while $\langle N_cN_s|M \rangle=\langle N_c|M\rangle\langle N_s|M \rangle$ because the occupation number of central and satellites are independent. Given the partition in centrals and satellites, the galaxy power spectrum can be written as
\begin{eqnarray}
\label{eq:57}
P_{gg}(k)&=&P_{cc}(k)+2P_{cs}(k)+P_{ss}(k)\\\nonumber
&=&\left[2P_{cs}(k)+P_{ss}(k)\right]_{1h}+\left[P_{cc}(k)+2P_{cs}(k)+P_{ss}(k)\right]_{2h}\\\nonumber
&=&P^{1h}_{gg}(k)+P^{2h}_{gg}(k)\, ,
\end{eqnarray} 
where the last two equations have been written as sums of the correspondent 1- and 2-halo terms. Notice that a halo can have at most one central galaxy and therefore $P_{cc}(k)=P_{cc}^{2h}(k)$. In analogy to~\eqref{eq:5} and~\eqref{eq:6} and using the decomposition in~\eqref{eq:56}, where the central-satellite term is multiplied by the NFW profile and the satellite-satellite is multiplied by the same term squared, the 1- and 2-halo terms of the galaxy power spectrum become
\begin{eqnarray}
\label{eq:58}
P_{gg}^{1h}(k)&=&\int_0^\infty dM \, n(M) \left[2\frac{\langle N_c N_s|M \rangle}{\bar{n}_g^2}u(k|M)+\frac{\langle N_s(N_s-1)|M \rangle}{\bar{n}_g^2}u^2(k|M)\right]\\\nonumber
&=&\int_0^\infty dM \, n(M) \left[2\frac{\langle N_c|M \rangle\langle N_s|M \rangle}{\bar{n}_g^2}u(k|M)+\frac{\langle N_s|M \rangle^2}{\bar{n}_g^2}u^2(k|M)\right]\\
\label{eq:59}
P_{gg}^{2h}(k)&=&\left\{ \int_0^\infty dM n(M) b(M)\left[\frac{\langle N_c|M \rangle}{\bar{n}_g} +\frac{\langle N_s|M \rangle}{\bar{n}_g} u(k|M) \right] \right\}^2 P^L(k)\, ,
\end{eqnarray}
where the convolution of NFW profiles in configuration-space has become multiplications in Fourier-space. Using the cold dark matter prescription (equations~\eqref{eq:30}-\eqref{eq:33}), we write these terms as a function of the peak height $\nu$:
\begin{eqnarray}
\label{eq:60}
P_{gg}^{1h}(k)&=&\int_0^\infty d\nu \, f(\nu)\frac{\bar{\rho}}{M} \left[2\frac{\langle N_c|M \rangle\langle N_s|M \rangle}{\bar{n}_g^2}u(k|M)+\frac{\langle N_s|M \rangle^2}{\bar{n}_g^2}u^2(k|M)\right]\\
\label{eq:61}
P_{gg}^{2h}(k)&=&\left\{ \int_0^\infty d\nu \, f(\nu)\, b(\nu)\frac{\bar{\rho}}{M} \left[\frac{\langle N_c|M \rangle}{\bar{n}_g} +\frac{\langle N_s|M \rangle}{\bar{n}_g} u(k|M) \right] \right\}^2 P^L(k)\, .
\end{eqnarray}
We compute the galaxy power spectrum for the massless $\Lambda$CDM and the $\sum m_\nu=0.3,\, 0.6$ eV massive neutrino cosmologies, using~\eqref{eq:60} and~\eqref{eq:61}. We calculate the mean occupation numbers $\langle N_c|M \rangle$ and $\langle N_s|M \rangle$ as described in~\eqref{eq:54}, using the HOD parameters $M_1,\,\alpha,\,M_{min}$ from Villaescusa-Navarro et al.~\cite{Paco_2013} (see table \ref{tab:ii}).  
\begin{figure}[tbp]
\centering 
\includegraphics[width=.50\textwidth,clip]{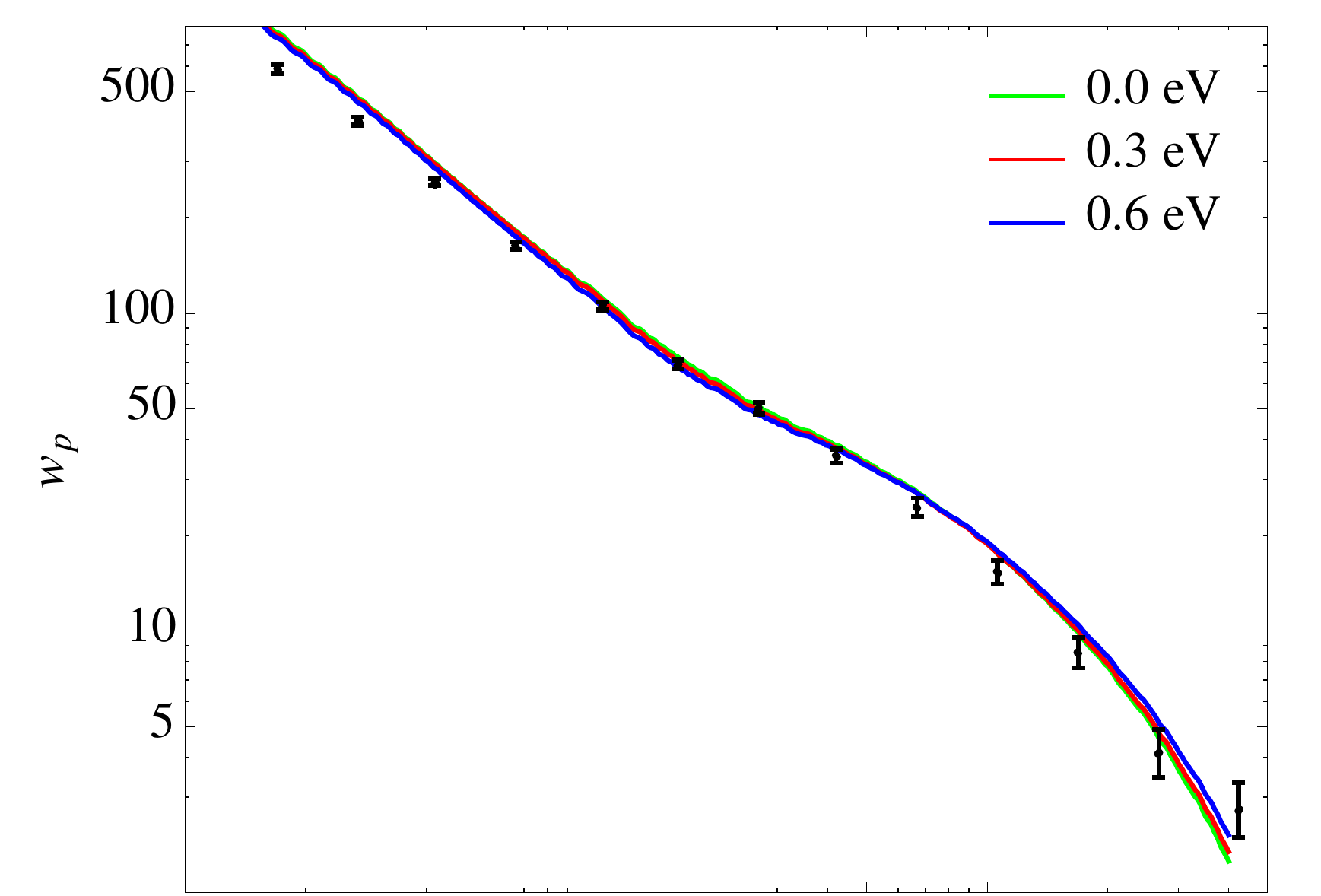}
\includegraphics[width=.50\textwidth,clip]{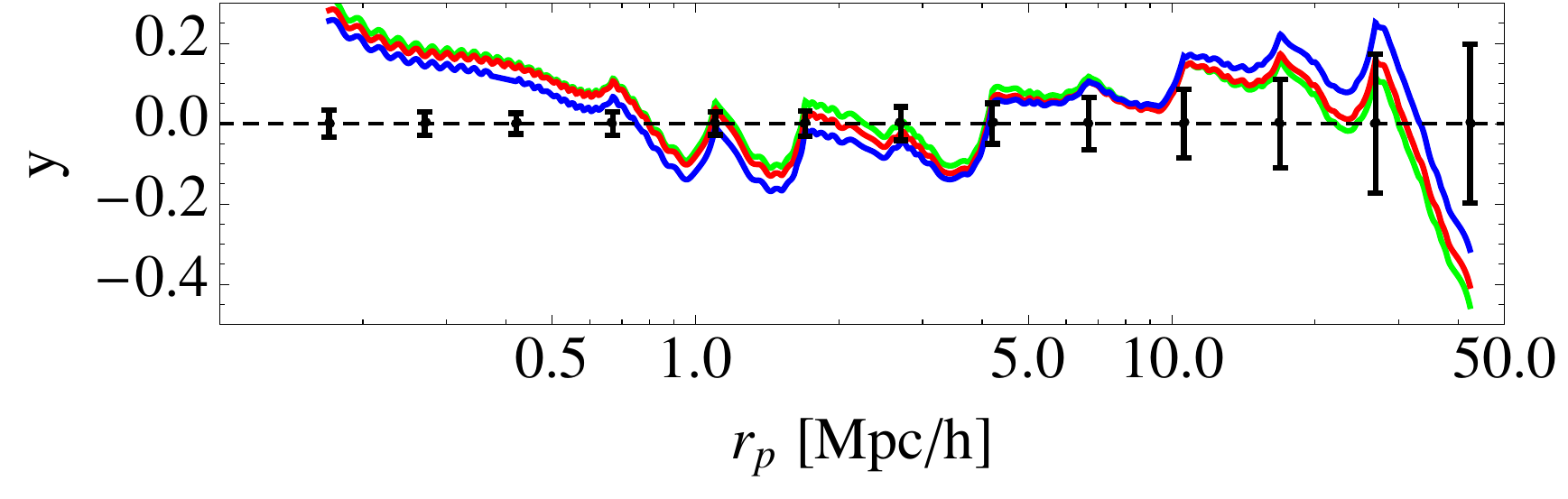}
\caption{\label{fig:9} Projected correlation function. Black dots are the $w_p$ measurements for galaxies with $M_r - 5 \log_{10} h = -21.0$ from Zehavi et al.~\cite{Zehavi_2004} and the error bars are the diagonal terms of the covariance matrix. Green, red and blue lines show the predictions from halo model for $\sum m_{\nu}=0.0,\,0.3,\,0.6$ eV cosmologies. The bottom panel shows the relative difference between them and the measurements.}
\end{figure}

Having calculated the galaxies power spectrum using the above formalism, we then compute the galaxies correlation function
\begin{equation}
\label{eq:62}
\xi_{gg}(r)=\int_0^{\infty} dk\, k^2 \, \frac{\sin(kr)}{kr}\,P_{gg}(k)\, .
\end{equation}
However, this quantity cannot be measured directly from galaxy surveys because of the unknown peculiar motion of the galaxies along the line-of-sight. What can be computed from observations is the redshift-space correlation function $\xi(r_p,r_{\pi})$, which is a function of the redshift-space separations parallel ($r_{\pi}$) and perpendicular ($r_{p}$) to the line-of-sight. In order to compare the model with observations we must consider the projected correlation function, $w_p(r_p)$, which is defined as
\begin{equation}
w_p(r_p)\equiv 2 \int_0^{r_{max}}dr_{\pi}\,\xi(r_p,r_{\pi})\, ,
\end{equation}
and it is related to the galaxy correlation function in configuration-space through~\cite{Davis_Peebles}
\begin{equation}
w_p(r_p)= \int_{r_p}^{\infty}dr\, \frac{2r}{\sqrt{r^2-r^2_p}}\,\xi_{gg}(r)\, .
\end{equation}
Figure~\ref{fig:9} shows the projected correlation function predicted by our model, together with the one measured by Zehavi et al.~\cite{Zehavi_2004}. As we see from the relative difference between model and observations in the bottom panel, both cosmologies reproduce very well the measurements for $r_p >1$ Mpc/$h$. This result confirms that the calibration of the HOD
parameters can be carried out, for massive neutrino cosmologies, using the above formalism together with our extension of the halo model. Since direct calibration of the HOD parameters with N-body simulations is difficult, CPU time consuming and its subject to resolution and cosmic variance, the above formalism is fast and precise, allowing us to explore a wider parameter space.

We conclude this section noticing that the effect of massive neutrinos on many cosmological observables, such as galaxy clustering, can be mimicked by varying the value $\sigma_8$ from a massless neutrino cosmology. This is the well known $\Omega_\nu-\sigma_8$ degeneracy (see for instance \cite{Marulli_2011,Fontanot_2014}). Our formalism is capable of reproducing such degeneracy at the same time it provides us with a physical insight.

\section{Conclusions}
\label{sec:conclusions}

The purpose of the present work has been to extend the halo model to
account for the effects of massive neutrinos. We have run a set of 8
large box-size N-body simulations containing massive neutrinos as
additional particles. Our simulation suite comprises four different
cosmological models with different neutrino masses: $\sum m_\nu=0.0$,
0.15, 0.30 and 0.60 eV. For each model we have run two different
simulations with two different box-sizes, in order to extract the
 power spectra over a wide range of wave numbers.

We have reviewed the standard framework of the halo model and used it
to compute the fully non-linear matter power spectrum for the massless
$\Lambda$CDM cosmology considered in this paper (see table
\ref{tab:i}). The comparison with the matter power spectrum from the
N-body simulations showed a very good agreement on large and small
scales, whereas a disagreement at the 20$\%$ (at $z=0$) and 30$\%$ (at
$z=1$) level is present on intermediate scales,
$k\sim0.2-2\,h$/Mpc. These scales represent the transition between the
1-halo and 2-halo terms, where the halo model is not very accurate.

We then focused on cosmologies with massive neutrinos, where the total
matter power spectrum can be expressed as the mass-weighted sum of
three different power spectra: CDM auto-power spectrum, neutrinos
auto-power spectrum and CDM-neutrinos cross-power spectrum. Thus, in
our extension of the halo model we need to model separately the
density field of both CDM and massive neutrinos.

The CDM density field is modeled in the same spirit the halo model
describes the distribution of matter in a massless neutrino cosmology:
all CDM is bound within ${\rm c}$-halos (CDM halos). A key ingredient is
to account for the fact that the clustering properties of the
${\rm c}$-halos depend only on the CDM field
\cite{Castorina_2013,Ichiki_Takada} (this is called the cold dark
matter prescription), thus, both the mass function and the halo bias
are computed using the linear CDM power spectrum. We find that halo
model is able to reproduce the cold dark matter power spectrum from
simulations on large and on small scales with $10\%$ accuracy. As the
standard halo model, on intermediate scales, $k\sim0.2-2~h$/Mpc, it
presents a disagreement at the $20\%$ level at $z=0$ and at the $30\%$
level at $z=1$, for the three massive neutrino cosmologies considered
here.

The neutrino density field can not be modeled in the same terms as the
CDM field, since the neutrinos large thermal velocities prevent their
clustering within ${\rm c}$-halos. In other words, the neutrino density
field can not be described as the distribution arising assuming that
all neutrinos are within ${\rm c}$-halos. Thus, we have modeled the neutrino
density field as the sum of a linear and a clustered component, differently from what present in literature (see~\cite{Abazajian_2005}). We
find that our model is capable of reproducing the CDM-neutrinos
cross-power spectra from simulations at the $30\%$ level until $k\sim
1\,h$/Mpc for the $\sum m_{\nu}=0.3$ eV cosmology and at the $40\%$
level on scales $k< 5\,h$/Mpc for the $\sum m_{\nu}=0.6$ eV model. The results for the neutrino
power spectrum are similar: there is a disagreement below $20\%$ until
$k\sim0.7~h$/Mpc for the $\sum m_{\nu}=0.3$ eV case and under $30\%$
until $k\sim1.5~h$/Mpc for the $\sum m_{\nu}=0.6$ eV case.  We
emphasize that in order to have a good description of both the cross
and the neutrinos auto-power spectrum we need to account for the tiny,
fully non-linear, clustering of neutrinos within ${\rm c}$-halos.

As stated above, the total matter power spectrum is a mass-weighted
sum of the cold dark matter, neutrino and cross power
spectra. However, we noticed that, while the clustered neutrino
component is important to determine the cross and the neutrino power
spectra at small scales, it is negligible in the computation of the
total matter power spectrum. Therefore, in order to compute the total
matter power spectrum in massive neutrinos cosmologies an excellent
assumption is that neutrinos follows linear theory and that the cross power spectrum is the linear one. Again, we find that our
model is capable of reproducing the total matter power spectrum from
simulations on large and small scales within a $10\%$, whereas in
intermediate scales, the most challenging for the halo model, it can
reproduce the results of the N-body simulations within a
$\sim20-30\%$.

We stress that the cold dark matter prescription must be taken into
account for computing the mass function and the halo bias. Our
analysis showed that in fact the agreement between the extended halo
model and N-body simulations is better when using the cold dark matter
prescription rather than the matter one.

We have computed the ratio between the matter power spectrum of a
massive and massless neutrinos cosmology,
$\Delta^2_{\nu}(k)/\Delta^2_{\Lambda \rm CDM}(k)$. Linear theory fails
to explain the spoon-shape trend present in that ratio around $k\sim 1
~h$/Mpc, whereas our model succeeds in doing it. In fact, the
disagreement between our predictions for the ratio
$\Delta^2_{\nu}(k)/\Delta^2_{\Lambda \rm CDM}(k)$ and simulations is
below $2\%,\,5\%,\,10\%$ for $\sum m_{\nu}=0.15,\, 0.3,\,0.6$ eV
cosmologies, at both $z=0$ and $z=1$ and over the whole range of
scales investigated. This result shows that our extension of the halo
model, even if it is able to reproduce the matter power spectrum only
with a $20-30\%$ accuracy, works much better when predicting the ratio
$\Delta^2_{\nu}(k)/\Delta^2_{\Lambda \rm CDM}(k)$. Moreover, we have
used the halo model to understand the spoon-shape in the ratio of the
two matter power spectra. We find that that feature is just a
consequence of the differences in the mass function between the two
cosmologies, and is not a unique feature of massive neutrino
cosmologies since is also present in the ratio between two massless
neutrino cosmologies with different values of $\sigma_8$.

We have also compared our model against {\sc HALOFIT}, when predicting the
ratio $\Delta^2_{\nu}(k)/\Delta^2_{\Lambda \rm CDM}(k)$. We conclude
that the disagreement is below $2\%$ for $\sum m_{\nu}=0.15$ eV, $4\%$
for $\sum m_{\nu}=0.3$ eV and $10\%$ for $\sum m_{\nu}=0.6$ eV.

Finally, we have investigated the clustering of galaxies, in massless and massive neutrinos cosmologies, using a simple HOD model and our halo model extension. We computed the projected correlation function of galaxies with magnitudes $M_r - 5 \log_{10} h = -21.0$ taking the HOD parameters calibrated by Villaescusa-Navarro et al. \cite{Paco_2013}. We find an excellent agreement between the galaxy clustering predicted by our model and the SDSS observations against with the values of the HOD parameters were calibrated. This result points out that our extension of the halo model can be used to calibrate the HOD parameters in cosmologies with massive neutrinos. Whereas the calibration of the HOD parameters directly from N-body simulations is computationally very demanding and is subjected to problems like resolution and cosmic variance, the halo model provides a fast, and very accurate framework to carry out this task.

Overall, the neutrino halo model presented in this paper is a simple
tool to quantitatively address non-linearities induced by neutrinos, which has also the big advantage
of offering physical insights on the relative interplay between neutrino
and cold dark matter fluids around virialized structures.

\acknowledgments We thank Ravi Sheth, Emanuele Castorina and Emiliano
Sefusatti for useful discussions.  N-body simulations have been run in
the Zefiro cluster (Pisa, Italy).  FVN and MV are supported by the ERC
Starting Grant ``cosmoIGM'' and partially supported by INFN IS PD51
"INDARK".  We acknowledge partial support from "Consorzio per la
Fisica - Trieste".

\bibliographystyle{JHEP}
\bibliography{Bibliography}

\end{document}